\documentclass[10pt]{book}
\usepackage{amsmath,amsopn,amsthm,amscd,amssymb,xypic,rotating}

\renewcommand{\familydefault}{cmtt}
\raggedright
\xyoption{all}

\newcommand{\VInsert}[2]   {\centerline{\immediate\pdfximage height #2 {#1}\pdfrefximage\pdflastximage}}
\newcommand{\HInsert}[2]   {\centerline{\immediate\pdfximage width #2  {#1}\pdfrefximage\pdflastximage}}

\newtheorem{theorem}{Theorem}[section]

\def\em{\sl}

\newcommand{\cb}       {\begin{tabbing}MM\=MM\=MM\=MM\=MM\=MM\=MM\=MM\=MM\=MM\= \kill}
\newcommand{\ce}         {\end{tabbing}}
\newcommand{\cmd}[1]      {\underline{{#1}}}

\newcommand{\tset}[1]      {\{{#1}\}}                     
\newcommand{\tbag}[1]      {\{\!\!|{#1}\}\!\!\!|}         
\newcommand{\tlist}[1]     {[{#1}]}                       

\newcommand{\double}       {\baselineskip 13pt}
\newcommand{\single}       {\baselineskip 11pt}

\newcommand{\separate}     {\vspace{0.3cm}\begin{center}*~~~~~~~~~~*~~~~~~~~~~*\end{center}\vspace{0.3cm}}

\newcommand{\dqt}[1]        {"{#1}"}
\newcommand{\bp}            {~~~~~~~~~~~~~~~~~}

\def\emph{\textsl}
\def\textbf{\pmb}

\newcommand{\LInsert}[2]   {{\immediate\pdfximage width #2  {#1}\pdfrefximage\pdflastximage}}

\newcommand{\ctitle}{A random walk on Area Restricted Search}
\newcommand{\cauthor}{Simone Santini}
\newcommand{\caffil}{Escuela Polit\'ecnica Superior\\Universidad Aut\'onoma de Madrid}
\newcommand{\cnumber}{\textbf{XXVI}}
\newcommand{\copyrgt}{2019}

\def\capstyle{\small}
\def\shortcite{\cite}
\def\dfeq{\stackrel{\triangle}{=}}
\def\eoe{\\\hfill(end of example)\\\bigskip}
\def\eor{\\\hfill(end of remark)}
\newcommand{\vct}[1]       {\mathbf{{#1}}}

\newcounter{myremark}
\newcounter{myexample}
\def\example{
\bigskip

\refstepcounter{myexample}%
\noindent \textbf{Example \Roman{myexample}:}\\
}

\def\remark{
\refstepcounter{myremark}%
\noindent \emph{Remark \arabic{myremark}:}
}

\renewcommand{\topfraction}{.99}
\renewcommand{\textfraction}{.01}

\begin{document}

\thispagestyle{empty}

$ $

\setlength{\unitlength}{10mm}
\parbox{22cm}{
\vspace{-1cm}
\hspace{-2.3cm}
\begin{picture}(22,2)(0,0)
\newsavebox{\densebox}
\newsavebox{\rarebox}
\savebox{\densebox}{
\multiput(0,0)(0.2,0){5}{
   \multiput(0,0)(0,0.2){5}{\circle*{0.0001}}
  }
}
\savebox{\rarebox}{
\multiput(0,0)(0.1,0){10}{
\multiput(0,0)(0,0.1){10}{\circle*{0.0001}}
}
}
\multiput(-1,0)(0,1){4}{
    \multiput(0,0)(1,0){22}{\usebox{\densebox}}
}

\put(2,2){\makebox(0,0)[l]{\Huge\sf cahiers d'informatique}}

\put(-1,-22){
\multiput(0,0)(1,0){22}{\usebox{\densebox}}
\multiput(0,6)(1,0){22}{\usebox{\densebox}}
\multiput(0,1)(0,4){2}{
    \multiput(0,0)(1,0){12}{\usebox{\densebox}}
    \multiput(12,0)(1,0){4}{\usebox{\rarebox}}
    \put(16,0){\usebox{\densebox}}
    \put(17,0){\usebox{\rarebox}}
    \multiput(18,0)(1,0){4}{\usebox{\densebox}}
}
\multiput(0,2)(0,1){3}{
   \multiput(0,0)(1,0){12}{\usebox{\densebox}}
   \put(12,0){\usebox{\rarebox}}
   \multiput(13,0)(1,0){4}{\usebox{\densebox}}
   \put(17,0){\usebox{\rarebox}}
   \multiput(18,0)(1,0){4}{\usebox{\densebox}}
}
}
\put(0,-14){\Large\sf\begin{turn}{90}cuadernosdeinform\'aticacomputingnotebooksquadernid'informatica\end{turn}}
\end{picture}
}

\vspace{4cm}
\begin{center}
\baselineskip 25pt

{\Huge\sc \ctitle}

\baselineskip 11pt

\vspace{2cm}

{\Large\sc \cauthor}

{\Large \sc \caffil}

\vspace{2cm}

{\Large\sc \cnumber}
\end{center}

\newpage

$ $ 

\vspace{16cm}
\begin{center}
\parbox{10cm}{\small The title of this series of technical reports constitute
my homage to the legendary magazine \emph{Cahiers du cin\'ema}, and to the 
role that it has played for many years in the creation of a widespread
critical sense regarding film.
}

\vspace{0.5cm}

{\small (C) Simone Santini, \copyrgt}
\end{center}

\newpage

\double
$ $

\vspace{10em}
\begin{center}
  \parbox{30em}{ These notes constituted the class material for part
    of the syllabus of the graduate course \dqt{Caracterizaci\'on de
      Redes y Topolog{\'\i}as Biol\'ogicas} read during the academic
    years 2017-18 and 2018-19. I gratefully acknowledge the
    contribution of all the students of these classes who, with their
    interaction, have contributed to improve these notes.}

  \vspace{10em}

  Author contact: \textbf{simone.santini@uam.es}
\end{center}

\newpage
$ $

\newpage
$ $

\vspace{2cm}

\section{\sc ARS, reward, and dopamine in the evolution of life}
\begin{figure}[bth]
  \begin{center}
    \VInsert{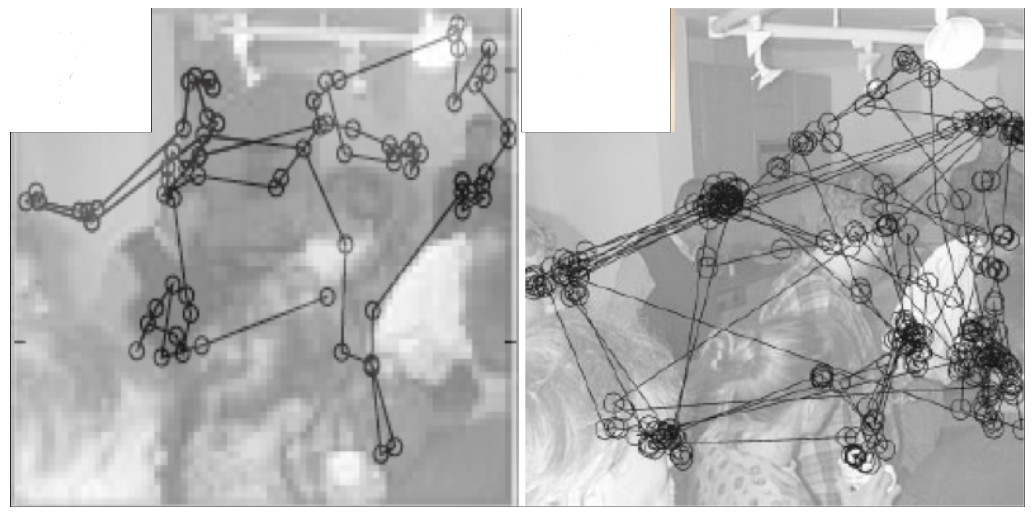}{0.3\columnwidth}
  \end{center}
  \caption{\capstyle Two images that, apparently, describe the same
    phenomenon. While the paths displayed in the two images are
    qualitatively similar, their origin is quite different. The path
    on the right is a recording of the saccadic eye movements of a
    person looking at the image to which it is superimposed. The path
    of the left is that of a spider monkey (a monkey common in
    the Yucatan peninsula of Mexico) looking for food. The two paths
    are example of a common behavior found throughout the animal
    kingdom: \emph{Area-Restricted Search}.}
  \label{wow}
\end{figure}

The starting point of these notes is a talk to which I heard few
years back in Milan, Italy. The speaker, Giuseppe Boccignone, showed
us the pair of images in Figure~\ref{wow}, which represent two paths
in two dimensions. It is not hard to see that the macroscopic
characteristics of these images are quite the same. Their most evident
macroscopic feature is that they are heavily clustered: there are a
bunch of points in a very restricted area and then suddenly the path
jumps to a different area where a new cluster of points is created.
The images are so similar that one would have little trouble believing
that they have been created by two instances of the same physical
phenomenon. Yet, much to my amazement, Boccignone told us that this
was not the case.  The figure on the right shows the saccadic eye
movements of a person looking at the picture that you can faintly see
in the background; the picture on the left is the path followed by a
spider monkey of the Yucatan peninsula while looking for food. It has
nothing to do with the picture and was superimposed to it only to make
the point more forcefully.

To find such a similarity in completely unrelated activities of two
different species is as striking as it would be to find out that the
ritualistic chant of a remote tribe in Papua has the same harmonic
structure as the Goldberg Variations. Just as in the case of the tribe
we would like to look for an explanation (maybe a previous contact
with some Bach-loving explorer that has been incorporated into the
rituals of the tribe), so in this case it is not too far-fetched to
start looking for some common underlying mechanism.

We are encouraged in our endeavor by the fact that the two behaviors
do have indeed something in common: they are both examples of
\emph{search}. Search for visual information in one case, search for
food in the other. So, we are on a hunt for a common mechanism that
guides \emph{search} in a wide variety of species under the most
diverse circumstances. The mechanism must be very general, since it
should apply not only to different species but also to very different
levels of abstraction (from search for food in physical space to
search for information in conceptual space).

The behavior that we observe in these two examples is commonly known
as \emph{ARS (Area Restricted Search)}, a strategy that consists in
\dqt{a concentration of searching effort around areas where rewards in
  the form of specific resources have been found in the past. When
  resources are encountered less frequently, behavior changes such
  that the search becomes less sensitive, but covers more space}
\shortcite{hills:06}.

\begin{figure}[bth]
  \HInsert{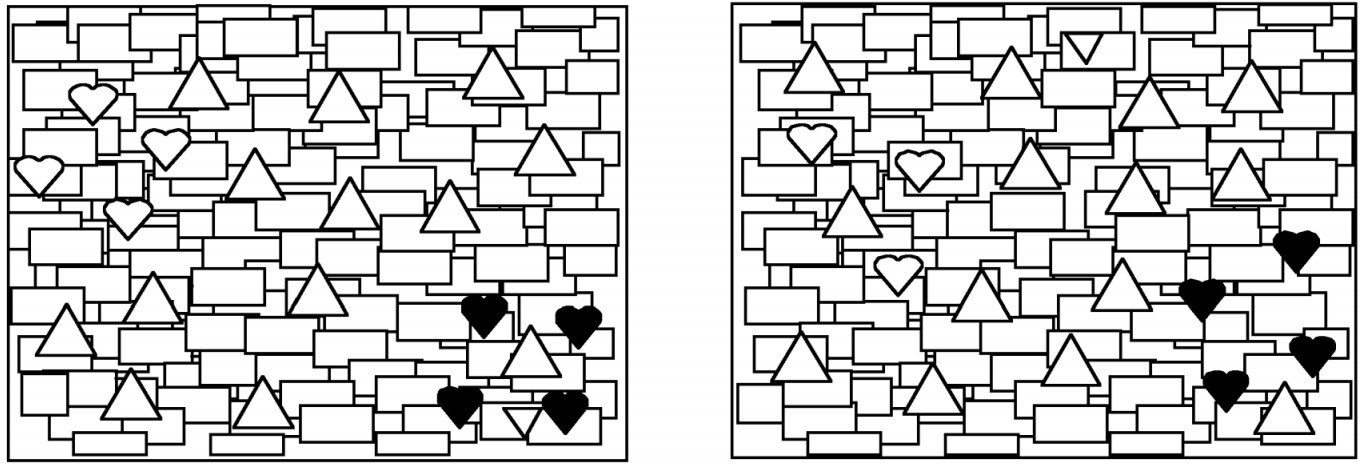}{\columnwidth}
  \caption{\capstyle Paying attention to where your eyes look, begin
    in the left figure and look for the upside-down triangle (there is
    only one). Once you have found it, move to the figure to the right
    and look for the upside-down triangle there. Go ahead and do that
    now, before you read the rest of the caption.  Where did you look
    first when you were looking for the triangle in the second figure?
    Did you look first near the black hearts? If you did, then you
    were performing ARS.  You focused the attention in the area where
    you expected (based on your previous experience) the reward (the
    upside-down triangle) to be found and then, once you found out
    that your \dqt{confidence} area did not have the resource you were
    looking for, you started a rapid scan of the rest of the image,
    until you found the sought-after triangle.}
  \label{ARS-exp}
\end{figure}

\begin{figure}
  \begin{center}
    {\small
    $\displaystyle
    \xymatrix@R=1em@C=4em{
      & & \mbox{metazoa} \ar@{-}[dr] \ar@{-}[dl] \\
      & \mbox{porifera} & & \mbox{eumetazoa} \ar@{-}[dr] \ar@{-}[dl] \\
      &        & \mbox{bilatera} \ar@{-}[dll] \ar@{-}[dl] \ar@{-}[d] \ar@{-}[drr] & & \mbox{radiata} \\
\mbox{lophotrochozoa} \ar@{-}[dd]& \mbox{platyzoa} & \mbox{ecydozoa} \ar@{-}[dr] \ar@{-}[dl] \ar@{-}[d] & & \mbox{deuterostomia} \ar@{-}[d] \\
                      &  {} \ar@{..}[r] & \mbox{nemotoda}\ar@{..}[r] & {} & \mbox{chordata} \ar@{-}[dr] \ar@{-}[dl] \ar@{-}[d] \\
(mollusks)      &       &   & {} \ar@{..}[r] & \mbox{craniata} \ar@{-}[dr] \ar@{-}[dl] \ar@{..}[r] & {} \\
      &       &   &  \mbox{vertebrata} & & \mbox{myxini}
    }
    $
    }
  \end{center}
  \caption{\capstyle A (very small) fragment of the taxonomy of \emph{metazoa}
    (i.e.\ animals). The group of \emph{porifera} is composed of
    animals without tissue, and consists pretty much of sponges and
    little else. The clade of the eumetazoa contains all other
    animals, and in this whole clade ARS has been
    observed.}
  \label{taxonomy}
\end{figure}
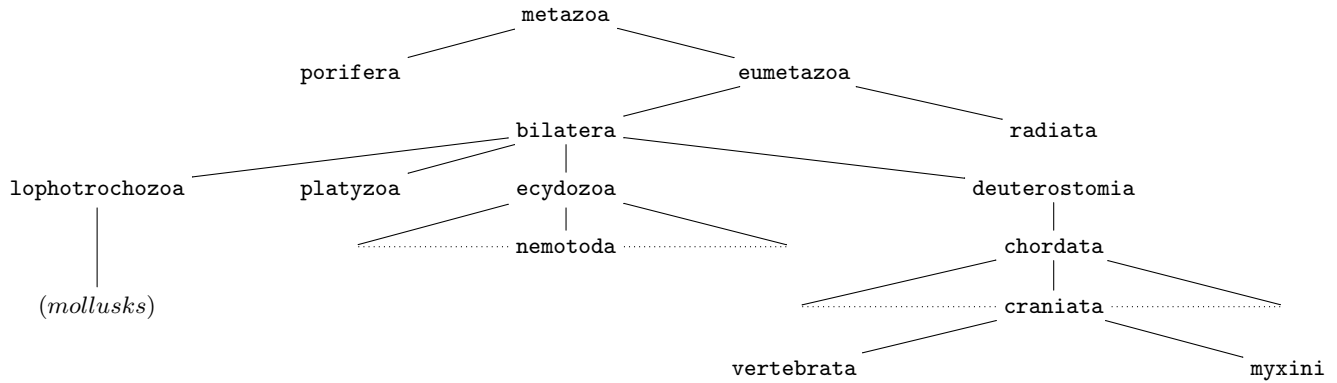

As we shall see, the same basic mechanism permits ARS in a variety of
cases and circumstances, from the foraging behavior of the nematode
\emph{C.elegans} to goal-directed cognition in people. You can have a
personal experience of ARS by looking at Figure~\ref{ARS-exp} and
following the instructions in the caption (read the caption before
looking at the pictures).
ARS is incredibly widespread. Some form of it has been found in all
major eumetazoan clades. To have an idea of what this entails, in
Figure~\ref{taxonomy} I have drawn a very partial taxonomy of the
animal kingdom.
The clade on the left of the root, the \emph{porifera} is composed of
animals that do not have a real tissue: sponges and little else. The
other clade, \emph{eumetazoa} contains all other animals, from worms
to mollusks to you and me. ARS can be observed, in some form or
another, in the whole eumetazoa clade. This broad presence indicates
that the mechanism behind ARS must have evolved quite early, since the
major divisions of the eumetazoa clade are very ancient, and it is
reasonable to assume that all forms of ARS derive from a mechanism
that was put in place before this division. ARS is, in other words,
one of the basic mechanisms of life.

ARS might even be more basic than the eumetazoa: there are molecular
mechanisms in protozoa that could be precursors of ARS. The most
primitive example is the \dqt{run and tumble} movement of
\emph{E.coli} and \emph{Salmonella typhimurium}. The movement of these
bacteria is controlled by a flagellar motor. \emph{Runs} consist of
forward motion (longish stretch of resource search), while
\emph{tumbles} are made of random turns that keep the bacteria more or
less in the same place (exploiting local resources while they
last). Receptor proteins in the membrane bind these behaviors to
external stimuli \shortcite{}.  The mechanism on which ARS is based
is, in its essential structure, fairly consistent across the whole
spectrum of eumetazoans. Its fully formed presence in organisms with
limited learning capabilities, such as \emph{C.elegans} suggests that
learning is not involved or, to the extent that it is (such as in
mammals), it is based on a fully formed pre-learning machinery. This
is what makes ARS so interesting: it is a basic mechanism that very
different forms of life have adopted as a basic strategy to solve such
diverse problems as looking for food in a Petri dish or trying to
prove a mathematical theorem. Its omnipresence derives from the
optimality of ARS as a search strategy in cases in which resources are
\dqt{clumpy} and the information about the locations of the
\dqt{clumps} is limited (we'll see that in the next section). In the
case of foraging animals, the resource is food, and the reward is
finding something to eat. In the case of somebody looking at an image,
the resource is information about its content, and the reward is
understanding what the image is about%
\footnote{Things are more complex in the case of looking at images:
  the actual paths that people's gaze follow depend on what they are
  looking for. Given an image of an interior, the actual paths are
  different if the subjects are asked e.g.\ \dqt{From what epoch is the
    interior?} or if they are asked e.g.\ \dqt{What are the people in
    the image doing?} This is not important for our present
  considerations: all these paths, different as they may be, exhibit
  the features of ARS.}%
. In all these cases, the animal or person moves locally around the
same clump as long as it is rewarding to do so (viz.\ as long as
locally one finds more food or new information), then starts moving
rapidly to explore quickly new territory in search of a new clump.
\begin{figure}[bthp]
  \begin{center}
    \setlength{\unitlength}{0.75em}
    \begin{picture}(56,6)(0,0)
      \put(0,0){
        \put(2,1){\line(2,1){2}}
        \put(2,6){\line(2,-1){2}}
        \put(4,2){\line(0,1){3}}
        \put(4,5){\line(2,1){2}}
        \put(6,6){\line(2,-1){2}}
        \put(8,5){\line(0,-1){3}}
        \put(8,2){\line(-2,-1){2}}
        \put(6,1){\line(-2,1){2}}
        \put(4.4,2.4){\line(0,1){2.2}}
        \put(6,5.6){\line(2,-1){1.6}}
        \put(6,1.4){\line(2,1){1.6}}
        \put(8,5){\line(2,1){2}}
        \put(10,6){\line(2,-1){2}}
        \put(12,5){\line(0,-1){2.5}}
        \put(12,1.8){\makebox(0,0){NH$_2$}}
        \put(1.8,1){\makebox(0,0)[r]{HO}}
        \put(1.8,6){\makebox(0,0)[r]{HO}}
        \put(6,-2){\makebox(0,0){dopamine}}
      }
      \put(20,0){
        \multiput(1,3)(4,0){3}{\line(2,1){2}}
        \multiput(3,4)(4,0){3}{\line(2,-1){2}}
        \multiput(0,0)(8,0){2}{%
          \put(2.8,4.2){\line(0,1){1.4}}
          \put(3.2,4.2){\line(0,1){1.4}}
          \put(3,6){\makebox(0,0)[b]{O}}
        }
        \put(9,3){\line(0,-1){2}}
        \put(8.75,1){\makebox(0,0)[lt]{NH$_3$ (+)}}
        \put(13.2,3){\makebox(0,0)[l]{O (-)}}
        \put(0.8,3){\makebox(0,0)[r]{(-) O}}
        \put(8.75,5.3){\makebox(0,0)[b]{H}}
        \put(9,3){\line(0,1){2}}
        \put(9,3){\line(-1,4){0.5}}
        \put(9,5){\line(-1,0){0.5}}
        \put(7,-2){\makebox(0,0){glutamate}}
      }
      \put(40,0){
        \multiput(2,3)(4,0){3}{\line(2,1){2}}
        \multiput(4,4)(4,0){2}{\line(2,-1){2}}
        \put(3.8,4.2){\line(0,1){1.4}}
        \put(4.2,4.2){\line(0,1){1.4}}
        \put(4,6){\makebox(0,0)[b]{O}}
        \put(1.8,3){\makebox(0,0)[r]{HO}}
        \put(12.2,4){\makebox(0,0)[l]{NH$_2$}}
        \put(8,-2){\makebox(0,0){GABA}}
      }
    \end{picture}
  \end{center}
  \caption{\capstyle Three neurotransmitters that will figure in our
    discussion. Simplifying things very (very!) much, glutamate has
    essentially an excitatory action: its release in a synapse will
    move the postsynaptic neuron closer to firing. GABA is, on the
    contrary, inhibitory, its release will make the postsynaptic
    neuron less likely to fire. Dopamine is a modulator of
    glutammate.}
  \label{dopamine}
\end{figure}
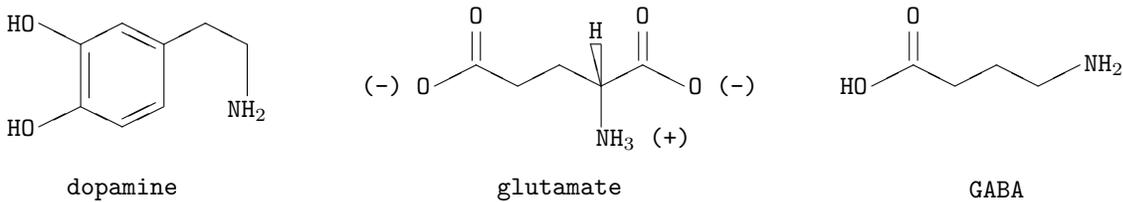

\begin{figure}[bthp]
  \begin{center}
    \VInsert{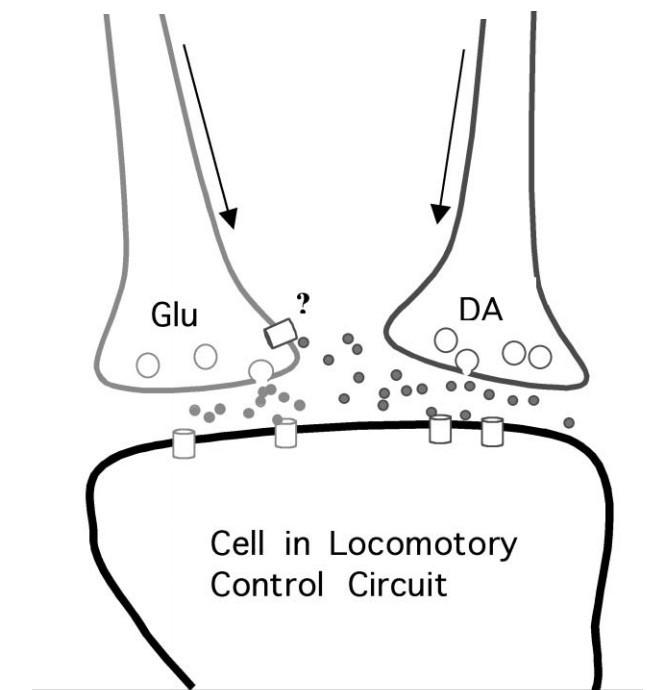}{15em}
  \end{center}
  \caption{\capstyle Dopaminergic action in \emph{C.elegans}. Glu and Da
    represent glutamatergic and dopaminergic presynaptic neurons,
    respectively. The release of dopamine from the dopaminergic
    neurons alters the postsynaptic neuron's response to glutammate. It
    is not known whether this is due to a presynaptic response of the
    glutamatergic neuron and/or to a postsynaptic response of the
    locomotory neurons (from \shortcite{hills:04}).}
  \label{elegans}
\end{figure}

The \dqt{reward} that one is after can be something very concrete
(food) or something very abstract (the \dqt{Eureka} moment of proving
a theorem) but in all cases the nervous system makes the reward
substantial by encoding it as the release of a very specific chemical:
\emph{dopamine} (figure~\ref{dopamine}).  The organism in which the
molecular basis of ARS is best understood is the nematode
\emph{C.elegans} \shortcite{hills:04}. The neural circuitry consists
of eight sensory neurons presynaptic to eight interneurons that
co\"ordinate forward and backward movements. The sensory neurons alter
the turning frequency by releasing dopamine on the interneurons,
modulating the reception of glutammate (Figure~\ref{elegans}).
External administration of dopamine increases the turning frequency,
while administration of a dopamine antagonist reduces it
\shortcite{hills:04}. A reasonable model of \emph{C.elegans} behavior
suggests that, while on food, the sensory neurons release dopamine,
which, via the action of the glutammate, leads to increased switching
behavior in the interneurons, resulting in more turns and,
consequently, in a trajectory that stays local. When off food, the
dopaminergic activity is reduced, and the interneurons reduce their
switching frequency, leading to less turns and more ground covered

Although \emph{C.elegans} is the only organism for which the
neuromolecular mechanism of ARS is well understood, there is strong
evidence of dopaminergic modultation of glutamatergic synapses
throughout the major clades of the eumetazoans
\shortcite{acerbo:02,cleland:97}. In insects, for example,
dopaminergic neurons in the abdominal ganglion are sparsely
distributed, but show large branching patterns, indicative of
neuromodulation \shortcite{nassel:96}. 

The relation between dopamine and ARS has been documented throughout
the invertebrates, especially in the fruit fly \emph{Drosophila
  melanogaster} \shortcite{bainton:00}, in crustaceans
\shortcite{harriswarrick:95}, in \emph{Aplysia} \shortcite{due:04},
etc.  In all these cases, ARS is limited to food search which is,
clearly, the search problem for which ARS first evolved.

In vertebrates, the modification of behavior by dopamine increases in
complexity and begins to involve behavior not directly related to
food. For example, in frogs and toads dopamine modulation is involved
in the visuomotor focus on preys \shortcite{buxbaumconradi:99};
similar dopaminergic involvement in visuomotor coordination can be
found in rats and humans \shortcite{barrett:04,dursun:99,evenden:93}.
This finding is significant in that it indicates a strong relation
between ARS and \emph{inhibition of return}: the fact that viewers show
significant latency in revisitig objects or regions of a scene that
have already been investigated, united to the lingering of saccadic
movements in regions of intertest \shortcite{tipper:94}. 

The important change in vertebrates is the extension of ARS-like
behavior to cover not only actions with an immediate reward, such as
the search for food, but also situations in which the reward is
projected or even in which the reward itself is a neural state. The
detachment from the immediate food rewards is what makes it possible
to adapt ARS to abstract functions such as \emph{goal-directed
  cognition}. It seems, in other words, that when new problems arose
that had the same abstract structure as search for food, animals,
rather than developing a new mechanism, co\"opted the dopaminergic
modulation that guided food search to work on the new problem.

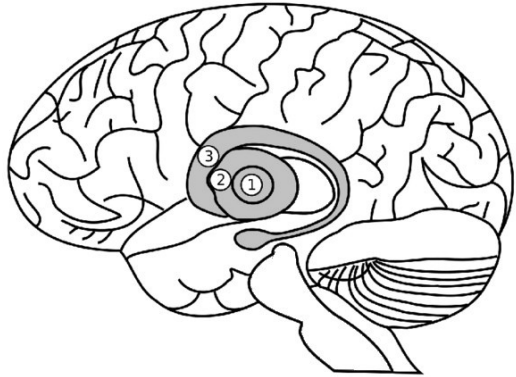
\begin{figure}
  \setlength{\unitlength}{1em}
  \begin{picture}(0,0)(0,0)
    \put(1,-4){\makebox(0,0)[l]{1. lateral medial}}
    \put(1,-6){\makebox(0,0)[l]{2. globus pallidus}}
    \put(1,-8){\makebox(0,0)[l]{3. striatum}}
  \end{picture}
  \begin{center}
    \VInsert{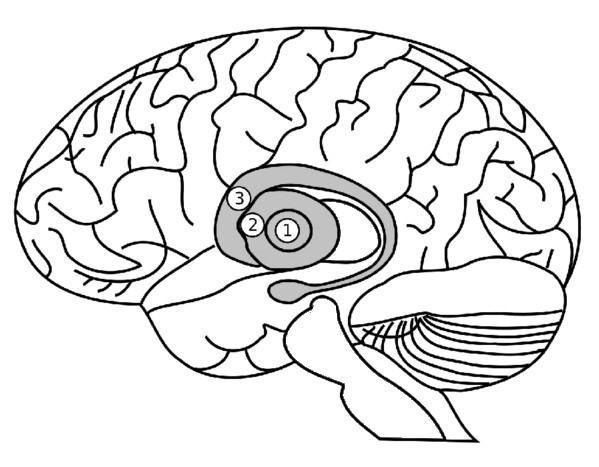}{15em}
  \end{center}
  \caption{\capstyle The basal Ganglia and, specifically the striatum
    have a large number of dopaminergic inputs from other parts of the
    brain, and are involved in ARS. While in simple animals like
    \emph{C.Elegans} ARS is in a simple pathway from sensory input to
    motor neuron, in mammals the input comes from other areas of the
    brain, and the output acts on the cortex, making it possible to
    use ARS for more abstract problems than the immediate search for
    food.}%
  \label{striatum}
\end{figure}

The most important neural structure associated with goal-directed
cognition is the \emph{basal ganglia} and, more specifically, the
\emph{striatum} \shortcite{delong:90,reiner:94} (Figure~\ref{striatum}). 
Information enters the basal ganglia through the striatum, and a great
number of the inputs to the system are dopaminergic
\shortcite{reiner:98}. The structure of the basal ganglia and much of
their connectivity are maintained across vertebrates
\shortcite{salas:03}. The major change from anamniotes (fish and
amphibians) to amniotes is the proliferation of dopaminergic neurons
that input to the striatum \shortcite{reiner:98}, while the structure
of the striatum stays pretty much the same.

\begin{figure}
  \begin{center}
    \setlength{\unitlength}{1.5em}
    \begin{picture}(20,8)(0,0)
      \put(0,3){
        \multiput(0,0)(0,2){2}{\line(1,0){20}}
        \multiput(0,0)(10,0){3}{\line(0,1){2}}
        \put(5,1){\makebox(0,0){{\large Focused}}}
        \put(15,1){\makebox(0,0){{\large Diffused}}}
        \put(0.1,0.1){\makebox(0,0)[lb]{{\small too much DA}}}
        \put(19.9,0.1){\makebox(0,0)[rb]{{\small too little DA}}}
        \put(6,2.2){\vector(1,0){8}}
        \multiput(8,2.7)(0.5,0){8}{\line(1,0){0.25}}
        \put(10,2.9){\makebox(0,0)[b]{schizophrenia}}
      }
      \put(0,2){
        \put(0,0){\vector(0,1){0.8}}
        \put(0,-0.1){\makebox(0,0)[t]{autism}}
      }
      \put(7,2){
        \put(0,0){\vector(0,1){0.8}}
        \put(0,-0.1){\makebox(0,0)[t]{addiction}}
      }
      \put(18,2){
        \put(0,0){\vector(0,1){0.8}}
        \put(0,-0.1){\makebox(0,0)[t]{parkinson's}}
      }
      \put(3,6){
        \put(0,0){\vector(0,-1){0.8}}
        \put(0,0.1){\makebox(0,0)[b]{OCD, TS}}
      }
      \put(16,6){
        \put(0,0){\vector(0,-1){0.8}}
        \put(0,0.1){\makebox(0,0)[b]{ADHD}}
      }        
    \end{picture}
  \end{center}
  \caption{\capstyle The continuous arrow at the top represents the normal
    temporal progression of dopaminergic activity in ARS, modulating
    behavior from focused (i.e.\ local) search to diffuse
    (i.e.\ global). The placement of the pathologies is qualitative
    and not based on a model but on the fact that they are treated
    either with dopamine or with dopamine antagonists. Although
    dopamine seems to be a factor in these diseases, there are clearly
    more factors at play. Schizophrenia is the most emblematic case in
    which the mechanism is not well understood, as reflected by its
    ambiguous positioning via the dashed line. OCD is
    \emph{Obsessive-Compulsive Disorder}, TS \emph{Tourette Syndrome},
    ADHD \emph{Attention Deficit Hyperactivity Disorder} (from
    \cite{hills:06}).}
  \label{diseases}
\end{figure}
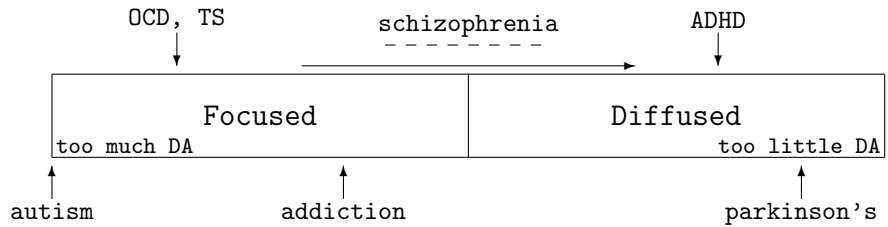

\begin{figure}
  \begin{center}
    \VInsert{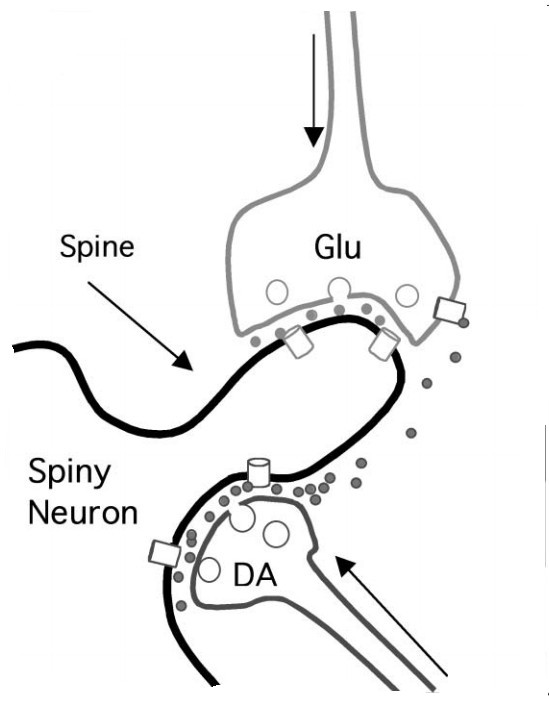}{18em}

    \vspace{-3em}
  \end{center}
  \caption{\capstyle The dopaminergic-glutammatergic interaction in the synapses
    of spiny neurons in the mammalian striatum. Glu and DA represent
    glutammatergic and dopaminergic pre-synaptic neurons, respectively
    (from \shortcite{hills:06}, redrawn from \shortcite{dani:04}).}
  \label{mammalian}
\end{figure}

The balance between glutammate and dopamine in the striatum is key to
the proper functioning of ARS-like activities, and an imbalance
between the two neurotransmitters is suspected in a number of
pathologies affecting goal-directed cognition, including Parkinson's,
schizophrenia, and addiction. Many of these conditions can be regarded
as radicalization of ARS in one direction or another (too local or too
global) due to imperfect dopamine control (see Figure~\ref{diseases}).
In the striatum, dopaminergic neurons modulate the glutammatergic
input at the tips of spiny neurons. The action of the dopaminergic
inputs appears to perform a neuromodulation of the strength of the
glutammatergic inputs (Figure~\ref{mammalian})
The mechanism is similar to that described for \emph{C.elegans} in
Figure~\ref{elegans}, but in the mammalian striatum the shape of
the spiny neuron has specialized for this function and the inputs,
several magnitudes higher in number, come primarily from connections
to cortical neurons rather than directly from sensory neurons as in
\emph{C.Elegans} \shortcite{white:86}. However, at the level of the
microcircuit, little has changed from nematodes to amniote
vertebrates. The origin of the dopaminergic input does, however, mark
a fundamental evolutionary shift in the activity range of ARS-like
behavior. While in \emph{C.elegans} or \emph{Drosophila} the afferent
dopaminergic signal is reliably related to the presence or absence of
food, in higher vertebrates the signal may represent the
\emph{expectation of a reward}. The critical transition here is from a
concrete (directly sensed) reward to its neural representation that
is, from a physical reward to the abstract idea of a reward
\shortcite{schultz:95}. As Hills puts it: \dqt{the evolutionary theory
  $[$of ARS$]$ is therefore completely consistent with the reward
  theory of dopamine, but adds the evolutionary hypothesis that the
  initial reward represented by the release of dopamine were
  food. Only later was this system co-opted to represent the
  expectation of a reward, which allows for goal-directed cognition}
\shortcite{hills:06}.

To conclude this brief \emph{excursus} of ARS, we consider a region of
relatively recent evolution that, outside of the basal ganglia, is
heavily involved in goal-directed behavior: the \emph{prefrontal
  cortex} (PFC). The PFC has clearly evolved much later than ARS;
nevertheless, it is heavily involved in goal-directed behavior via
massive connections to the striatum \shortcite{odonnell:95}. Dopamine
has been shown to be a factor in the sustained activation of the PFC
\shortcite{seamans:98,wang:04}. Most models of PFC see the r\^ole of
dopamine as holding objects in attention long enough for appropriate
behavior to be activated \shortcite{braver:00}. Consistently with ARS,
already known solutions mediated by the PFC are most typically tried
when a problem has to be solved in a new situation
\shortcite{eichenbaum:01}.

The context in which goal-directed cognition takes place includes
external and internal stimuli; ARS depends in part on the alignment of
external stimuli with previous expectations. This is likely to be
controlled by the connections between the PFC and the \emph{Nucleus
  Accumbens} (NAcc) in the striatum, which modulates attention, eye
movements, and the maintenance of working memory
\cite{bertolucci:90,floresco:99,schultz:04}; dopamine has been
identified as one of the main influences in the modulation of NAcc
activity \cite{floresco:96}: novel stimuli lead to increase in
dopamine in the NAcc and in the PFC \cite{berns:01}.

So, in the evolution of vertebrates, we see a progressive extension of
the r\^ole of ARS, from the dopaminergic control of visuomotor control
in frogs and toads to the similarity mediated maintenance of ideas in
working memory \cite{schultz:95}. ARS appears therefore to be one of
the fundamental strategies in the animal kingdom, co-opted and adapted
to a number of situations, from the \dqt{run and tumble} behavior of
\emph{E.coli} to the way we focus on and later abandon ideas when we
think about a problem.

\separate

This brief explanation of the evolutionary basis of ARS has been
centered on its molecular mechanism, especially on the r\^ole of
dopamine as neuromodulator. From now on, however, our focus will
change: we shall try to understand the \emph{exterior} characteristics
of the behavior. ARS leads to a well identified patterns of motion
either in the physical space (in the case of foraging), in the visual
space (scanning an image), or in any number of abstract spaces. We
shall study mathematically these patterns of motion and try to
characterize them. Our methods will be based mostly on the study of
random walks, of diffusion, and on the kinds of anomalous diffusion to
which ARS leads.

\section{\sc Optimality of ARS}
\label{genetic}
The evolutionary success of ARS entails, according to the theory of
natural selection, that ARS is an optimal strategy---if not globally,
at least locally---for a large set of problems. In abstract terms, we
have a space with certain resources placed in different parts of it;
we need a strategy to navigate this space collecting the greatest
amount of resources. This must be done without information on the
placement of the resources. (If we can sense from afar where the
resources are located, we simply walk there and get them: no search
strategy is necessary.) The nature of the resources can be the most
diverse: in the case of foraging (to which we shall mostly make
reference), the resource is food; in the case of saccadic movements,
it is the visual information that we get from the visual field, and so
on. ARS is optimal if the resources are \dqt{patchy,} that is, if they
are organized in resource-rich patches separated by areas of small or
zero resource concentration. In the model that we shall develop in
this section, we assume that the resources are consumed in the course
of the activity, and that they are not replenished while the activity
goes on. Resources are consumed simply by moving on top of them
(assuming that, after walking on them, they would be consumed with a
certain probability would not substantially change the model).
In the example of food, this means that we have food distributed in
patches (a grove, a pond, a herd, a school of fish...) and the forager
moves inside the patch and between patches eating what it finds. We
make a number of hypotheses. Firstly, we assume that the food doesn't
move around or if it does (as is the case of animal preys) its
movement is not significant and food can be modeled as
static. Secondly, we assume that the forager will eat all the food it
can find as soon as it finds it (its eyesight is perfect and its
appetite endless). Finally, food doesn't grow back: once it has been
eaten at a particular location, that location will remain barren for
the rest of the forager's activity.

In the case of saccades, we assume (as is often the case) that there
are patchy areas in the visual field that are rich in information
useful to interpret the scene (relevant or telling objects, faces,
etc.). We also assume that once we have analyzed the information in a
given area of the visual field, that information is remembered and it
is not necessary to analyze it again. This is equivalent to the
hypothesis that the food doesn't grow back once it has been eaten.

\separate

In this section, we want to check whether ARS emerges as an optimal
solution to the foraging problem. We shall do this by implementing a
genetic algorithm based on a competition among individuals whose
characteristics are encoded in a string of bits called a \emph{gene}
\cite{mitchell:98}.

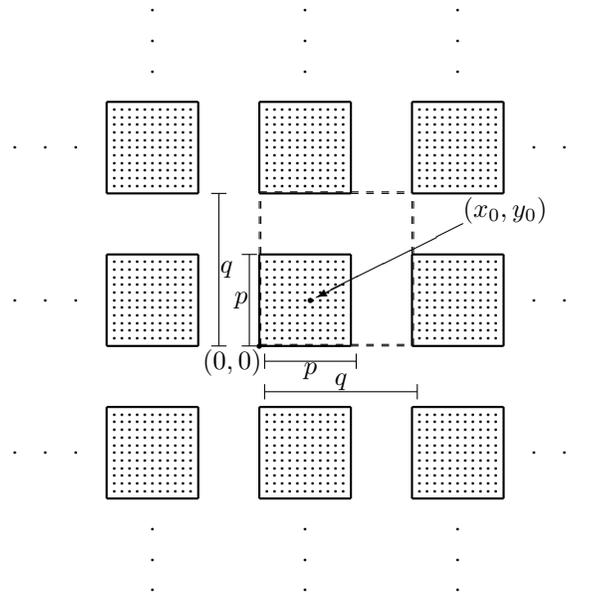
\begin{figure}[btp]
  \begin{center}
    \setlength{\unitlength}{1.1em}
    \begin{picture}(21,21)(-3,-3)
      \newsavebox{\patch}
      \savebox{\patch}{
        \thicklines
        \multiput(0,0)(3,0){2}{\line(0,1){3}}
        \multiput(0,0)(0,3){2}{\line(1,0){3}}
        \thinlines
        \multiput(0.25,0.25)(0.25,0){11}{
          \multiput(0,0)(0,0.25){11}{\circle*{0.0001}}
        }          
      }
      \multiput(0,0)(0,5){3}{
        \multiput(0,0)(5,0){3}{\usebox{\patch}}%
      }
      \multiput(5,5)(0,5){2}{
        \multiput(0,0)(0.5,0){10}{%
          \multiput(0,0)(0,0.05){2}{\line(1,0){0.25}}}%
      }
      \multiput(5,5)(5,0){2}{
        \multiput(0,0)(0,0.5){10}{%
          \multiput(0,0)(0.05,0){2}{\line(0,1){0.25}}}%
      }
      \multiput(-1,1.5)(0,5){3}{%
        \multiput(0,0)(-1,0){3}{\circle*{0.1}}
      }
      \multiput(14,1.5)(0,5){3}{%
        \multiput(0,0)(1,0){3}{\circle*{0.1}}
      }
      \multiput(1.5,-1)(5,0){3}{%
        \multiput(0,0)(0,-1){3}{\circle*{0.1}}
      }
      \multiput(1.5,14)(5,0){3}{%
        \multiput(0,0)(0,1){3}{\circle*{0.1}}
      }
      \put(5,5){%
        \circle*{0.15}%
        \put(-0.1,-0.1){\makebox(0,0)[tr]{$(0,0)$}}
        \multiput(0,-0.75)(3,0){2}{\line(0,1){0.5}}%
        \put(0,-0.5){\line(1,0){3}}%
        \put(1.5,-0.55){\makebox(0,0)[t]{$p$}}
        \multiput(0,-1.75)(5,0){2}{\line(0,1){0.5}}%
        \put(0,-1.5){\line(1,0){5}}%
        \put(2.5,-1.45){\makebox(0,0)[b]{$q$}}
        \multiput(-0.75,0)(0,3){2}{\line(1,0){0.5}}%
        \put(-0.5,0){\line(0,1){3}}%
        \put(-0.55,1.5){\makebox(0,0)[r]{$p$}}
        \multiput(-1.75,0)(0,5){2}{\line(1,0){0.5}}%
        \put(-1.5,0){\line(0,1){5}}%
        \put(-1.45,2.5){\makebox(0,0)[l]{$q$}}
        \put(1.5,1.5){\circle*{0.2}}
        \put(6.5,4){\vector(-2,-1){4.8}}
        \put(6.5,4){\makebox(0,0)[lb]{$(x_0,y_0)$}}
      }
    \end{picture}
  \end{center}
  \caption{\capstyle The environment for the application of genetic
    algorithms to the evolution of ARS. The dotted squares represent
    the patches composed of $p\times{p}$ pellets of food, where $p$ is a
    program parameter. The distance between the patches, $q$, is
    determined by the desired density of food, $\rho$ (also a program
    parameter) through the relation
    $q=\lceil{p/\sqrt{\rho}}\rceil$. In the implementation, patches
    are generated dynamically the first time that the forager walks on
    them so that the foraging field is virtually infinite. }
  \label{feedenvironment}
\end{figure}

Individuals move around a \emph{foraging areas} under the guidance of
their gene and collect food. Their \emph{score}, which determines
their fitness for survival, is the amount of food they have collected.
The world in which these individuals move is a regular grid of patches
of food of $p\times{p}$ ($p\in{\mathbb{N}}$) \emph{pellets}, separated
by barren areas without food (see Figure~\ref{feedenvironment}). Each
pellet is an atomic unit of food, that it, it is either not consumed or
consumed entirely. The consumption of each pellet increases  the
survival fitness by one unit.

Each patch is separated from the other by being placed in the
lower-left corner of a larger square of $q\times{q}$ units
($q\in{\mathbb{N}}$, $q>p$) called a \emph{plot}. The density of the
food is $\rho=p^2/q^2$. The plots are created dynamically at run time
as the walker steps on them for the first time, so that the foraging
field is virtually infinite.
In all the tests discussed below, $p$ is kept fixed ($p=16$), $\rho$
is a parameter that varies from $\rho=0.01$ to $\rho=0.95$, and $q$ is
determined as $q=\lceil{p/\sqrt{\rho}}\rceil$. Each individual does a
random walk (specified by certain parameters, as described below)
starting at $(x_0,y_0)=(p/2,p/2)$ that is, in the center of the patch
whose lower-left corner is the origin.

\subsection{The walk parameters}
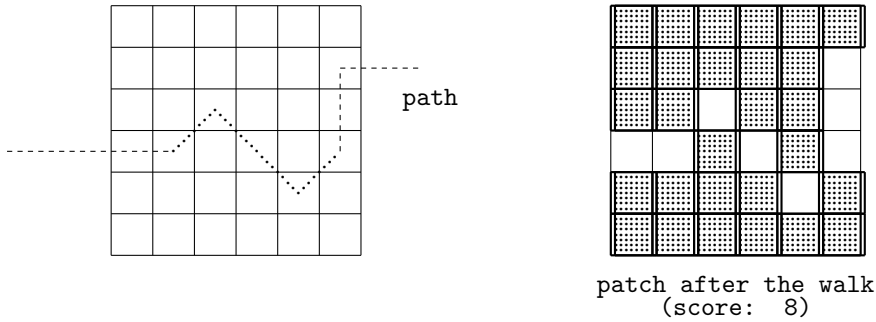
\begin{figure}[tbhp]
  \begin{center}
    \setlength{\unitlength}{1.5em}
    \begin{picture}(25,7)(0,-1)
      \savebox{\patch}{
        \thicklines
        \multiput(0,0)(1,0){2}{\line(0,1){1}}
        \multiput(0.1,0)(1,0){2}{\line(0,1){1}}
        \multiput(0,0)(0,1){2}{\line(1,0){1}}
        \multiput(0.1,0)(0,1){2}{\line(1,0){1}}
        \thinlines
        \multiput(0.125,0.125)(0.125,0){7}{
          \multiput(0,0)(0,0.125){7}{\circle*{0.0001}}
        }          
      }
      \multiput(2,0)(0,1){7}{\line(1,0){6}}
      \multiput(2,0)(1,0){7}{\line(0,1){6}}
      \multiput(-0.5,2.5)(0.25,0){16}{\line(1,0){0.125}}
      \multiput(3.5,2.5)(0.125,0.125){8}{\circle*{0.001}}
      \multiput(4.5,3.5)(0.125,-0.125){16}{\circle*{0.001}}
      \multiput(6.5,1.5)(0.125,0.125){8}{\circle*{0.001}}
      \multiput(7.5,2.5)(0,0.25){8}{\line(0,1){0.125}}
      \multiput(7.5,4.5)(0.25,0){8}{\line(1,0){0.125}}
      \put(9,4){\makebox(0,0)[tl]{path}}
      \put(14,0){
        \multiput(0,0)(0,1){7}{\line(1,0){6}}
        \multiput(0,0)(1,0){7}{\line(0,1){6}}
        \multiput(0,0)(1,0){6}{\usebox{\patch}}
        \put(0,1){\usebox{\patch}}
        \put(1,1){\usebox{\patch}}
        \put(2,1){\usebox{\patch}}
        \put(3,1){\usebox{\patch}}
        \put(5,1){\usebox{\patch}}
        \put(2,2){\usebox{\patch}}
        \put(4,2){\usebox{\patch}}
        \put(0,3){\usebox{\patch}}
        \put(1,3){\usebox{\patch}}
        \put(3,3){\usebox{\patch}}
        \put(4,3){\usebox{\patch}}
        \multiput(0,4)(1,0){5}{\usebox{\patch}}
        \multiput(0,5)(1,0){6}{\usebox{\patch}}
        \put(3,-0.5){\makebox(0,0)[t]{patch after the walk}}
        \put(3,-1){\makebox(0,0)[t]{(score: 8)}}
      }
    \end{picture}
  \end{center}
  \caption{\capstyle A walk of an individual through a patch of food: each
    square crossed by the individual is \dqt{eaten,} and is removed
    from the patch. In this case, the individual passes over 8
    patches: after the walk, its score is eight, and eight pellets are
    removed from the patch.}
  \label{eating}
\end{figure}

Each time the walker walks on a position containing a pellet, it
\dqt{eats} it, incrementing its score (which determines its
evolutionary fitness) by one. The pellet is removed, so that further
visits to the location will not provide any food
(Figure~\ref{eating}). The movement of each individual is a random
walk whose statistical features depend on whether the individual is
currently eating (status: on-food) or whether it has been without food
for some time (status: off-food). The individual doesn't go
\dqt{off-food} immediately as soon as it steps on a location with no
food: the individual has memory, so that it gradually changes its
status from on-food to off-food during a certain number of time
steps. The amount of time without food that it takes to go to the
status off-food is controlled by a parameter in the gene of the
individual.
The general behavior of the random walk is the same regardless of
whether the individual is on-food or off-food (the only thing that
changes in the two cases is the numerical value of the
parameters). Consider a generic situation in which the parameters are
$(\alpha_0,l_0)$. The individual is coming from a direction $\theta$,
being currently at location $(x,y)$. The individual chooses a
deviation angle $\alpha$, selected with a Gaussian distribution
centered at $\alpha_0$, and a length $l$ selected with an exponential
distribution with average $l_0$, and performs a jump in a direction at
an angle $\alpha$ from its current direction, and for a length $l$
(Figure~\ref{jump}). The parameters $(\alpha_0,l_0)$ characterize the
statistics of the jump, they are encoded in the individual's gene, and
they take different values depending on whether the individual is
on-food or off-food. When the individual is on-food, the jumps are
done according to the parameters $(\alpha_f,l_f)$, while when the
individual has been for some time off-food, the jumps are done
according to $(\alpha_n,l_n)$. The switch from the on-food parameters
to the off-food is done gradually when the individual is in an area
without food.  The gene defines a memory threshold $\tau$ to switch
from the on-food to the off-food behavior. If $t$ is the number of
step that the individual has been off-food ($t=0$ if the individual is
on-food) then the parameters for the next jump will be
\begin{equation}
        }
      \end{picture}
  \end{center}
  \caption{\capstyle Generation of a single jump of the random walk. In (a), a
    walker is coming to point $(x,y)$ following a trajectory with
    direction $\theta$. An angle $\alpha$ is chosen from the Gaussian
    angle distribution in (b), centered around the angle $\alpha_0$:
    the direction of the new jump is $\alpha$ from the previous
    direction (viz.\ $\theta+\alpha$ absolute direction). The length
    of the jump is drawn from the exponential distribution with
    average $l_0$ in (c).}
  \label{jump}
\end{figure}

The parameters of the jump are reset to $(\alpha_f,l_f)$ as soon as
the individual finds food. That is, there is a lingering memory that
food was around there even if currently no food is found---a memory
that fades away in a time $\tau$---but the absence of food is
forgotten as soon as new food is found. Any moral or philosophical
conclusion, be it positive or negative, that can be drawn from this
hypothesis is beyond the scope of these notes.

\subsection{Gene definition}
Each individual is therefore characterized by five parameters:
$(\alpha_f,l_f,\alpha_n,l_n,\tau)$. We represent each one as a 8-bit
value (these values are scaled in order to compute the actual values
of the parameters) and collect them in a 40-bit \dqt{gene.} We try to
keep related parameters in nearby positions of the gene (this is
believed to speed up the convergence of the algorithm). Calling $A_F$,
$L_F$, $A_N$, $L_N$, and $T$ the 8-bit representations of the
parameters we have the genetic representation of an individual in
Figure~\ref{gene}.
\begin{figure}[tbh]
  \begin{center}
    \setlength{\unitlength}{1.5em}
    \begin{picture}(25,3)(0,2)
      \multiput(0,4)(0,1){2}{\line(1,0){25}}
      \multiput(0,4)(5,0){6}{\line(0,1){1}}
      \multiput(0,3)(5,0){6}{\line(0,1){0.75}}
      \put(0,2.8){\makebox(0,0)[t]{0}}
      \put(5,2.8){\makebox(0,0)[t]{7}}
      \put(10,2.8){\makebox(0,0)[t]{15}}
      \put(15,2.8){\makebox(0,0)[t]{23}}
      \put(20,2.8){\makebox(0,0)[t]{31}}
      \put(25,2.8){\makebox(0,0)[t]{39}}
      \put(2.5,4.5){\makebox(0,0){$A_F$}}
      \put(7.5,4.5){\makebox(0,0){$L_F$}}
      \put(12.5,4.5){\makebox(0,0){$T$}}
      \put(17.5,4.5){\makebox(0,0){$A_N$}}
      \put(22.5,4.5){\makebox(0,0){$L_N$}}
      \put(0,2){\vector(1,0){1}}
      \put(1.1,2){\makebox(0,0)[l]{bits}}
    \end{picture}
  \end{center}
  \caption{\capstyle The genetic representation of an individual. The five 8-bit
    parameters are scaled to provide the jump paramterers
    $(\alpha_f,l_f,\alpha_n,l_n,\tau)$.}
  \label{gene}
\end{figure}
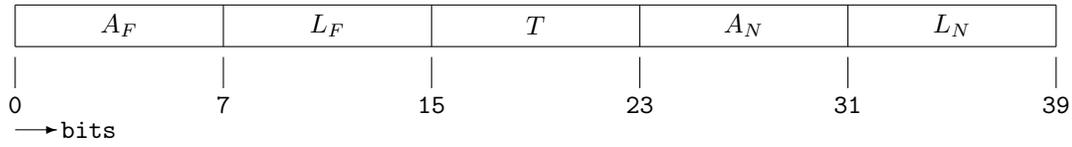
The jump parameters are derived from these 8-bit integers as
\begin{equation}
  \label{normal}
  \begin{array}{ccc}
    \displaystyle \alpha_f=2\pi\frac{A_F}{256} & \displaystyle l_f = \frac{L_F}{4} & \displaystyle \tau = \frac{T}{10} \\
    & \\
    \displaystyle \alpha_n=2\pi\frac{A_N}{256} & \displaystyle l_n = \frac{L_N}{4} 
  \end{array}
\end{equation}
The scaling factors (except those for $\alpha_f$ and $\alpha_n$, which
are derived from geometric considerations) have been determined by
trial and error.
\begin{figure}[tbhp]
  \begin{center}
    \setlength{\unitlength}{1.5em}
    \begin{picture}(20,12)(0,0)
      \put(4,1){\line(0,1){11}}
      \put(11,1){\line(0,1){11}}
      \put(4,0.8){\makebox(0,0)[t]{a}}
      \put(11,0.8){\makebox(0,0)[t]{b}}
      \put(0,2){
        \thicklines
        \multiput(0,0)(0,1){2}{\line(1,0){14}}
        \multiput(0,0)(14,0){2}{\line(0,1){1}}
        \thinlines
        \put(2,0.5){\makebox(0,0){$B_1$}}
        \put(7.5,0.5){\makebox(0,0){$A_2$}}
        \put(12.5,0.5){\makebox(0,0){$B_3$}}
        \put(16,0.5){\makebox(0,0)[l]{offspring 2}}
      }
      \put(0,4){
        \thicklines
        \multiput(0,0)(0,1){2}{\line(1,0){14}}
        \multiput(0,0)(14,0){2}{\line(0,1){1}}
        \thinlines
        \put(2,0.5){\makebox(0,0){$A_1$}}
        \put(7.5,0.5){\makebox(0,0){$B_2$}}
        \put(12.5,0.5){\makebox(0,0){$A_3$}}
        \put(16,0.5){\makebox(0,0)[l]{offspring 1}}
      }
      \put(0,7){
        \thicklines
        \multiput(0,0)(0,1){2}{\line(1,0){14}}
        \multiput(0,0)(14,0){2}{\line(0,1){1}}
        \thinlines
        \put(2,0.5){\makebox(0,0){$B_1$}}
        \put(7.5,0.5){\makebox(0,0){$B_2$}}
        \put(12.5,0.5){\makebox(0,0){$B_3$}}
        \put(16,0.5){\makebox(0,0)[l]{parent B}}
      }
      \put(0,9){
        \thicklines
        \multiput(0,0)(0,1){2}{\line(1,0){14}}
        \multiput(0,0)(14,0){2}{\line(0,1){1}}
        \thinlines
        \put(2,0.5){\makebox(0,0){$A_1$}}
        \put(7.5,0.5){\makebox(0,0){$A_2$}}
        \put(12.5,0.5){\makebox(0,0){$A_3$}}
        \put(16,0.5){\makebox(0,0)[l]{parent A}}
      }
    \end{picture}
  \end{center}
  \caption{\capstyle Double cut for the generation of offsprings. Two parents
    generate two offprings, mixing the bits of their genes as
    represented.}
  \label{offsprings}
\end{figure}

\subsection{The algorithm}
The genetic algorithm is pretty standard. A \emph{generation} is a set
of individuals. Each individual is placed in the environment in the
same initial position, and does a random walk of predetermined length,
according to the parameters encoded in its gene, and collecting
pellets of food as specified above. The environment is restored
between individuals, so that each one has the same initial supply of
food (this entails that there is no competition among the
individuals). A point is scored for each pellet that is eaten. As a
result, after all individuals have executed a random walk, individual
number $k$, characterized by gene $\gamma_k$ has a score $s_k$, with
$k=1,\ldots,G$, where $G$ is the number of individuals in a
generation.

There are several methods to create the following generation of
individuals. Since the performance of the algorithms seems to have
little dependence in the specific method used, we use one of the
simplest, based on the creation of an intermediate \emph{gene
  pool}. The gene pool is a set ${\mathcal{P}}$ of $P$ individuals
possibly replicated (generally $|P|=G$: the pool has the same size
as the generations) such that the number of \dqt{copies} of an
individual in the pool is proportional to its score. An easy algorithm
for generating a pool is the \emph{tournament}: we do $P$ comparisons
of pairs of individuals taken at random from the generation: the
individual with the highest score goes into the pool:

\parbox{\columnwidth}{
  {\tt
    \cb
    \> $P$ $\leftarrow$ $\emptyset$ \\
    \> \cmd{for} k=1 \cmd{to} $P$ \cmd{do} \\
    \> \> i $\leftarrow$ rnd(1,G) \\
    \> \> j $\leftarrow$ rnd(1,G) \\
    \> \> \cmd{if} $s_i\ge{s_j}$ \cmd{then} \\
    \> \> \> ${\mathcal{P}}$ $\leftarrow$ ${\mathcal{P}} \cup \{\gamma_i\}$ \\
    \> \> \cmd{else} \\
    \> \> \> ${\mathcal{P}}$ $\leftarrow$ ${\mathcal{P}} \cup \{\gamma_j\}$ \\
    \> \> \cmd{fi} \\
    \> \cmd{od}
    \ce
  }
}

In order to build the next generation, pairs of genes are taken at
random from the pool (with uniform distribution) and crossed to create
two new individuals that will go into the next generation (this
requires that $G$ be even). We use the method of the \emph{double cut}
to cross the genes. Two values $a,b\in[0,39]$ are chosen randomly. The
two offspring are then generated as in Figure~\ref{offsprings}, in
which we assume $a<b$. We also define a small mutation probability:
for each new gene, with a (small) probability $p$, we pick a random
bit and flip it.
Note that this method doesn't guarantee that the best individual of a
generation will pass unchanged to the next, so we actually use the
crossing to create $G-2$ individuals to which we add the two best
performers of the previous generation.

\subsection{Results}
Figure~\ref{paths} shows typical paths from the best individual for
various values of the density of food,
while Table~\ref{params} shows the value of the parameters for the
same individual.
\begin{figure}[btp]
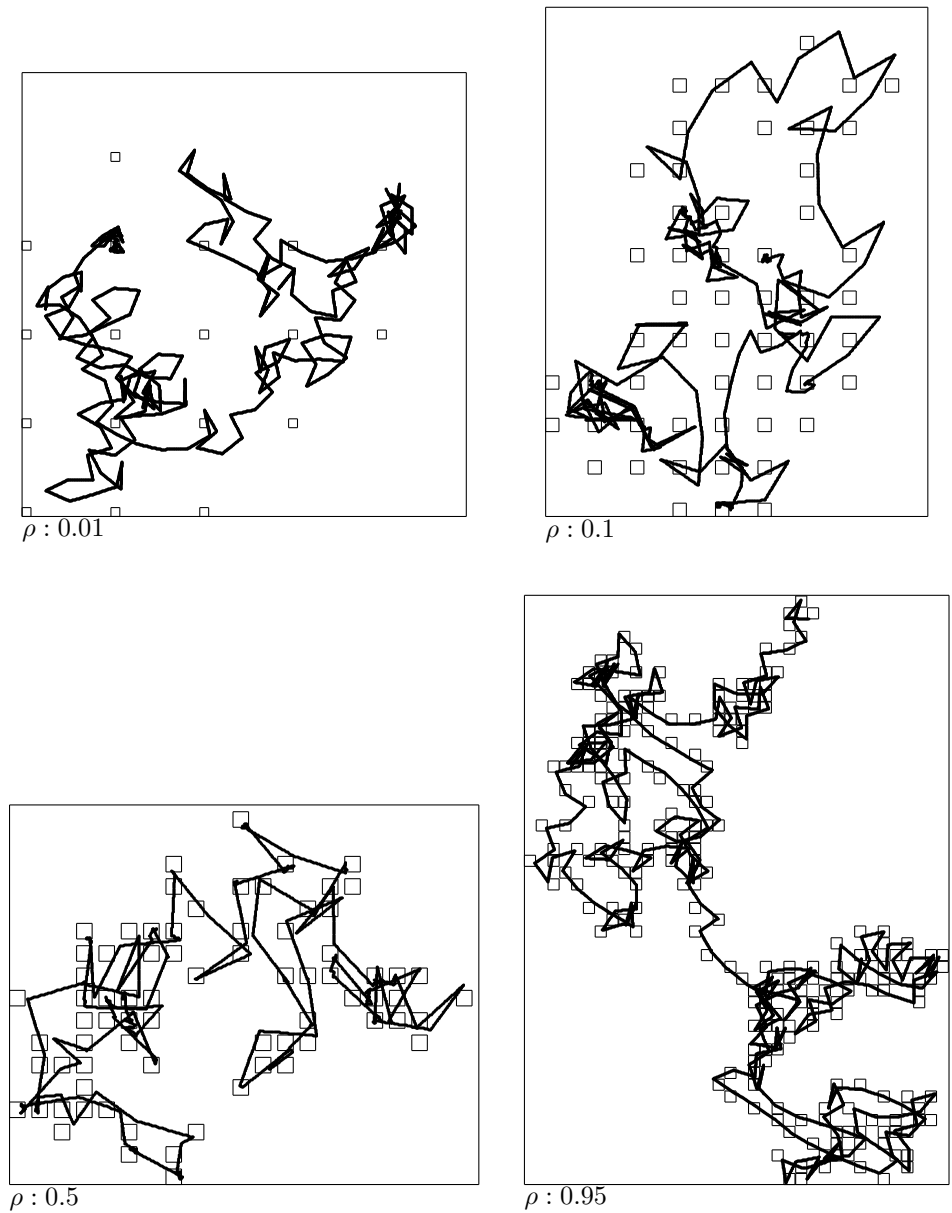

  \setlength{\unitlength}{15em}
  \begin{center}

  \end{center}
  \caption{\capstyle A typical path of the best individual for various values of
    the density. Note the different macroscopic behavior for high
    density and for low density. In the case of high density, the path
    is practically always contained in some patch, and the off-food
    status is practically unused. The low density situation shows the
    characteristics of ARS.}
  \label{paths}
\end{figure}
\begin{table}
  \begin{center}
    \begin{tabular}{|l|c|c|c|c|c|}
      \hline
      \hline
      $\delta$ & $\alpha_f$ & $l_f$ & $\tau$ & $\alpha_n$ & $l_n$ \\
      \hline
      \hline
      0.95 & 0.00 & 136.94 &     57 &   34.25 &   33.50 \\
      \hline
      0.5 &  199.06 &  15.53 &      0 &    2.00 &   60.50 \\
      \hline
      0.1 &  309.18 &   1.75 &     18 &  285.18 &   58.00 \\
      \hline
      0.01 & 227.29 &  1.50 &    37 & 345.88 &  45.50 \\
      \hline
    \end{tabular}
  \end{center}
  \caption{\capstyle The values of $A_F,L_F,T,A_N,L_F$ for the best individual
    for different values of the food density. The values are
    normalized as in (\ref{normal}) to obtain the walk parameters
    $(\alpha_f,l_f,\tau,\alpha_n,l_n)$.}
  \label{params}
\end{table}
\begin{figure}[tbp]
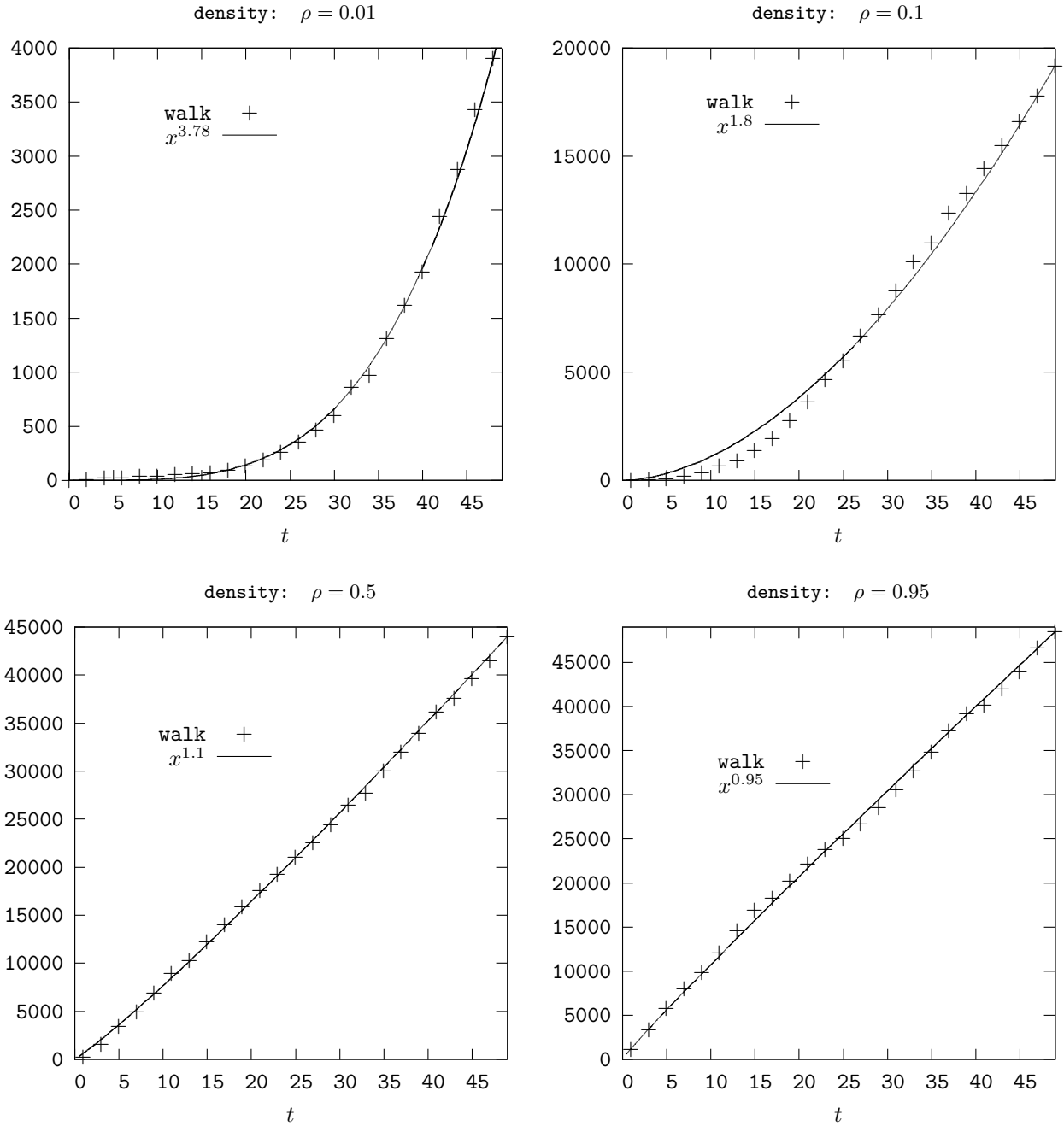

  \begin{center}
    %
  \end{center}
  \caption{\capstyle The behavior of $\langle{X^2(t)}\rangle$ as a function of
    $t$ for various values of the resource density $\rho$. Together
    with the curve, the figure shows its best approximation as
    $\langle{X^2(t)}\rangle\sim{x^\nu}$. The coefficient $\nu$ of the
    best approximation varies with $\rho$.
    On the significance of this variation, see the text.}
  \label{variances}
\end{figure}
We note that in the case of high food density the random walk does not
exhibit the characteristics of ARS (this observation will be made more
formal later on), simply because the individual is always, or almost
always, on food. As the density decreases, the behavior becomes more
characteristic of ARS.

In order to study the characteristics of these walks, we consider the
square average of the displacement from the initial position:
\begin{equation}
  \langle X^2 \rangle \dfeq \langle (x-x_0)^2 + (y-y_0)^2 \rangle.
\end{equation}
Let the individual be fixed (viz.\ let it be the best performing
individual). We execute, with this individual, $N$ random walks on the
environment with the prescribed density, each of length $T$. Let 
\begin{equation}
  w^i = [p_0^i,\ldots,p_t^i,\ldots,p_T^i]
\end{equation}
be the $i$th random walk, where $p_t^i=(x_t^i-x_0,y_t^i-y_0)$. We are
interested in knowing how far the individual has gone, on average,
from its initial position, at time $t$. That is, we are interested in
studying the function
\begin{equation}
  \langle X^2(t) \rangle = \frac{1}{N} \sum_{i=1}^N (p_t^i)^2
\end{equation}
Figure~\ref{variances} shows the behavior of $\langle{X^2(t)}\rangle$
as a function of $t$ for various values of the density $\rho$, 
together with the best approximation of the form
$\langle{X^2(t)}\rangle\sim{x^\nu}$, where the exponent $\nu$ depends
on $\rho$: 

\begin{center}
  \begin{tabular}{|c|c|c|c|c|}
    \hline
    $\rho$ & 0.01 & 0.1 & 0.5 & 0.9\\
    \hline
    $\nu$  & 3.78 & 1.8 & 1.1 & 0.95 \\
    \hline
  \end{tabular}
\end{center}

Figure~\ref{varplot} shows the behavior of the
exponent $\nu$ as a function of $\rho$. As density approaches 1 or 0,
that is, as the environment becomes more homogeneous (either with a
lot of food or very little food), the exponent approaches 1, that is,
we approach a situation in which
$\langle{X^2(t)}\rangle\sim{x^1}$. This, as we shall see, is the
behavior characteristic of Brownian motion, as well as of several
other types of random walks. This should not come as a surprise: when
food can be found everywhere, the individuals have no particular
reason to modify their behavior, and will simply move to and fro in an
haphazard manner: they will do a random walk. The same is true if
there is very little food. The patches are so small and far apart that
the in-patch behavior will last for a very short time and will not
change significantly the characteristics of the walk, which will be a
random walk from patch to patch looking for resources.

When the food is patchy, on the other hand, the random walk of the
individual doesn't follow the standard Brownian model. In the next
section I shall consider the foraging walk more closely from a
mathematical point of view. We shall begin, in the next section by
considering the switching from the on-patch behavior to the off-patch,
without taking onto account the spatial characteristics of the
environment. Then, in the following section, we shall study random
walks in search of a model that fits the characteristics of a forager
on patchy resources. We shall see that such a model is given by the
so-called \emph{Levy walks}.
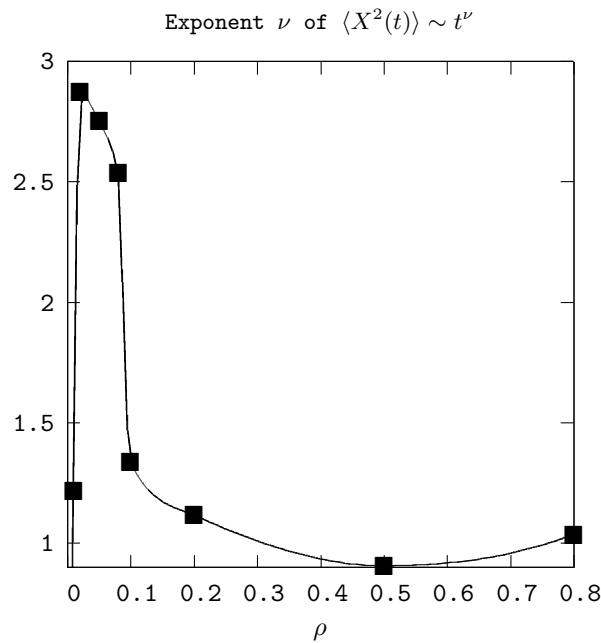
\begin{figure}[thbp]
  \begin{center}
\setlength{\unitlength}{0.240900pt}
\ifx\plotpoint\undefined\newsavebox{\plotpoint}\fi
\sbox{\plotpoint}{\rule[-0.200pt]{0.400pt}{0.400pt}}%
\begin{picture}(1050,1050)(0,0)
\sbox{\plotpoint}{\rule[-0.200pt]{0.400pt}{0.400pt}}%
\put(162.0,169.0){\rule[-0.200pt]{4.818pt}{0.400pt}}
\put(142,169){\makebox(0,0)[r]{ 1}}
\put(937.0,169.0){\rule[-0.200pt]{4.818pt}{0.400pt}}
\put(162.0,358.0){\rule[-0.200pt]{4.818pt}{0.400pt}}
\put(142,358){\makebox(0,0)[r]{ 1.5}}
\put(937.0,358.0){\rule[-0.200pt]{4.818pt}{0.400pt}}
\put(162.0,547.0){\rule[-0.200pt]{4.818pt}{0.400pt}}
\put(142,547){\makebox(0,0)[r]{ 2}}
\put(937.0,547.0){\rule[-0.200pt]{4.818pt}{0.400pt}}
\put(162.0,737.0){\rule[-0.200pt]{4.818pt}{0.400pt}}
\put(142,737){\makebox(0,0)[r]{ 2.5}}
\put(937.0,737.0){\rule[-0.200pt]{4.818pt}{0.400pt}}
\put(162.0,926.0){\rule[-0.200pt]{4.818pt}{0.400pt}}
\put(142,926){\makebox(0,0)[r]{ 3}}
\put(937.0,926.0){\rule[-0.200pt]{4.818pt}{0.400pt}}
\put(162.0,131.0){\rule[-0.200pt]{0.400pt}{4.818pt}}
\put(162,90){\makebox(0,0){ 0}}
\put(162.0,906.0){\rule[-0.200pt]{0.400pt}{4.818pt}}
\put(261.0,131.0){\rule[-0.200pt]{0.400pt}{4.818pt}}
\put(261,90){\makebox(0,0){ 0.1}}
\put(261.0,906.0){\rule[-0.200pt]{0.400pt}{4.818pt}}
\put(361.0,131.0){\rule[-0.200pt]{0.400pt}{4.818pt}}
\put(361,90){\makebox(0,0){ 0.2}}
\put(361.0,906.0){\rule[-0.200pt]{0.400pt}{4.818pt}}
\put(460.0,131.0){\rule[-0.200pt]{0.400pt}{4.818pt}}
\put(460,90){\makebox(0,0){ 0.3}}
\put(460.0,906.0){\rule[-0.200pt]{0.400pt}{4.818pt}}
\put(560.0,131.0){\rule[-0.200pt]{0.400pt}{4.818pt}}
\put(560,90){\makebox(0,0){ 0.4}}
\put(560.0,906.0){\rule[-0.200pt]{0.400pt}{4.818pt}}
\put(659.0,131.0){\rule[-0.200pt]{0.400pt}{4.818pt}}
\put(659,90){\makebox(0,0){ 0.5}}
\put(659.0,906.0){\rule[-0.200pt]{0.400pt}{4.818pt}}
\put(758.0,131.0){\rule[-0.200pt]{0.400pt}{4.818pt}}
\put(758,90){\makebox(0,0){ 0.6}}
\put(758.0,906.0){\rule[-0.200pt]{0.400pt}{4.818pt}}
\put(858.0,131.0){\rule[-0.200pt]{0.400pt}{4.818pt}}
\put(858,90){\makebox(0,0){ 0.7}}
\put(858.0,906.0){\rule[-0.200pt]{0.400pt}{4.818pt}}
\put(957.0,131.0){\rule[-0.200pt]{0.400pt}{4.818pt}}
\put(957,90){\makebox(0,0){ 0.8}}
\put(957.0,906.0){\rule[-0.200pt]{0.400pt}{4.818pt}}
\put(162.0,131.0){\rule[-0.200pt]{0.400pt}{191.515pt}}
\put(162.0,131.0){\rule[-0.200pt]{191.515pt}{0.400pt}}
\put(957.0,131.0){\rule[-0.200pt]{0.400pt}{191.515pt}}
\put(162.0,926.0){\rule[-0.200pt]{191.515pt}{0.400pt}}
\put(559,29){\makebox(0,0){$\rho$}}
\put(559,988){\makebox(0,0){\small Exponent $\nu$ of $\langle{X^2(t)}\rangle\sim{t^\nu}$}}
\put(172,250){\makebox(0,0){$\blacksquare$}}
\put(182,877){\makebox(0,0){$\blacksquare$}}
\put(212,832){\makebox(0,0){$\blacksquare$}}
\put(242,750){\makebox(0,0){$\blacksquare$}}
\put(261,297){\makebox(0,0){$\blacksquare$}}
\put(361,213){\makebox(0,0){$\blacksquare$}}
\put(659,133){\makebox(0,0){$\blacksquare$}}
\put(957,182){\makebox(0,0){$\blacksquare$}}
\multiput(170.59,131.00)(0.488,39.654){13}{\rule{0.117pt}{30.150pt}}
\multiput(169.17,131.00)(8.000,538.422){2}{\rule{0.400pt}{15.075pt}}
\multiput(178.59,732.00)(0.488,9.475){13}{\rule{0.117pt}{7.300pt}}
\multiput(177.17,732.00)(8.000,128.848){2}{\rule{0.400pt}{3.650pt}}
\multiput(186.59,873.72)(0.488,-0.560){13}{\rule{0.117pt}{0.550pt}}
\multiput(185.17,874.86)(8.000,-7.858){2}{\rule{0.400pt}{0.275pt}}
\multiput(194.59,863.47)(0.488,-0.956){13}{\rule{0.117pt}{0.850pt}}
\multiput(193.17,865.24)(8.000,-13.236){2}{\rule{0.400pt}{0.425pt}}
\multiput(202.59,848.26)(0.488,-1.022){13}{\rule{0.117pt}{0.900pt}}
\multiput(201.17,850.13)(8.000,-14.132){2}{\rule{0.400pt}{0.450pt}}
\multiput(210.59,832.47)(0.488,-0.956){13}{\rule{0.117pt}{0.850pt}}
\multiput(209.17,834.24)(8.000,-13.236){2}{\rule{0.400pt}{0.425pt}}
\multiput(218.59,817.26)(0.488,-1.022){13}{\rule{0.117pt}{0.900pt}}
\multiput(217.17,819.13)(8.000,-14.132){2}{\rule{0.400pt}{0.450pt}}
\multiput(226.59,800.23)(0.488,-1.352){13}{\rule{0.117pt}{1.150pt}}
\multiput(225.17,802.61)(8.000,-18.613){2}{\rule{0.400pt}{0.575pt}}
\multiput(234.59,776.53)(0.488,-2.211){13}{\rule{0.117pt}{1.800pt}}
\multiput(233.17,780.26)(8.000,-30.264){2}{\rule{0.400pt}{0.900pt}}
\multiput(242.59,709.02)(0.485,-13.002){11}{\rule{0.117pt}{9.871pt}}
\multiput(241.17,729.51)(7.000,-150.511){2}{\rule{0.400pt}{4.936pt}}
\multiput(249.59,530.22)(0.488,-15.352){13}{\rule{0.117pt}{11.750pt}}
\multiput(248.17,554.61)(8.000,-208.612){2}{\rule{0.400pt}{5.875pt}}
\multiput(257.59,333.75)(0.488,-3.730){13}{\rule{0.117pt}{2.950pt}}
\multiput(256.17,339.88)(8.000,-50.877){2}{\rule{0.400pt}{1.475pt}}
\multiput(265.59,285.68)(0.488,-0.890){13}{\rule{0.117pt}{0.800pt}}
\multiput(264.17,287.34)(8.000,-12.340){2}{\rule{0.400pt}{0.400pt}}
\multiput(273.59,272.09)(0.488,-0.758){13}{\rule{0.117pt}{0.700pt}}
\multiput(272.17,273.55)(8.000,-10.547){2}{\rule{0.400pt}{0.350pt}}
\multiput(281.59,260.51)(0.488,-0.626){13}{\rule{0.117pt}{0.600pt}}
\multiput(280.17,261.75)(8.000,-8.755){2}{\rule{0.400pt}{0.300pt}}
\multiput(289.00,251.93)(0.494,-0.488){13}{\rule{0.500pt}{0.117pt}}
\multiput(289.00,252.17)(6.962,-8.000){2}{\rule{0.250pt}{0.400pt}}
\multiput(297.00,243.93)(0.671,-0.482){9}{\rule{0.633pt}{0.116pt}}
\multiput(297.00,244.17)(6.685,-6.000){2}{\rule{0.317pt}{0.400pt}}
\multiput(305.00,237.93)(0.671,-0.482){9}{\rule{0.633pt}{0.116pt}}
\multiput(305.00,238.17)(6.685,-6.000){2}{\rule{0.317pt}{0.400pt}}
\multiput(313.00,231.94)(1.066,-0.468){5}{\rule{0.900pt}{0.113pt}}
\multiput(313.00,232.17)(6.132,-4.000){2}{\rule{0.450pt}{0.400pt}}
\multiput(321.00,227.94)(1.066,-0.468){5}{\rule{0.900pt}{0.113pt}}
\multiput(321.00,228.17)(6.132,-4.000){2}{\rule{0.450pt}{0.400pt}}
\multiput(329.00,223.95)(1.579,-0.447){3}{\rule{1.167pt}{0.108pt}}
\multiput(329.00,224.17)(5.579,-3.000){2}{\rule{0.583pt}{0.400pt}}
\multiput(337.00,220.95)(1.579,-0.447){3}{\rule{1.167pt}{0.108pt}}
\multiput(337.00,221.17)(5.579,-3.000){2}{\rule{0.583pt}{0.400pt}}
\multiput(345.00,217.95)(1.579,-0.447){3}{\rule{1.167pt}{0.108pt}}
\multiput(345.00,218.17)(5.579,-3.000){2}{\rule{0.583pt}{0.400pt}}
\multiput(353.00,214.95)(1.579,-0.447){3}{\rule{1.167pt}{0.108pt}}
\multiput(353.00,215.17)(5.579,-3.000){2}{\rule{0.583pt}{0.400pt}}
\multiput(361.00,211.94)(1.066,-0.468){5}{\rule{0.900pt}{0.113pt}}
\multiput(361.00,212.17)(6.132,-4.000){2}{\rule{0.450pt}{0.400pt}}
\multiput(369.00,207.95)(1.579,-0.447){3}{\rule{1.167pt}{0.108pt}}
\multiput(369.00,208.17)(5.579,-3.000){2}{\rule{0.583pt}{0.400pt}}
\multiput(377.00,204.94)(1.066,-0.468){5}{\rule{0.900pt}{0.113pt}}
\multiput(377.00,205.17)(6.132,-4.000){2}{\rule{0.450pt}{0.400pt}}
\multiput(385.00,200.95)(1.579,-0.447){3}{\rule{1.167pt}{0.108pt}}
\multiput(385.00,201.17)(5.579,-3.000){2}{\rule{0.583pt}{0.400pt}}
\multiput(393.00,197.95)(1.355,-0.447){3}{\rule{1.033pt}{0.108pt}}
\multiput(393.00,198.17)(4.855,-3.000){2}{\rule{0.517pt}{0.400pt}}
\multiput(400.00,194.94)(1.066,-0.468){5}{\rule{0.900pt}{0.113pt}}
\multiput(400.00,195.17)(6.132,-4.000){2}{\rule{0.450pt}{0.400pt}}
\multiput(408.00,190.95)(1.579,-0.447){3}{\rule{1.167pt}{0.108pt}}
\multiput(408.00,191.17)(5.579,-3.000){2}{\rule{0.583pt}{0.400pt}}
\multiput(416.00,187.95)(1.579,-0.447){3}{\rule{1.167pt}{0.108pt}}
\multiput(416.00,188.17)(5.579,-3.000){2}{\rule{0.583pt}{0.400pt}}
\multiput(424.00,184.95)(1.579,-0.447){3}{\rule{1.167pt}{0.108pt}}
\multiput(424.00,185.17)(5.579,-3.000){2}{\rule{0.583pt}{0.400pt}}
\multiput(432.00,181.95)(1.579,-0.447){3}{\rule{1.167pt}{0.108pt}}
\multiput(432.00,182.17)(5.579,-3.000){2}{\rule{0.583pt}{0.400pt}}
\multiput(440.00,178.95)(1.579,-0.447){3}{\rule{1.167pt}{0.108pt}}
\multiput(440.00,179.17)(5.579,-3.000){2}{\rule{0.583pt}{0.400pt}}
\multiput(448.00,175.95)(1.579,-0.447){3}{\rule{1.167pt}{0.108pt}}
\multiput(448.00,176.17)(5.579,-3.000){2}{\rule{0.583pt}{0.400pt}}
\multiput(456.00,172.95)(1.579,-0.447){3}{\rule{1.167pt}{0.108pt}}
\multiput(456.00,173.17)(5.579,-3.000){2}{\rule{0.583pt}{0.400pt}}
\multiput(464.00,169.95)(1.579,-0.447){3}{\rule{1.167pt}{0.108pt}}
\multiput(464.00,170.17)(5.579,-3.000){2}{\rule{0.583pt}{0.400pt}}
\put(472,166.17){\rule{1.700pt}{0.400pt}}
\multiput(472.00,167.17)(4.472,-2.000){2}{\rule{0.850pt}{0.400pt}}
\multiput(480.00,164.95)(1.579,-0.447){3}{\rule{1.167pt}{0.108pt}}
\multiput(480.00,165.17)(5.579,-3.000){2}{\rule{0.583pt}{0.400pt}}
\put(488,161.17){\rule{1.700pt}{0.400pt}}
\multiput(488.00,162.17)(4.472,-2.000){2}{\rule{0.850pt}{0.400pt}}
\multiput(496.00,159.95)(1.579,-0.447){3}{\rule{1.167pt}{0.108pt}}
\multiput(496.00,160.17)(5.579,-3.000){2}{\rule{0.583pt}{0.400pt}}
\put(504,156.17){\rule{1.700pt}{0.400pt}}
\multiput(504.00,157.17)(4.472,-2.000){2}{\rule{0.850pt}{0.400pt}}
\put(512,154.17){\rule{1.700pt}{0.400pt}}
\multiput(512.00,155.17)(4.472,-2.000){2}{\rule{0.850pt}{0.400pt}}
\put(520,152.17){\rule{1.700pt}{0.400pt}}
\multiput(520.00,153.17)(4.472,-2.000){2}{\rule{0.850pt}{0.400pt}}
\put(528,150.17){\rule{1.700pt}{0.400pt}}
\multiput(528.00,151.17)(4.472,-2.000){2}{\rule{0.850pt}{0.400pt}}
\put(536,148.17){\rule{1.700pt}{0.400pt}}
\multiput(536.00,149.17)(4.472,-2.000){2}{\rule{0.850pt}{0.400pt}}
\put(544,146.17){\rule{1.700pt}{0.400pt}}
\multiput(544.00,147.17)(4.472,-2.000){2}{\rule{0.850pt}{0.400pt}}
\put(552,144.17){\rule{1.700pt}{0.400pt}}
\multiput(552.00,145.17)(4.472,-2.000){2}{\rule{0.850pt}{0.400pt}}
\put(560,142.67){\rule{1.686pt}{0.400pt}}
\multiput(560.00,143.17)(3.500,-1.000){2}{\rule{0.843pt}{0.400pt}}
\put(567,141.17){\rule{1.700pt}{0.400pt}}
\multiput(567.00,142.17)(4.472,-2.000){2}{\rule{0.850pt}{0.400pt}}
\put(575,139.67){\rule{1.927pt}{0.400pt}}
\multiput(575.00,140.17)(4.000,-1.000){2}{\rule{0.964pt}{0.400pt}}
\put(583,138.67){\rule{1.927pt}{0.400pt}}
\multiput(583.00,139.17)(4.000,-1.000){2}{\rule{0.964pt}{0.400pt}}
\put(591,137.17){\rule{1.700pt}{0.400pt}}
\multiput(591.00,138.17)(4.472,-2.000){2}{\rule{0.850pt}{0.400pt}}
\put(599,135.67){\rule{1.927pt}{0.400pt}}
\multiput(599.00,136.17)(4.000,-1.000){2}{\rule{0.964pt}{0.400pt}}
\put(615,134.67){\rule{1.927pt}{0.400pt}}
\multiput(615.00,135.17)(4.000,-1.000){2}{\rule{0.964pt}{0.400pt}}
\put(623,133.67){\rule{1.927pt}{0.400pt}}
\multiput(623.00,134.17)(4.000,-1.000){2}{\rule{0.964pt}{0.400pt}}
\put(607.0,136.0){\rule[-0.200pt]{1.927pt}{0.400pt}}
\put(647,132.67){\rule{1.927pt}{0.400pt}}
\multiput(647.00,133.17)(4.000,-1.000){2}{\rule{0.964pt}{0.400pt}}
\put(631.0,134.0){\rule[-0.200pt]{3.854pt}{0.400pt}}
\put(671,132.67){\rule{1.927pt}{0.400pt}}
\multiput(671.00,132.17)(4.000,1.000){2}{\rule{0.964pt}{0.400pt}}
\put(655.0,133.0){\rule[-0.200pt]{3.854pt}{0.400pt}}
\put(703,133.67){\rule{1.927pt}{0.400pt}}
\multiput(703.00,133.17)(4.000,1.000){2}{\rule{0.964pt}{0.400pt}}
\put(679.0,134.0){\rule[-0.200pt]{5.782pt}{0.400pt}}
\put(726,134.67){\rule{1.927pt}{0.400pt}}
\multiput(726.00,134.17)(4.000,1.000){2}{\rule{0.964pt}{0.400pt}}
\put(711.0,135.0){\rule[-0.200pt]{3.613pt}{0.400pt}}
\put(742,135.67){\rule{1.927pt}{0.400pt}}
\multiput(742.00,135.17)(4.000,1.000){2}{\rule{0.964pt}{0.400pt}}
\put(750,136.67){\rule{1.927pt}{0.400pt}}
\multiput(750.00,136.17)(4.000,1.000){2}{\rule{0.964pt}{0.400pt}}
\put(758,137.67){\rule{1.927pt}{0.400pt}}
\multiput(758.00,137.17)(4.000,1.000){2}{\rule{0.964pt}{0.400pt}}
\put(734.0,136.0){\rule[-0.200pt]{1.927pt}{0.400pt}}
\put(774,138.67){\rule{1.927pt}{0.400pt}}
\multiput(774.00,138.17)(4.000,1.000){2}{\rule{0.964pt}{0.400pt}}
\put(782,139.67){\rule{1.927pt}{0.400pt}}
\multiput(782.00,139.17)(4.000,1.000){2}{\rule{0.964pt}{0.400pt}}
\put(790,140.67){\rule{1.927pt}{0.400pt}}
\multiput(790.00,140.17)(4.000,1.000){2}{\rule{0.964pt}{0.400pt}}
\put(798,141.67){\rule{1.927pt}{0.400pt}}
\multiput(798.00,141.17)(4.000,1.000){2}{\rule{0.964pt}{0.400pt}}
\put(806,143.17){\rule{1.700pt}{0.400pt}}
\multiput(806.00,142.17)(4.472,2.000){2}{\rule{0.850pt}{0.400pt}}
\put(814,144.67){\rule{1.927pt}{0.400pt}}
\multiput(814.00,144.17)(4.000,1.000){2}{\rule{0.964pt}{0.400pt}}
\put(822,145.67){\rule{1.927pt}{0.400pt}}
\multiput(822.00,145.17)(4.000,1.000){2}{\rule{0.964pt}{0.400pt}}
\put(830,147.17){\rule{1.700pt}{0.400pt}}
\multiput(830.00,146.17)(4.472,2.000){2}{\rule{0.850pt}{0.400pt}}
\put(838,148.67){\rule{1.927pt}{0.400pt}}
\multiput(838.00,148.17)(4.000,1.000){2}{\rule{0.964pt}{0.400pt}}
\put(846,150.17){\rule{1.700pt}{0.400pt}}
\multiput(846.00,149.17)(4.472,2.000){2}{\rule{0.850pt}{0.400pt}}
\put(854,152.17){\rule{1.700pt}{0.400pt}}
\multiput(854.00,151.17)(4.472,2.000){2}{\rule{0.850pt}{0.400pt}}
\put(862,154.17){\rule{1.700pt}{0.400pt}}
\multiput(862.00,153.17)(4.472,2.000){2}{\rule{0.850pt}{0.400pt}}
\put(870,156.17){\rule{1.500pt}{0.400pt}}
\multiput(870.00,155.17)(3.887,2.000){2}{\rule{0.750pt}{0.400pt}}
\put(877,158.17){\rule{1.700pt}{0.400pt}}
\multiput(877.00,157.17)(4.472,2.000){2}{\rule{0.850pt}{0.400pt}}
\put(885,160.17){\rule{1.700pt}{0.400pt}}
\multiput(885.00,159.17)(4.472,2.000){2}{\rule{0.850pt}{0.400pt}}
\put(893,162.17){\rule{1.700pt}{0.400pt}}
\multiput(893.00,161.17)(4.472,2.000){2}{\rule{0.850pt}{0.400pt}}
\put(901,164.17){\rule{1.700pt}{0.400pt}}
\multiput(901.00,163.17)(4.472,2.000){2}{\rule{0.850pt}{0.400pt}}
\multiput(909.00,166.61)(1.579,0.447){3}{\rule{1.167pt}{0.108pt}}
\multiput(909.00,165.17)(5.579,3.000){2}{\rule{0.583pt}{0.400pt}}
\put(917,169.17){\rule{1.700pt}{0.400pt}}
\multiput(917.00,168.17)(4.472,2.000){2}{\rule{0.850pt}{0.400pt}}
\multiput(925.00,171.61)(1.579,0.447){3}{\rule{1.167pt}{0.108pt}}
\multiput(925.00,170.17)(5.579,3.000){2}{\rule{0.583pt}{0.400pt}}
\multiput(933.00,174.61)(1.579,0.447){3}{\rule{1.167pt}{0.108pt}}
\multiput(933.00,173.17)(5.579,3.000){2}{\rule{0.583pt}{0.400pt}}
\put(941,177.17){\rule{1.700pt}{0.400pt}}
\multiput(941.00,176.17)(4.472,2.000){2}{\rule{0.850pt}{0.400pt}}
\multiput(949.00,179.61)(1.579,0.447){3}{\rule{1.167pt}{0.108pt}}
\multiput(949.00,178.17)(5.579,3.000){2}{\rule{0.583pt}{0.400pt}}
\put(766.0,139.0){\rule[-0.200pt]{1.927pt}{0.400pt}}
\put(162.0,131.0){\rule[-0.200pt]{0.400pt}{191.515pt}}
\put(162.0,131.0){\rule[-0.200pt]{191.515pt}{0.400pt}}
\put(957.0,131.0){\rule[-0.200pt]{0.400pt}{191.515pt}}
\put(162.0,926.0){\rule[-0.200pt]{191.515pt}{0.400pt}}
\end{picture}
  \end{center}
  \caption{\capstyle Exponent of the best approximation
    $\langle{X^2(t)}\rangle\sim{x^\nu}$ as a function of $\rho$. The
    squares are the values that have been calculated from the
    simulation (in each case the best gene as resulting from the
    genetic algorithm has been used), the continuous line is a spline
    interpolation \cite{press:86}. The value $\nu\sim{1}$ is
    characteristic of standard Brownian motion. For very low and high
    $\rho$, the walk is essentially Brownian: when the density is very
    low, the forager wanders long distances and spends comparatively
    little time on each patch. When $\rho\sim{1}$, there is no ARS, as
    the food is everywhere. The region $0.1<\rho<0.4$ is that in which
    ARS is clearly taking place.}
  \label{varplot}
\end{figure}

\section{\sc Should I stay of Should I go?}
In the next section, I shall analyze the global characteristics of ARS
exploration considering it as a random walk such as those that emerge
from our genetic experiment. Before that, in this section I shall
consider a more basic problem. Suppose that you are in a patch.  For a
while, you stay there happily eating, as there is plenty of
food. After a while, the food begins to dwindle, the resources of the
patch begin to be exhausted. When is it a good time to leave? You are
confronted with two contrasting criteria. On the one hand, staying
implies that you can continue eating without having to make a possibly
long journey without food before you find another patch. In the long
run, you want to spend more time in a patch and less between
patches. On the other hand, the new patch that you will find has lots
and lots of food so it might be a good idea for you to move now to
greener pastures instead of half starving in this half barren
patch. When is it a good time to leave? This is the question I want to
answer in this section.  I shall consider a very simple model: I
analyze only the time that an individual spends on a patch ($t_p$) and
the time that it spends between patches ($t_b$), and how to optimize
them for maximum foraging efficiency. In this, I follow essentially
the techniques developed for \emph{optimal foraging theory}
\cite{stephens:86}.

Suppose that a forager searches for food for a certain (long) amount of time. It
spends a total time $T_b$ looking for the next patch, and a total time
$T_p$ staying on a patch and eating (all the symbols used in this
section are shown in Table~\ref{forasymbols})
\begin{table}[bht]
  \begin{center}
    \begin{tabular}{|l|p{25em}|}
      \hline
      \multicolumn{1}{|c|}{Symbol} &  \multicolumn{1}{|c|}{Meaning} \\
      \hline
      $T_b$  & Total time spent looking for a patch, \\
      $T_p$  & total time spent on a patch, \\
      $t_b$  & average time spent looking for a patch, \\
      $t_p$  & average time spent on a patch, \\
      $G$    & total resource gain, \\
      $g$    & average gain per patch, \\
      $g(t)$ & gain per patch as a function of the time spent on a patch, \\
      $R$    & rate of reward: average gain per time unit, \\
      $\pi$  & ${=g/t_p}$, profitability of a patch (resource per unit time when on the patch), \\
      $\lambda$ & average number of patches found per unit time, \\
      $P$    & number of types of patches, \\
      $p_i$  & probability of using a resource of type $i$. \\
      \hline
    \end{tabular}
  \end{center}
  \caption{\capstyle Symbols used in this section.}
  \label{forasymbols}
\end{table}
If the total gain of the activity is $G$, then the rate of reward
(reward per unit time), is
\begin{equation}
  \label{totalgain}
  R = \frac{G}{T_b+T_p}
\end{equation}

This equation is inconvenient as it depends on the total times $T_p$
and $T_b$ and on the total gain $G$ (if the time spent goes to
infinity, both the numerator and the denominator go to infinity). One
can derive a more convenient equation, independent on the actual
foraging time, by considering the average on-patch time $t_p$, the
average time between patches $t_b$ and the average gain per patch
$g$. The rate at which patches are discovered, that is, the number of
patches discovered per unit time, is $\lambda=1/t_b$. The total number
of patches discovered during foraging is therefore $\lambda{T_b}$. The
total gain and the total time spent on patches depend on this number,
that is:
\begin{equation}
  \begin{aligned}
    G &= \lambda T_b g \\
    T_p &= \lambda T_b t_p 
  \end{aligned}
\end{equation}
Introducing these values in (\ref{totalgain}) we have
\begin{equation}
  R = \frac{\lambda T_b g}{T_b + \lambda T_b t_p} = \frac{\lambda g}{1 + \lambda t_p}
\end{equation}
or, in terms of average times
\begin{equation}
  R = \frac{g/t_b}{1 + t_p/t_b} = \frac{g}{t_b+t_p}
\end{equation}
This equation is known as the \emph{Holling disk} equation \cite{holling:59}%
\footnote{The name \dqt{disk equation} has nothing to do with the
  properties of the equation. Holling developed his model by studying
  the behavior of a blindfolded researcher assistant who was given the
  task of picking up randomly scattered sandpaper disks.}%
. Define the profitability of a patch as $\pi=g/t_p$, that is, the
gain per unit of time spent on the patch. With this definition we have:
\begin{equation}
  R = \frac{\pi}{{\displaystyle 1+\frac{t_b}{t_p}}}
\end{equation}
When the patches become more and more dense, then $t_b\rightarrow{0}$, and 
\begin{equation}
  \label{pilimit}
  \lim_{t_b\rightarrow{0}} \frac{\pi}{{\displaystyle 1+\frac{t_b}{t_p}}} = \pi
\end{equation}
This simple model can be extended in several ways. One very useful one
is to consider that a patch has \emph{diminishing returns}: as the
forager spends time on a path, its resources become depleted, so it
becomes harder to get rewards, and the profitability of the patch is
reduced.

That is, the reward $g$ is a function of $t$, $g(t)$, that tells us
how much reward one accumulates while foraging on a patch for a time
equal to $t$. The profitability is also a function of time:
$\pi(t)=g(t)/t$. Physical considerations place certain constraints on
these functions. The gain is positive and cumulative (you never lose
what you have gained), which implies $g(t)\ge{0}$ and
$g'(t)\ge{0}$. The first inequality also entails $\pi(t)\ge{0}$. On
the other hand, it is reasonable to assume that, as time goes by and
the resources become depleted, it will take longer and longer to amass
the same amount of reward; this entails $\pi'(t)\le{0}$. These two
relations imply $\lim_{t\rightarrow\infty}\pi(t)=C\ge{0}$. The
condition $\pi'(t)\le{0}$ imposes conditions on $g'(t)$. From
$\pi(t)=g(t)/t$, we have
\begin{equation}
  \pi'(t) = \frac{1}{t^2}\Bigl[g'(t)t-g(t)\Bigr] \le 0
\end{equation}
that is
\begin{equation}
  \label{gprime}
  g'(t)\le \frac{g(t)}{t} = \pi(t)
\end{equation}
I assume certain regularity conditions. In particular, that $\pi(t)$
decreases without \dqt{bumps}, that is, that $\pi'$ is monotonically
increasing, which entails that $\pi''\ge{0}$. Similarly, I assume
$g''(t)\le{0}$. Note that in the most common case the patch will
become depleted, that is,
\begin{equation}
  \lim_{t\rightarrow\infty} g(t) = g_\infty > 0
\end{equation}
but the analysis applies to the more general case in which $g(t)$ goes
to infinity slower than a linear function.

Given the average between-patches time $t_b$, we are interested in
finding the optimal time that the forager should spend on a patch
($t_p$) to maximize the reward $R$. The idea is that if you spend too
little time on a patch, then you don't take full advantage of its
resources, and spend comparatively too much time without patches, in
an area where you have no reward: your rate of gain will be reduced.

On the other hand, since the resources get depleted as we stay on a
patch, if you spend too much time there you shall waste your time on a
depleted patch that won't yield too much, while it would be more
convenient to invest some time ($t_b$) to find a new patch with better
yield. Given the equality
\begin{equation}
  R(t) = \frac{g(t)}{t_b+t}
\end{equation}
compute the derivative
\begin{equation}
  \frac{\partial R}{\partial t} = \frac{g'(t)(t_b+t)-g(t)}{(t_b+t)^2}
\end{equation}
It is $\partial{R}/\partial{t}=0$ if $g'(t)(t_b+t)-g(t)=0$, that is, if
\begin{equation}
  g'(t)=\frac{g(t)}{t_b+t}=R
\end{equation}
This result is known as the \emph{Charnov's Marginal Value Theorem}
\cite{charnov:76}.

This equation has a simple geometric interpretation. The average gain
$R$ results in a straight line in a $t$-$g$ diagram
(Figure~\ref{patchdiag}).
\begin{figure}
  \begin{center}
    \setlength{\unitlength}{1em}
\begin{picture}(17.746531,10.000000)(-6.760408,0)
\thicklines
\put(-6.760408,0){\line(1,0){17.746531}}
\put(0,0){\line(0,1){10.000000}}
\thinlines
\put(0.000000,0.000000){\circle*{0.000001}}
\put(0.036620,0.072973){\circle*{0.000001}}
\put(0.073241,0.145414){\circle*{0.000001}}
\put(0.109861,0.217326){\circle*{0.000001}}
\put(0.146482,0.288714){\circle*{0.000001}}
\put(0.183102,0.359580){\circle*{0.000001}}
\put(0.219722,0.429929){\circle*{0.000001}}
\put(0.256343,0.499765){\circle*{0.000001}}
\put(0.292963,0.569091){\circle*{0.000001}}
\put(0.329584,0.637912){\circle*{0.000001}}
\put(0.366204,0.706230){\circle*{0.000001}}
\put(0.402825,0.774050){\circle*{0.000001}}
\put(0.439445,0.841375){\circle*{0.000001}}
\put(0.476065,0.908208){\circle*{0.000001}}
\put(0.512686,0.974554){\circle*{0.000001}}
\put(0.549306,1.040415){\circle*{0.000001}}
\put(0.585927,1.105796){\circle*{0.000001}}
\put(0.622547,1.170700){\circle*{0.000001}}
\put(0.659167,1.235131){\circle*{0.000001}}
\put(0.695788,1.299091){\circle*{0.000001}}
\put(0.732408,1.362584){\circle*{0.000001}}
\put(0.769029,1.425614){\circle*{0.000001}}
\put(0.805649,1.488184){\circle*{0.000001}}
\put(0.842269,1.550298){\circle*{0.000001}}
\put(0.878890,1.611958){\circle*{0.000001}}
\put(0.915510,1.673168){\circle*{0.000001}}
\put(0.952131,1.733932){\circle*{0.000001}}
\put(0.988751,1.794252){\circle*{0.000001}}
\put(1.025371,1.854132){\circle*{0.000001}}
\put(1.061992,1.913575){\circle*{0.000001}}
\put(1.098612,1.972584){\circle*{0.000001}}
\put(1.135233,2.031163){\circle*{0.000001}}
\put(1.171853,2.089314){\circle*{0.000001}}
\put(1.208474,2.147041){\circle*{0.000001}}
\put(1.245094,2.204347){\circle*{0.000001}}
\put(1.281714,2.261234){\circle*{0.000001}}
\put(1.318335,2.317706){\circle*{0.000001}}
\put(1.354955,2.373767){\circle*{0.000001}}
\put(1.391576,2.429418){\circle*{0.000001}}
\put(1.428196,2.484663){\circle*{0.000001}}
\put(1.464816,2.539505){\circle*{0.000001}}
\put(1.501437,2.593946){\circle*{0.000001}}
\put(1.538057,2.647991){\circle*{0.000001}}
\put(1.574678,2.701641){\circle*{0.000001}}
\put(1.611298,2.754899){\circle*{0.000001}}
\put(1.647918,2.807769){\circle*{0.000001}}
\put(1.684539,2.860253){\circle*{0.000001}}
\put(1.721159,2.912354){\circle*{0.000001}}
\put(1.757780,2.964075){\circle*{0.000001}}
\put(1.794400,3.015418){\circle*{0.000001}}
\put(1.831020,3.066387){\circle*{0.000001}}
\put(1.867641,3.116984){\circle*{0.000001}}
\put(1.904261,3.167212){\circle*{0.000001}}
\put(1.940882,3.217073){\circle*{0.000001}}
\put(1.977502,3.266570){\circle*{0.000001}}
\put(2.014123,3.315706){\circle*{0.000001}}
\put(2.050743,3.364484){\circle*{0.000001}}
\put(2.087363,3.412905){\circle*{0.000001}}
\put(2.123984,3.460973){\circle*{0.000001}}
\put(2.160604,3.508691){\circle*{0.000001}}
\put(2.197225,3.556060){\circle*{0.000001}}
\put(2.233845,3.603083){\circle*{0.000001}}
\put(2.270465,3.649764){\circle*{0.000001}}
\put(2.307086,3.696104){\circle*{0.000001}}
\put(2.343706,3.742105){\circle*{0.000001}}
\put(2.380327,3.787771){\circle*{0.000001}}
\put(2.416947,3.833104){\circle*{0.000001}}
\put(2.453567,3.878106){\circle*{0.000001}}
\put(2.490188,3.922779){\circle*{0.000001}}
\put(2.526808,3.967126){\circle*{0.000001}}
\put(2.563429,4.011150){\circle*{0.000001}}
\put(2.600049,4.054853){\circle*{0.000001}}
\put(2.636669,4.098237){\circle*{0.000001}}
\put(2.673290,4.141304){\circle*{0.000001}}
\put(2.709910,4.184056){\circle*{0.000001}}
\put(2.746531,4.226497){\circle*{0.000001}}
\put(2.783151,4.268628){\circle*{0.000001}}
\put(2.819772,4.310452){\circle*{0.000001}}
\put(2.856392,4.351971){\circle*{0.000001}}
\put(2.893012,4.393186){\circle*{0.000001}}
\put(2.929633,4.434101){\circle*{0.000001}}
\put(2.966253,4.474717){\circle*{0.000001}}
\put(3.002874,4.515037){\circle*{0.000001}}
\put(3.039494,4.555062){\circle*{0.000001}}
\put(3.076114,4.594796){\circle*{0.000001}}
\put(3.112735,4.634239){\circle*{0.000001}}
\put(3.149355,4.673395){\circle*{0.000001}}
\put(3.185976,4.712265){\circle*{0.000001}}
\put(3.222596,4.750851){\circle*{0.000001}}
\put(3.259216,4.789156){\circle*{0.000001}}
\put(3.295837,4.827181){\circle*{0.000001}}
\put(3.332457,4.864929){\circle*{0.000001}}
\put(3.369078,4.902401){\circle*{0.000001}}
\put(3.405698,4.939600){\circle*{0.000001}}
\put(3.442319,4.976528){\circle*{0.000001}}
\put(3.478939,5.013186){\circle*{0.000001}}
\put(3.515559,5.049576){\circle*{0.000001}}
\put(3.552180,5.085701){\circle*{0.000001}}
\put(3.588800,5.121562){\circle*{0.000001}}
\put(3.625421,5.157162){\circle*{0.000001}}
\put(3.662041,5.192501){\circle*{0.000001}}
\put(3.698661,5.227583){\circle*{0.000001}}
\put(3.735282,5.262409){\circle*{0.000001}}
\put(3.771902,5.296981){\circle*{0.000001}}
\put(3.808523,5.331300){\circle*{0.000001}}
\put(3.845143,5.365369){\circle*{0.000001}}
\put(3.881763,5.399190){\circle*{0.000001}}
\put(3.918384,5.432763){\circle*{0.000001}}
\put(3.955004,5.466092){\circle*{0.000001}}
\put(3.991625,5.499177){\circle*{0.000001}}
\put(4.028245,5.532021){\circle*{0.000001}}
\put(4.064865,5.564626){\circle*{0.000001}}
\put(4.101486,5.596992){\circle*{0.000001}}
\put(4.138106,5.629122){\circle*{0.000001}}
\put(4.174727,5.661018){\circle*{0.000001}}
\put(4.211347,5.692681){\circle*{0.000001}}
\put(4.247968,5.724113){\circle*{0.000001}}
\put(4.284588,5.755315){\circle*{0.000001}}
\put(4.321208,5.786290){\circle*{0.000001}}
\put(4.357829,5.817039){\circle*{0.000001}}
\put(4.394449,5.847564){\circle*{0.000001}}
\put(4.431070,5.877865){\circle*{0.000001}}
\put(4.467690,5.907946){\circle*{0.000001}}
\put(4.504310,5.937807){\circle*{0.000001}}
\put(4.540931,5.967450){\circle*{0.000001}}
\put(4.577551,5.996877){\circle*{0.000001}}
\put(4.614172,6.026089){\circle*{0.000001}}
\put(4.650792,6.055088){\circle*{0.000001}}
\put(4.687412,6.083875){\circle*{0.000001}}
\put(4.724033,6.112452){\circle*{0.000001}}
\put(4.760653,6.140821){\circle*{0.000001}}
\put(4.797274,6.168983){\circle*{0.000001}}
\put(4.833894,6.196939){\circle*{0.000001}}
\put(4.870514,6.224691){\circle*{0.000001}}
\put(4.907135,6.252241){\circle*{0.000001}}
\put(4.943755,6.279589){\circle*{0.000001}}
\put(4.980376,6.306738){\circle*{0.000001}}
\put(5.016996,6.333689){\circle*{0.000001}}
\put(5.053617,6.360444){\circle*{0.000001}}
\put(5.090237,6.387003){\circle*{0.000001}}
\put(5.126857,6.413368){\circle*{0.000001}}
\put(5.163478,6.439541){\circle*{0.000001}}
\put(5.200098,6.465523){\circle*{0.000001}}
\put(5.236719,6.491315){\circle*{0.000001}}
\put(5.273339,6.516919){\circle*{0.000001}}
\put(5.309959,6.542336){\circle*{0.000001}}
\put(5.346580,6.567568){\circle*{0.000001}}
\put(5.383200,6.592615){\circle*{0.000001}}
\put(5.419821,6.617480){\circle*{0.000001}}
\put(5.456441,6.642163){\circle*{0.000001}}
\put(5.493061,6.666667){\circle*{0.000001}}
\put(5.529682,6.690991){\circle*{0.000001}}
\put(5.566302,6.715138){\circle*{0.000001}}
\put(5.602923,6.739109){\circle*{0.000001}}
\put(5.639543,6.762905){\circle*{0.000001}}
\put(5.676163,6.786527){\circle*{0.000001}}
\put(5.712784,6.809976){\circle*{0.000001}}
\put(5.749404,6.833255){\circle*{0.000001}}
\put(5.786025,6.856364){\circle*{0.000001}}
\put(5.822645,6.879304){\circle*{0.000001}}
\put(5.859266,6.902077){\circle*{0.000001}}
\put(5.895886,6.924683){\circle*{0.000001}}
\put(5.932506,6.947125){\circle*{0.000001}}
\put(5.969127,6.969403){\circle*{0.000001}}
\put(6.005747,6.991518){\circle*{0.000001}}
\put(6.042368,7.013472){\circle*{0.000001}}
\put(6.078988,7.035265){\circle*{0.000001}}
\put(6.115608,7.056900){\circle*{0.000001}}
\put(6.152229,7.078377){\circle*{0.000001}}
\put(6.188849,7.099697){\circle*{0.000001}}
\put(6.225470,7.120861){\circle*{0.000001}}
\put(6.262090,7.141871){\circle*{0.000001}}
\put(6.298710,7.162728){\circle*{0.000001}}
\put(6.335331,7.183433){\circle*{0.000001}}
\put(6.371951,7.203986){\circle*{0.000001}}
\put(6.408572,7.224389){\circle*{0.000001}}
\put(6.445192,7.244644){\circle*{0.000001}}
\put(6.481813,7.264751){\circle*{0.000001}}
\put(6.518433,7.284711){\circle*{0.000001}}
\put(6.555053,7.304525){\circle*{0.000001}}
\put(6.591674,7.324195){\circle*{0.000001}}
\put(6.628294,7.343721){\circle*{0.000001}}
\put(6.664915,7.363105){\circle*{0.000001}}
\put(6.701535,7.382347){\circle*{0.000001}}
\put(6.738155,7.401449){\circle*{0.000001}}
\put(6.774776,7.420411){\circle*{0.000001}}
\put(6.811396,7.439235){\circle*{0.000001}}
\put(6.848017,7.457922){\circle*{0.000001}}
\put(6.884637,7.476473){\circle*{0.000001}}
\put(6.921257,7.494888){\circle*{0.000001}}
\put(6.957878,7.513168){\circle*{0.000001}}
\put(6.994498,7.531315){\circle*{0.000001}}
\put(7.031119,7.549330){\circle*{0.000001}}
\put(7.067739,7.567214){\circle*{0.000001}}
\put(7.104359,7.584966){\circle*{0.000001}}
\put(7.140980,7.602590){\circle*{0.000001}}
\put(7.177600,7.620084){\circle*{0.000001}}
\put(7.214221,7.637451){\circle*{0.000001}}
\put(7.250841,7.654692){\circle*{0.000001}}
\put(7.287462,7.671806){\circle*{0.000001}}
\put(7.324082,7.688796){\circle*{0.000001}}
\put(7.360702,7.705661){\circle*{0.000001}}
\put(7.397323,7.722404){\circle*{0.000001}}
\put(7.433943,7.739024){\circle*{0.000001}}
\put(7.470564,7.755523){\circle*{0.000001}}
\put(7.507184,7.771902){\circle*{0.000001}}
\put(7.543804,7.788161){\circle*{0.000001}}
\put(7.580425,7.804302){\circle*{0.000001}}
\put(7.617045,7.820324){\circle*{0.000001}}
\put(7.653666,7.836230){\circle*{0.000001}}
\put(7.690286,7.852020){\circle*{0.000001}}
\put(7.726906,7.867694){\circle*{0.000001}}
\put(7.763527,7.883255){\circle*{0.000001}}
\put(7.800147,7.898701){\circle*{0.000001}}
\put(7.836768,7.914035){\circle*{0.000001}}
\put(7.873388,7.929257){\circle*{0.000001}}
\put(7.910008,7.944368){\circle*{0.000001}}
\put(7.946629,7.959369){\circle*{0.000001}}
\put(7.983249,7.974260){\circle*{0.000001}}
\put(8.019870,7.989042){\circle*{0.000001}}
\put(8.056490,8.003717){\circle*{0.000001}}
\put(8.093111,8.018284){\circle*{0.000001}}
\put(8.129731,8.032746){\circle*{0.000001}}
\put(8.166351,8.047101){\circle*{0.000001}}
\put(8.202972,8.061352){\circle*{0.000001}}
\put(8.239592,8.075499){\circle*{0.000001}}
\put(8.276213,8.089543){\circle*{0.000001}}
\put(8.312833,8.103484){\circle*{0.000001}}
\put(8.349453,8.117324){\circle*{0.000001}}
\put(8.386074,8.131062){\circle*{0.000001}}
\put(8.422694,8.144700){\circle*{0.000001}}
\put(8.459315,8.158239){\circle*{0.000001}}
\put(8.495935,8.171679){\circle*{0.000001}}
\put(8.532555,8.185021){\circle*{0.000001}}
\put(8.569176,8.198265){\circle*{0.000001}}
\put(8.605796,8.211413){\circle*{0.000001}}
\put(8.642417,8.224465){\circle*{0.000001}}
\put(8.679037,8.237422){\circle*{0.000001}}
\put(8.715657,8.250284){\circle*{0.000001}}
\put(8.752278,8.263052){\circle*{0.000001}}
\put(8.788898,8.275727){\circle*{0.000001}}
\put(8.825519,8.288310){\circle*{0.000001}}
\put(8.862139,8.300800){\circle*{0.000001}}
\put(8.898760,8.313200){\circle*{0.000001}}
\put(8.935380,8.325509){\circle*{0.000001}}
\put(8.972000,8.337729){\circle*{0.000001}}
\put(9.008621,8.349859){\circle*{0.000001}}
\put(9.045241,8.361900){\circle*{0.000001}}
\put(9.081862,8.373854){\circle*{0.000001}}
\put(9.118482,8.385721){\circle*{0.000001}}
\put(9.155102,8.397500){\circle*{0.000001}}
\put(9.191723,8.409194){\circle*{0.000001}}
\put(9.228343,8.420803){\circle*{0.000001}}
\put(9.264964,8.432327){\circle*{0.000001}}
\put(9.301584,8.443767){\circle*{0.000001}}
\put(9.338204,8.455123){\circle*{0.000001}}
\put(9.374825,8.466397){\circle*{0.000001}}
\put(9.411445,8.477588){\circle*{0.000001}}
\put(9.448066,8.488697){\circle*{0.000001}}
\put(9.484686,8.499726){\circle*{0.000001}}
\put(9.521307,8.510674){\circle*{0.000001}}
\put(9.557927,8.521542){\circle*{0.000001}}
\put(9.594547,8.532331){\circle*{0.000001}}
\put(9.631168,8.543041){\circle*{0.000001}}
\put(9.667788,8.553673){\circle*{0.000001}}
\put(9.704409,8.564227){\circle*{0.000001}}
\put(9.741029,8.574704){\circle*{0.000001}}
\put(9.777649,8.585105){\circle*{0.000001}}
\put(9.814270,8.595430){\circle*{0.000001}}
\put(9.850890,8.605680){\circle*{0.000001}}
\put(9.887511,8.615855){\circle*{0.000001}}
\put(9.924131,8.625955){\circle*{0.000001}}
\put(9.960751,8.635982){\circle*{0.000001}}
\put(9.997372,8.645936){\circle*{0.000001}}
\put(10.033992,8.655817){\circle*{0.000001}}
\put(10.070613,8.665626){\circle*{0.000001}}
\put(10.107233,8.675363){\circle*{0.000001}}
\put(10.143853,8.685029){\circle*{0.000001}}
\put(10.180474,8.694625){\circle*{0.000001}}
\put(10.217094,8.704151){\circle*{0.000001}}
\put(10.253715,8.713607){\circle*{0.000001}}
\put(10.290335,8.722994){\circle*{0.000001}}
\put(10.326956,8.732313){\circle*{0.000001}}
\put(10.363576,8.741564){\circle*{0.000001}}
\put(10.400196,8.750747){\circle*{0.000001}}
\put(10.436817,8.759863){\circle*{0.000001}}
\put(10.473437,8.768913){\circle*{0.000001}}
\put(10.510058,8.777896){\circle*{0.000001}}
\put(10.546678,8.786815){\circle*{0.000001}}
\put(10.583298,8.795668){\circle*{0.000001}}
\put(10.619919,8.804456){\circle*{0.000001}}
\put(10.656539,8.813180){\circle*{0.000001}}
\put(10.693160,8.821841){\circle*{0.000001}}
\put(10.729780,8.830438){\circle*{0.000001}}
\put(10.766400,8.838973){\circle*{0.000001}}
\put(10.803021,8.847445){\circle*{0.000001}}
\put(10.839641,8.855856){\circle*{0.000001}}
\put(10.876262,8.864205){\circle*{0.000001}}
\put(10.912882,8.872493){\circle*{0.000001}}
\put(10.949502,8.880721){\circle*{0.000001}}
\put(-4.506939,0){\line(3,2){15.493061}}
\multiput(5.493061,-1)(0,0.766667){12}{\line(0,1){0.383333}}
\put(5.493061,-0.8){\makebox(0,0)[t]{$t_p$}}
\put(-4.506939,-0.8){\makebox(0,0)[t]{$-t_b$}}
\put(10.986123,8.444444){\makebox(0,0)[tl]{$g(t)$}}
\put(10.986123,10.328708){\makebox(0,0)[tl]{$R$}}
\put(-3.605551,0.600925){\line(0,1){1}}
\put(-3.605551,1.600925){\line(1,0){1.500000}}
\put(-3.805551,0.800925){\makebox(0,0)[br]{$R$}}
\end{picture}

  \end{center}
  \caption{\capstyle A graphical illustration of Charnov's Marginal Value
    Theorem. The patch time $t_w$ that maximizes the rate of gain $R$
    occurs when the line with slope $R$ is tangent to the function
    $g(t)$, that is, when the instantaneous rate of gain on the patch
    equals the average rate of gain $R$.}
  \label{patchdiag}
\end{figure}
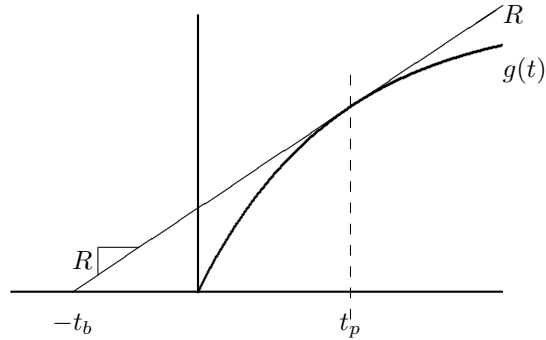
The patch gain is a curve that stays at zero for a time $t_b$ and then
grows as $g(t)$; the optimal $t_p$ occurs when the slope of the curve
$g(t)$ is equal to the average rate of gain. Note that there are two
ways in which the environment can change so that $R$ increases. First,
the profitability of the patch may increase, that is, the curve $g(t)$
can be pulled upward (Figure~\ref{increase_patch}.a). Second, the
patches may become dense, reducing $t_b$
(Figure~\ref{increase_patch}.b). As $t_b\rightarrow{0}$ the rate $R$
approaches $\pi$ (the profitability of the patch), as per (\ref{pilimit}).
\begin{figure}
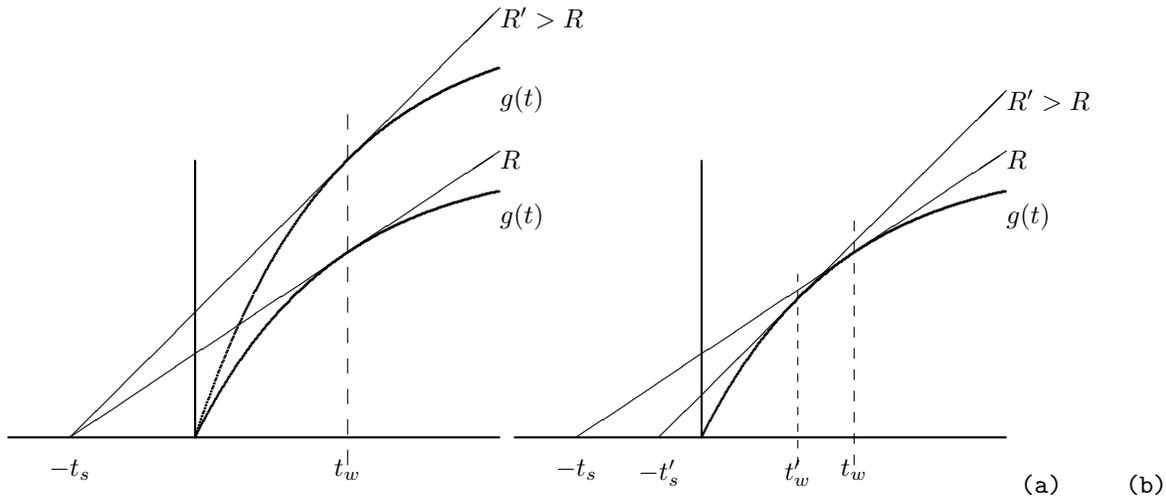

  \begin{center}
    \setlength{\unitlength}{1em}

  \end{center}
  \caption{\capstyle As a consequence of Charnov's Marginal Value
    Theorem, there are two ways to increase the average rate of gain
    $R$: one can either increase the profitability of a patch,
    i.e.\ raising the gain curge $g(t)$ as in (a), or make the patches
    more dense, thereby decreasing the average beyween-patches time
    $t_b$, as in (b).}
  \label{increase_patch}
\end{figure}
\example
Suppose we are looking for a low-priced hotel in Paris. We have
several web sites available, and our strategy is to log in to one,
start checking prices looking for the cheapest price for a while, then
move to another one looking for a new price or, simply, stop looking
and accept the lowest price we have found. The question is: how long
should we stay on the site and keep looking before we move on?

When we are on a page, the important events are the prices that we
look at so, for the sake of convenience, we take the time that it
takes to move from one hotel to the next one on the same page as the
time unit so that at time $t=n$ we have looked at $n+1$ prices (we
look at the first price at $t=0$). The actual length depends of what
we are looking for: if we look just for the best price, the interval
is very small; if we look for price subject to certain constraint
(hotel with a bar, with a sauna, etc.), it will take longer. In any
case, we look at a new hotel (one the same page) per unit of time. In
these units, let $t_b$ be the time that it takes to get set on a page
(including typing the address, logging in, etc.)

We begin by determinig the expected value of the minimum price that we
have observed if we have observed $n$ prices. In this simple example
we shall favor simplicity over plausibility and we shall assume that
the prices of teh hotels are uniformly distributed in the interval
$[\mu-a,\mu+a]$. 

To begin with, we shall answer an even simpler question: given $n$
observations $X=\{x_1,\ldots,x_n\}$ of $n$ random variables
independent and uniformly distributed in $[-1,1]$, which is the
expected value of $\min(X)$? The cumulative distribution and the
density for each of the $x_i$ are
\begin{equation}
  \Phi(x) = 
  \begin{cases}
    0 & x \le 1 \\
    \frac{x+1}{2} & -1<x<1 \\
    1 & x\ge 1
  \end{cases}
\end{equation}
and
\begin{equation}
  \phi(x) = 
  \begin{cases}
    \frac{1}{2} & -1\le{x}\le{1} \\
    0           & \mbox{otherwise}
  \end{cases}
\end{equation}
respectively. The probability density for the minimum is given by
(\ref{mindense}):
\begin{equation}
  \phi_{\min}(x) = n[1 - \Phi(x)]^{n-1}\phi(x) = 
  \begin{cases}
    \frac{n}{2^n}[1-x]^{n-1} & (-1\le{x}\le{1}) \\
    0                      & \mbox{otherwise}
  \end{cases}
\end{equation}
and its expected value is
\begin{equation}
  \begin{aligned}
    m(n) &= \int_{-1}^{1} u \phi_{\min}(u) du = \frac{n}{2^n} \int_{-1}^{1} u(1-u)^{n-1} du \\
         &= \frac{n}{2^n} \left[ \int_{-1}^{1} (1-u)^{n-1}du - \int_{-1}^{1}(1-u)^ndu\right] \\
         &= \frac{n}{2^n} \left[ - \int_{2}^{0} u^{n-1}du + \int_{2}^{0}u^n du\right] \\
         &= \frac{n}{2^n} \left[ \frac{2^n}{n} - \frac{2^{n+1}}{n+1} \right] \\
         &= \frac{1-n}{1+n}
  \end{aligned}
\end{equation}

Scaling and shifting one obtains, for variables distributed in $[\mu-a,\mu+a]$:
\begin{equation}
  m(n) = \mu - a \frac{1-n}{1+n}
\end{equation}
Note that if we only observe one price, the expected value of the
minimum is $\mu$. We we observe more prices, the expected value
decreases, with
\begin{equation}
  \lim_{t\rightarrow\infty} m(t) = \mu-a
\end{equation}

Let us consider that we start exploring at the time when we observe
the first price, and that everything that comes before that is
preparation that is included in the time $t_b$. At time $t$, we have
looked at $t+1$ prices, and the expected value for the minimum price
is $m(t+1)$. The gain that we have obtained is the difference between
the first price that we saw (expected value equal to $\mu$) and the
current one:
\begin{equation}
  \label{gdef}
  g(t) = m(1) = m(t+1) = a \frac{t}{t+2}
\end{equation}
Up to now, I have considered $t$ as a discrete variable (the number
of prices observed); from now on we regard it as a continuous
variable, so I can take the derivative:
\begin{equation}
  g'(t) = \frac{2a}{(t+2)^2}
\end{equation}
I now apply Charnov's Marginal value theorem: the optimal value to
spend on the site is given by
\begin{equation}
  g'(t) = \frac{2a}{(t+2)^2} = \frac{at}{t+2} \frac{1}{t+t_b} = \frac{g(t)}{t+t_b}
\end{equation}
which has solution $\tau_1=\sqrt{2t_b}$%
\footnote{Dimensionally, the equation is sound: the factor $t+2$ in
  (\ref{gdef}) entails that the term 2 has the dimensions of a time.}%
. The corresponding rate of reward is
\begin{equation}
  \label{t1bum}
  R_1 = g'(\tau_1) = \frac{2a}{(\sqrt{2t_b}+2)^2} = \frac{a}{(\sqrt{t_b}+\sqrt{2})^2}.
\end{equation}
Note that the time that we spend on a site grown sub-linearly with
the time we spend looking for the site: if we spend twice as long
looking for the web page, the time we should spend on the web page
grows like $\sqrt{2}$, that is, we should stay about 40\% longer.

~~\eoe

\example
The considerations of the previous example are valid for the first
\dqt{patch}, that is, for the first site that we visit. Suppose now
that we want to visit a second site (the time necessary to do the
switch is assumed to be $t_b$). Now we already have a minimum, the one
that we found in the first patch, namely
\begin{equation}
  m_1 \dfeq \mu - a\frac{\tau_1}{\tau_1+2} = \mu - a \frac{\sqrt{t_b}}{\sqrt{t_b}+\sqrt{2}} \dfeq \mu - a\alpha
\end{equation}
with 
\begin{equation}
  \alpha \dfeq \frac{\sqrt{t_b}}{\sqrt{t_b}+\sqrt{2}} < 1
\end{equation}
While we explore the second site, as long as the minimum price that we
find there is greater than $m_1$, our gain is zero. Assume that the
second patch has the same average price as the first, but a larger
spread, that is, in the second patch the prices are uniformly
distributed in $[\mu-b,\mu+b]$, with $b>a$. Also, define
$\gamma=b/a>1$ (I shall need it later).

The current minimum after we have explored the second site for a time
$t$ is
\begin{equation}
  \tilde{m}(t) = \mu - b\frac{t}{t+2}
\end{equation}
The expected minimum in the second site is the same as the optimal
price in the first at a time $t_0$ such that $\tilde{m}(t_0)=m_1$,
that is
\begin{equation}
  t_0 = \frac{2a\alpha}{b-a\alpha} = \frac{2\alpha}{\gamma-\alpha}
\end{equation}
The gain in the second site is given by the savings we obtain over the
previous minimum, that is
\begin{equation}
  g_2(t) = 
  \begin{cases}
    0 & t<t_0 \\
    b\frac{t}{t+2} - a\alpha & t \ge t_0
  \end{cases}
\end{equation}
and
\begin{equation}
  g_2^\prime(t) = 
  \begin{cases}
    0 & t<t_0 \\
    b\frac{2b}{(t+2)^2} - a\alpha & t \ge t_0
  \end{cases}
\end{equation}
The situation is depicted schematically in figure~\ref{patchdiag2}.
\begin{figure}
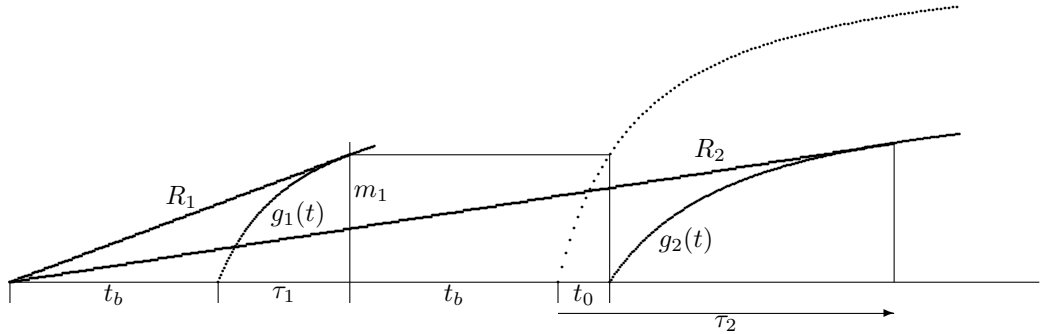

  \begin{center}
    \setlength{\unitlength}{1.5em}

  \end{center}
  \caption{\capstyle A graphical illustration of the Charnov's
    Marginal Value Theorem for two sites. When we get to the second
    site, we already have a minimum $m_1$ from the first site so that
    the gain $g_2(t)$ only is greater than zero starting from
    $t_0$. The second patch has prices distributed in $[\mu-b,\mu+b]$:
    for a large enough value of $\gamma=b/a$ the rate of gain $R_2$
    that we obtain by exploring the second site is greater than the
    rate $R_1$ obtained in the first site.}
  \label{patchdiag2}
\end{figure}
When we arrive at the second site, a time $T\dfeq{2t_b}+\tau_1$ has
already elapsed, so if we stay on the site a time $\tau$, the rate of
gain is:
\begin{equation}
  R_2(\tau) = \frac{g_2(\tau)}{T+\tau} = \frac{g_2(\tau)}{2t_b+\tau_1+\tau}
\end{equation}
To find the maximum of $R_2(\tau)$ we proceed as in the previous
example, setting
\begin{equation}
  g_2^\prime(\tau) = \frac{g_2(\tau)}{T+\tau}
\end{equation}
that is
\begin{equation}
  \frac{2b}{(\tau+2)^2} = \frac{1}{T+\tau} \Bigl[\frac{\tau b}{\tau+2} - a\alpha\Bigr]
\end{equation}
or
\begin{equation}
  \frac{2}{(\tau+2)^2} = \frac{1}{T+\tau} \Bigl[\frac{\tau}{\tau+2} - \frac{\alpha}{\gamma}\Bigr]
\end{equation}
that is
\begin{equation}
  2(T+\tau) = (\tau+2)\Bigl[\tau - \frac{\alpha}{\gamma}(\tau+2)\Bigr]
\end{equation}
Rearranging the terms, we obtain the equation
\begin{equation}
  (\gamma-\alpha)\tau^2-4\alpha\tau-(4\alpha+2\gamma{T})=0
\end{equation}
whose only positive solution is given by
\begin{equation}
  \begin{cases}
    \displaystyle \Delta = 8\gamma\Bigl[2\alpha+(\gamma-\alpha)T\Bigr] \\
    \displaystyle \tau_2 = \frac{4\alpha+\sqrt{\Delta}}{2(\gamma-\alpha)}
  \end{cases}
\end{equation}
The corresponding maximum rate of gain is
\begin{equation}
  R_2 = g_2^\prime(\tau_2) = \frac{2b}{(\tau_2+2)^2}
\end{equation}
Using the second patch is convenient if we improve our gain rate, that
is, if $R_2>R_1$, where $R_1$ is as in (\ref{t1bum}). This imposes a
condition on $b/a$, that is, on $\gamma$: it is convenient to use the
second patch if $\gamma$ is large enough that we can find a better
price that offsets the extra time spent in searching. The limit
condition $R_2=R_1$ yields
\begin{equation}
  \gamma = \left[ \frac{\tau_2+2}{\sqrt{2t_b}+2}\right]^2
\end{equation}
The value of $\tau_2$ depends on $\gamma$ in such a way that we can't
find a closed form solution to this equation. We can, however,
determine its limits. For $t_b\rightarrow{0}$, we have
$\alpha\rightarrow{0}$, $T\rightarrow{0}$, $\Delta\rightarrow{0}$, and
$\tau_2\rightarrow{0}$, therefore
\begin{equation}
  \lim_{t_b\rightarrow{0}} \gamma = 1
\end{equation}
For $t_b\rightarrow\infty$, $\alpha\rightarrow{1}$, $T\sum2t_b$, and 
\begin{equation}
  \begin{aligned}
    \Delta &\sim 16\gamma(\gamma-1)t_b \\
    \tau &\sim 2\frac{\sqrt{\gamma(\gamma-1)}}{\gamma-1}\sqrt{t_b}
  \end{aligned}
\end{equation}
The equation
\begin{equation}
  \label{gamma}
  \gamma = \left[ \frac{2\sqrt{\gamma(\gamma-1)}}{\sqrt{2}(\gamma-1)} \right]^2 = \frac{2\gamma}{\gamma-1}
\end{equation}
has solution $\gamma=3$, therefore
\begin{equation}
  \lim_{t_b\rightarrow\infty} \gamma = 3
\end{equation}

The behavior of $\gamma$ can be found solving (\ref{gamma})
numerically, which, given the stability of the equation, can be done
with a simple iteration. The result is shown in figure~\ref{gammafig}
\begin{figure}
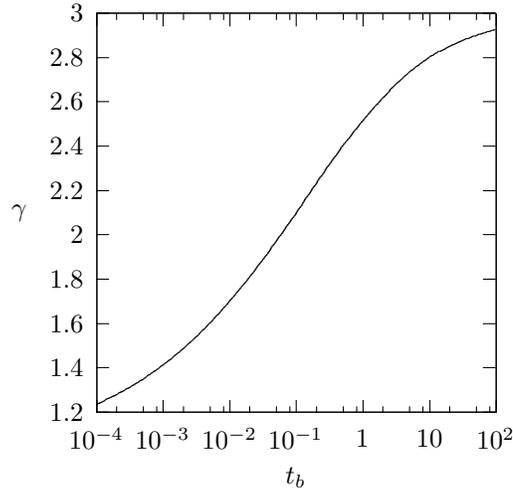

  \begin{center}
    \setlength{\unitlength}{20em}
\setlength{\unitlength}{0.240900pt}
\ifx\plotpoint\undefined\newsavebox{\plotpoint}\fi

  \end{center}
  \caption{The behavior of $\gamma$ as a function of $t_b$ (the
    horizontal axis is logarithmic). For a given $t_b$, if the ratio
    of the spreads of the two patches, $b/a$ is greater than the
    $\gamma$ shown in this graph, then it is convenient to explore the
    second patch.}
  \label{gammafig}
\end{figure}
For $\frac{b}{a}>\gamma(t_b)$ and switch time $t_b$, it is convenient
to explore the second patch.

\section{\sc Walking in Continuous Space}
We have seen, in the first section, the neurological basis of ARS, and
its widespread presence to solve many apparently unrelated problems in
the animal kingdom. All these problems have a common abstract
structure: that of foraging in a patchy environment, that is, in an
environment in which the resources are distributed in clumps: there
are concentrated areas in which the resource is present, separated by
stretches in which no (or little) resource is found.

Some fiddling with genetic algorithms convinced us that ARS is indeed
an optimal strategy for this kind of problems. The problem that I
should like to consider now is how to characterize this behavior at a large scale. I have shown in the previous section how to decide
when to leave a patch and venture in search of another, now I am interested in analyzing the global behavior that
results from these decisions. One
result, which we have already glimpsed at the end of section \ref{genetic} is
to see ARS as a type of \emph{random walk}. 
In this section, we shall study the paths produced by ARS as random
walks in a continuum (viz.\ in ${\mathbb{R}}^n$, usually in
${\mathbb{R}}^2$). As we shall see, an important parameter in this
exploration is the exponent $\nu$ of the curves in
Figure~\ref{variances}, that is, in the relation
$\langle{X^2(t)}\rangle\sim{t^\nu}$; the fact that $\nu>1$ makes ARS
walk of a peculiar kind, known as \emph{Levy walks}.

\subsection{Random Walks and Diffusion Processes}
A random walk is the description of the motion of a point (sometimes
called, in reference to the physics in which random walks were first
studied, a \emph{particle} or, in reference to ecology, an
\emph{individual}) subject to forces that can be modeled as a
stochastic process. Random walks can be described at three levels of
detail, given by the schema of Figure~\ref{walkonsunshine}).
\begin{figure}
  \begin{center}
    \begin{tabular}{cccp{6em}}
      Input & Description Level & Output & Math, formalism \\
      \hline
      \parbox{8em}{individual fluctuations} & \framebox{\parbox{8em}{\centerline{microscopic}}} & trajectories & Langevin equations \\
      \parbox{8em}{averaged fluctuations} & \framebox{\parbox{8em}{\centerline{mesoscopic}}} & prob. density & Master equation \\
      \parbox{8em}{macroscopic parameters} & \framebox{\parbox{8em}{\centerline{macroscopic}}} & prob. density & Fokker-Planck equations \\
    \end{tabular}
    \setlength{\unitlength}{1em}
    \begin{picture}(0,0)(0,0)
      \put(0,-4){\vector(0,1){8}}
      \put(0.2,-2){\makebox(0,0)[l]{complexity}}
    \end{picture}
  \end{center}
  \caption{\capstyle Three levels of description of random walks: the mesoscopic
    entails considering the time span of each motion small with
    respect to the time of the whole phenomenon, and approximating
    time functions with their first derivatives; the macroscopic
    entails making the same approximation on space. This entails that
    the master equation (mesocsopic) is a differential equation in time
    and an integral equation in space, while the Fokker-Planck
    equation (macroscopic) is differential in time and space. (Adapted
    from \cite{mendez:14}.)}
  \label{walkonsunshine}
\end{figure}
We shall consider these levels of description one by one, especially
as they apply to the best known model of random walk: \emph{Brownian
  motion}.

\subsection{Microscopic description--Langevin equations}
Langevin's equations were originally studied for what become the
prototype of diffusive random walks, namely \emph{Brownian motion}
\cite{hida:80,karatzas:12}. In 1827, botanist R. Brown discovered,
during microscopic observations, that particles of pollen suspended in
water exhibited an incessant and irregular motion
\cite{powles:78}. Vitalist explanations were soon discarded since
mineral (viz.\ non-living) particles exhibited the same
phenomenon. Brownian motion occurs when the mass of the particle of
pollen is larger than the mass of the molecules of the liquid, so that
the continuous collisions drive the particles in a chaotic way (Figure~\ref{brownian}). 
The first theoretical explanation of this phenomenon was given in 1905
by Albert Einstein at a macroscopic level \cite{einstein:05}, and we
shall consider his approach shortly. In 1906, Paul Langevin offered a
microscopic model of Brownian motion based on stochastic differential
equations. Langevin's equations for a one-dimensional Brownian motion
(the case that is commonly studied) are:
\begin{equation}
  \begin{aligned}
    \frac{dx}{dt} &= v \\
    m\frac{dv}{dt} &= -\gamma v + \sigma \xi(t)
  \end{aligned}
\end{equation}
where $m$ is the particle mass, $\gamma$ is the friction coefficient,
and $\xi(t)$ is the force resulting from the impact with the
molecules. Langevin assumed that $\xi(t)$ is a Gaussian, uncorrelated
stochastic process with zero mean, that is $\langle\xi(t)\rangle=0$,
$\langle\xi(t)\xi(t')\rangle=\delta(t-t')$. He also considered the
limit of strong friction $m|dv/dt|\ll|\gamma{v}|$ (more on this
hypothesis later) so the equations become
\begin{equation}
  \gamma \frac{dx}{dt} = \sigma\xi(t)
\end{equation}
and, defining the \emph{diffusion coefficient}
$D=\frac{\sigma^2}{2\gamma^2}$, 
\begin{equation}
  \frac{dx}{dt} = \sqrt{2D} \xi(t)
\end{equation}
or
\begin{equation}
  \label{intme}
  dx = \sqrt{2D}\xi(t) dt = \sqrt{2D}dW(t)
\end{equation}
where $W(t)$ is a Wiener process (see section
\ref{gausswie}). Integrating (\ref{intme}) we get
\begin{equation}
  x(t) = x(0) + \sqrt{2D}w(t)
\end{equation}
So, Brownian motion is the motion of a particle whose displacement is
a Wiener process and whose velocity is an uncorrelated Gaussian
process. The hypothesis of strong friction is key to obtain
velocity as an uncorrelated stochastic process: if inertial phenomena
are present, then the velocities at two different instants in time are
correlated.

From the solution of this equation we can determine macroscopic
quantities such as the mean position $\langle{X}\rangle$ and the mean
square displacement $\langle{X^2}\rangle-\langle{X}\rangle^2$:
\begin{equation}
  \label{lookieloopsie}
  \begin{aligned}
    \langle X \rangle &= \langle X(0) \rangle + \sqrt{2D} \langle W(t) \rangle = \langle X(0) \rangle \\
    \langle{X^2}\rangle-\langle{X}\rangle^2 &= 2D\langle W^2(t) \rangle = 2Dt
  \end{aligned}
\end{equation}
From the second equation we see that
$\langle{X^2}\rangle\sim{t}$. This is an equation that we have already
encountered: it characterizes the ARS walk for very low $\rho$ and for
$\rho\sim{1}$ (figure~\ref{variances}). It is a behavior typical of
\emph{diffusive processes}, and we shall meet it quite a few times in
the following.
\begin{figure}[tbhp]
  \begin{center}
    \setlength{\unitlength}{1.5em}
    \begin{picture}(15,14)(0,0)
      \newsavebox{\bump}
      \savebox{\bump}{
        \thicklines
        \put(0,0){\circle{1}}
        \thinlines
        \put(0,0){\circle{1.2}}
        \put(0,0){\circle{1.4}}
      }
      \put(4.5,5.5){\usebox{\bump}}
      \put(5.5,2.5){\usebox{\bump}}
      \put(8.5,8.5){\usebox{\bump}}
      \put(9.5,6.5){\usebox{\bump}}
      \put(6.5,7.5){\usebox{\bump}}
      \put(7.5,4.5){\usebox{\bump}}
      \put(13,4){\usebox{\bump}}
      \put(10.5,2.5){\usebox{\bump}}
      \put(12.5,6){\usebox{\bump}}
      \put(10,11.5){\usebox{\bump}}
      \put(6.5,10.5){\usebox{\bump}}
      \put(8.5,12.5){\usebox{\bump}}
      \put(0.5,1){\line(1,1){4}}
      \put(4.5,5){\line(1,-2){1}}
      \put(5.5,3){\line(3,5){3}}
      \put(8.5,8){\line(1,-2){0.5}}
      \put(9,7){\line(-4,1){2}}
      \put(7,7.5){\line(1,-5){0.5}}
      \put(7.5,5){\line(5,-1){5}}
      \put(12.5,4){\line(-3,-2){1.5}}
      \put(11,3){\line(1,3){1}}
      \put(12,6){\line(-2,5){2}}
      \put(10,11){\line(-6,-1){3}}
      \put(7,10.5){\line(1,2){1}}
      \put(8,12.5){\line(-1,2){0.5}}
      \put(3.5,1.5){\circle{1}}
      \put(7.5,0.5){\circle{1}}
      \put(8.5,2.5){\circle{1}}
      \put(2,7){\circle{1}}
      \put(2.5,4.5){\circle{1}}
      \put(4,9){\circle{1}}
      \put(2.5,11.5){\circle{1}}
      \put(4.5,12.5){\circle{1}}
      \put(12.5,10.5){\circle{1}}
      \put(13.5,8){\circle{1}}
    \end{picture}
  \end{center}
  \caption{\capstyle A schematic illustration of Brownian motion: a
    particle of pollen \dqt{bumps} on the molecules of the liquid in
    which it is suspended, creating an irregular random trajectory.}
  \label{brownian}
\end{figure}
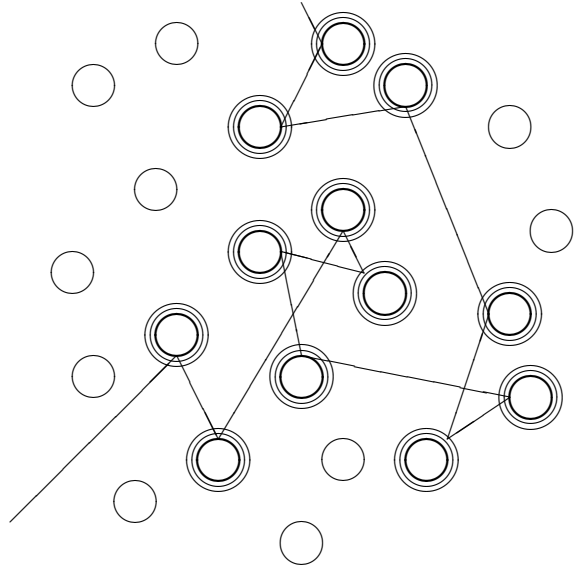

\subsection{Mesoscopic description--Master equation}
The \emph{master equation} is an ensemble equation that expresses the
probability $P(x,t)$ that a particle be at position $x$ at time
$t$. It is an integro-differential equation, expressing the time
derivative of $P(x,t)$ as a balance of the probability of arriving at
$x$ and the probability of leaving the position $x$ once we are there.

Consider a stationary Markov process. For this kind of process, the
probability $P(x,t|z,t')$ depends only on $t-t'$, so we can define
$P_\tau(x,z)=P(x,t+\tau|z,t)=P(x,\tau|z,0)$.

Consider now $P(x,t+\tau)$, that is, the probability that the walking
particle will be in position $x$ at time $t+\tau$. The particle is in $x$
at time $t+\tau$ if there is a position $z$ such that the particle was
in $z$ at time $t$ and has moved from $z$ to $x$ in the time interval
$\tau$ (if $z=x$, this is the probability that the particle were
already in $x$ and hasn't moved). This event, for a specific $z$, has
a probability $P(z,t)P_\tau(x,z)$. Integrating over all possible $z$,
we obtain
\begin{equation}
  P(x,t+\tau) = \int_{-\infty}^\infty\!\!\!\!P(z,t)P_\tau(x,z)\,dz
\end{equation}
If $\tau\ll{t}$, we can approximate $P(x,t+\tau)$ as 
\begin{equation}
  P(x,t+\tau) = P(x,t) + \frac{\partial P}{\partial t} \tau + O(\tau^2)
\end{equation}
Let $\omega(x|z)$ be the transition probability per unit time from $z$
to $x$, that is, $\omega(x|z)\tau$ is the probability that the
particle go from $z$ to $x$ in a time $\tau$. If the particle is in
$x$ at time $t$, then
\begin{equation}
  \int_{-\infty}^{\infty}\!\!\!\!\omega(z|x)\tau\,dz
\end{equation}
is the probability that it will move somewhere else, and 
\begin{equation}
  1 - \int_{-\infty}^{\infty}\!\!\!\!\omega(z|x)\tau\,dz
\end{equation}
is the probability that it will stay in $x$. Balancing the probability
of arriving at $x$ and that of not moving if we are already there, we
obtain
\begin{equation}
  P(x,t+\tau) = P(x,t)\left( 1 - \int_{-\infty}^{\infty}\!\!\!\!\omega(z|x)\tau\,dz \right) + \int_{-\infty}^{\infty}\!\!\!\!\omega(x|z)P(z,t)\tau\,dz
\end{equation}
The first terms gives us the probability that the particle were in $x$
at time $t$ and did not move in the interval $[t,t+\tau]$, while the
second is the probability that the particle were in a different
position at $t$ and that it moved to $x$ in the interval $[t,t+\tau]$.
Rearranging and taking the limit for $\tau\rightarrow{0}$, we obtain
the \emph{master equation}
\begin{equation}
  \frac{\partial}{\partial t} P(x,t) =  \int_{-\infty}^{\infty}\!\!\!\!\omega(x|z)P(z,t)\,dz - \int_{-\infty}^{\infty}\!\!\!\!\omega(z|x)P(x,t)\tau\,dz
\end{equation}
If $X$ is a discrete stochastic process, then, calling $\omega_{nm}$
the prbability of moving from position $x_n$ to position $x_m$ in unit
time, the equation becomes
\begin{equation}
  \frac{\partial}{\partial t} P(n,t) = \sum_m \omega_{mn}P(m,t) - \sum_m \omega_{nm}P(n,t)
\end{equation}

~~\eoe

\example
As an example, consider a counting process that transitions from $n$
to $n+1$ with probability $\lambda$ at each instant, that is,
$\omega_{n,n+1}=\lambda$ and $\omega_{nm}=0$ for $m\ne{n+1}$. Then the
master equation reads
\begin{equation}
  \frac{\partial}{\partial t} P(n,t) = \lambda\bigr[ P(n-1,t) - P(n,t) \bigr]
\end{equation}

~~\eoe

This type of equation can be solved through the use of the
$z$-transform of the sequence $P(n,t)$, defined as
\begin{equation}
  \label{this}
  F(z,t) = {\mathcal{Z}}[P(n,t)] = \sum_{n=0}^\infty z^n P(n,t)
\end{equation}
Then
\begin{equation}
  \sum_{n=0}^\infty z^n \frac{\partial}{\partial t} P(n,t) = \sum_{n=0}^\infty \frac{\partial}{\partial t} (z^n P(n,t)) = 
  \frac{\partial}{\partial t} F(z,t)
\end{equation}
and
\begin{equation}
  \sum_{n=0}^\infty z^n P(n-1,t) = z \sum_{n=1}^\infty z^{n-1} P(n-1,t) = z \sum_{n=0}^\infty z^n P(n,t) = z F(z,t)
\end{equation}
so that
\begin{equation}
  \sum_{n=0}^\infty z^n \lambda\bigr[ P(n-1,t) - P(n,t) \bigr] = 
  \lambda \Bigl[ \sum_{n=0}^\infty z^n P(n-1,t) - \sum_{n=0}^\infty z^n P(n,t) \Bigr] = 
  \lambda(z-1)F(z,t)
\end{equation}
resulting in
\begin{equation}
  \label{ohohoh}
  \frac{\partial}{\partial t} F(z,t) = \lambda(z-1)F(z,t)
\end{equation}
If $P(n,0)=\delta_{n,0}$, it is easy to check that $F(z,0)=1$. With
this initial condition, (\ref{ohohoh}) can be easily integrated
yielding
\begin{equation}
  F(z,t) = \exp(\lambda(z-1)t) = \sum_{n=0}^\infty z^n \frac{(\lambda t)^n}{n!} e^{-\lambda t}
\end{equation}
Comparing with (\ref{this}) we have that $P(n,t)$ follows a Poisson
distribution
\begin{equation}
  P(n,t) = \frac{(\lambda t)^n}{n!} e^{-\lambda t}
\end{equation}

~~\eoe

\subsection{Macroscopic level--Fokker-Planck equations}

\subsubsection{Diffusion}
I shall introduce the macroscopic level of description in a slightly
more general setting that needed here before seeing how it related to Brownian motion:
as \emph{diffusion}. As I mentioned, the macroscopic level
consists in considering that the characteristic magnitudes of the walk
(time and distance between collisions) are much smaller than the
magnitudes we are considering. This means that we can characterize the
problem using a continuous \emph{population density}
$\rho(\vct{x},t)$: the number of particles in a unit volume around
$\vct{x}$ at time $t$ (Figure~\ref{diffmodel}).
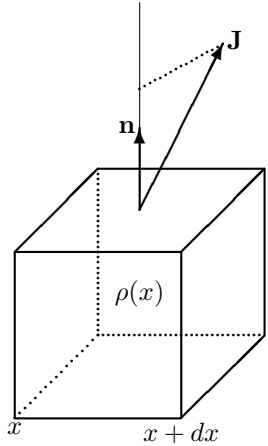
\begin{figure}
  \begin{center}
    \setlength{\unitlength}{1.5em}
    \begin{picture}(6,10)(0,0)
      \thicklines
      \multiput(0,0)(0,4){2}{\line(1,0){4}}
      \multiput(0,0)(4,0){2}{\line(0,1){4}}
      \put(4,4){\line(1,1){2}}
      \put(4,0){\line(1,1){2}}
      \put(0,4){\line(1,1){2}}
      \put(2,6){\line(1,0){4}}
      \put(6,6){\line(0,-1){4}}
      \multiput(0,0)(0.125,0.125){16}{\circle*{0.000001}}
      \multiput(2,2)(0.125,0){32}{\circle*{0.000001}}
      \multiput(2,2)(0,0.125){32}{\circle*{0.000001}}
      \put(3,5){\vector(0,1){2}}
      \put(3,5){\vector(1,2){2}}
      \put(2.9,7){\makebox(0,0)[r]{$\mathbf{n}$}}
      \put(5.1,9.1){\makebox(0,0)[l]{$\mathbf{J}$}}
      \put(3,3){\makebox(0,0){$\mathbf{\rho}(x)$}}
      \put(0,-0.1){\makebox(0,0)[t]{$x$}}
      \put(4,-0.1){\makebox(0,0)[t]{$x+dx$}}
      \thinlines
      \put(3,5){\line(0,1){5}}
      \multiput(5,9)(-0.125,-0.0675){17}{\circle*{0.000001}}
    \end{picture}
  \end{center}
  \caption{\capstyle Schematic view of the model for the derivation of the
    diffusion equation; $\rho(\vct{x},t)$ is the local density of
    particle, $\vct{J}(\vct{x},t)$ is the population flow: the number
    of particles that move in a given direction, at a given point and
    at a given time.}
  \label{diffmodel}
\end{figure}
We can approximate the local density of particles with a continuous
field, and take the limit for space and time going to zero. In
addition to the density, we define the population flow
$\vct{J}(\vct{x},t)$, which is a vector pointing in the direction of
movement and indicating how many particles move per unit time in a
surface patch of unitary area. Considering a closed volume $V$ with a
closed surface $S$, if there is no generation or annihilation of
particles, the variation in density is due to the particles that enter
and leave through the surface. So, we have:
\begin{equation}
  \label{booze}
  \frac{\partial}{\partial t}\int_V\!\!\!\rho(\vct{x},t)\,dV = 
  - \oint_S \vct{J}(\vct{x},t)\cdot\vct{n}\,dS
\end{equation}
where $\vct{n}$ is the normal to the surface at $\vct{x}$. By the
divergence theorem:
\begin{equation}
  \oint_S \vct{J}(\vct{x},t)\cdot\vct{n}\,dS = \int_V\!\!\!\nabla\cdot\vct{J}\,dV
\end{equation}
Applying this theorem to (\ref{booze}) we have
\begin{equation}
  \int_V \Bigl[ \frac{\partial \rho(\vct{x},t)}{\partial t} + \nabla\cdot\vct{J}(\vct{x},t) \Bigr]\,dV = 0
\end{equation}
Since the volume $V$ is arbitrary, we get the continuity equation
\begin{equation}
  \label{fock}
  \frac{\partial \rho(\vct{x},t)}{\partial t} + \nabla\cdot\vct{J}(\vct{x},t) = 0
\end{equation}
In order to get a solvable equation in $\rho$, we need to determine how
$\vct{J}$ emerges as a consequence of variations of the population
density, that is, how $\vct{J}$ relates to $\rho$. Such an expression
is called a \emph{constitutive equation}. One common constitutive
equation, known as \emph{Fick's law}, assume that the flow is
proportional to the local population gradient, that is
\begin{equation}
  \label{fick}
  \vct{J}(\vct{x},t) = - D \nabla\rho(\vct{x},t)
\end{equation}
The minus sign takes into account the fact that the flow goes from
regions of high density to regions of low density. Introducing
(\ref{fick}) into (\ref{fock}) one gets the diffusion equation
\begin{equation}
  \label{fuck}
  \frac{\partial\rho}{\partial t} = \nabla \cdot (D\nabla\rho)
\end{equation}
If $D$ is a constant  independent of $\vct{x}$ then (\ref{fuck}) turns
into
\begin{equation}
  \label{diff}
  \frac{\partial\rho}{\partial t} = D \nabla^2\rho
\end{equation}
and, in the one-dimensional case
\begin{equation}
  \label{diffone}
  \frac{\partial\rho}{\partial t} = D \frac{\partial^2 \rho}{\partial x^2}
\end{equation}
This \emph{diffusion equation} has, in principle, nothing to do with
Brownian motion or random walks: it has been derived considering a
completely different problem, namely the diffusion of a fluid into
space under the action of the gradient of its density. Yet,
surprisingly, it turns out that this equation does indeed describe
Brownian motion. In particular, it describes the evolution of the
probability of finding a Brownian walker in $x$ at time $t$.

\subsubsection{Fokker-Planck equation}
The \emph{Fokker-Planck equation} is a partial differential equation
in time and space that describes Brownian motion at a macroscopic
level. This makes it a macroscopic equation, since the use of
differential operators entails that we are considering times and
distances much greater than the time and space between changes in
direction of a particle. In this section we present the Einstein
derivation of the Fokker-Planck equations considering, for the sake of
simplicity, the one-dimensional case \cite{einstein:05}.

The motion of a particle in a Brownian motion can be interpreted as a
series of jumps that can have an arbitrary length $z$. Let the jump lengths
be distributed according to a PDF $\phi(z)$, and let them be
i.i.d. The density of individuals at position $x$ at time $t+\tau$ is
given by those individuals that were at a position $z$ at time $t$ and
that have jumped to $x$ after waiting a time $\tau$. Since $z$ is
arbitrary, we integrate over all possible $z$, obtaining a form of
non-Markovian Chapman-Kolmogorov equation
\begin{equation}
  \label{albert}
  \rho(x,t+\tau) = \int_{-\infty}^\infty\!\!\!\!\!\rho(x-z,t)\phi(z)\,dz
\end{equation}
Note that this equation is continuous in space and discrete in
time. In particular, the PDF $\phi(z)$, which in ecology is called the
\emph{dispersion kernel} is continuous, meaning that we are
making an implicit assumption of a large number of
individuals/particles. If we now take the macroscopic limit, that is,
is we consider that $\tau$ and $z$ are both small with respect to the
scale of interest, then we can use a Taylor expansion in $t$ and $z$:
\begin{equation}
  \begin{aligned}
    \rho(x,t+\tau) &= \sum_{n=0}^\infty \frac{\tau^n}{n!} \frac{\partial^n \rho}{\partial \tau^n} \\
    \rho(x-z,t)    &= \sum_{n=0}^\infty \frac{(-z)^n}{n!} \frac{\partial^n \rho}{\partial z^n} 
  \end{aligned}
\end{equation}
Inserting into (\ref{albert}), one gets:
\begin{equation}
  \rho(x,t) + \tau \frac{\partial \rho}{\partial \tau} + \cdots = 
  \rho(x,t)\int_{-\infty}^\infty\!\!\!\!\!\phi(z)\,dz - \frac{\partial \rho}{\partial x}\int_{-\infty}^\infty\!\!\!\!\!z\phi(z)\,dz
  + \frac{\partial^2 \rho}{\partial x^2}\int_{-\infty}^\infty\!\frac{z^2}{2!}\phi(z)\,dz + \cdots
\end{equation}
The kernel $\phi(z)$ is a PDF, therefore
$\int_{-\infty}^\infty\phi(z)\,dz=1$. Moreover, if the movements are
isotropic, that is, there is no preferential direction of movement,
then $\phi(z)=\phi(-z)$, and $\int_{-\infty}^\infty{z^n}\phi(z)\,dz=0$
for $n$ odd. So, we have:
\begin{equation}
  \rho(x,t) + \tau \frac{\partial \rho}{\partial \tau} + O(\tau^2) = 
  \rho(x,t) + \frac{\partial^2 \rho}{\partial x^2}\int_{-\infty}^\infty\!\frac{z^2}{2!}\phi(z)\,dz + O(z^4)
\end{equation}
Or, simplifying the common term and dividing by $\tau$,
\begin{equation}
  \frac{\partial \rho}{\partial \tau} = 
  \frac{\partial^2 \rho}{\partial x^2}\int_{-\infty}^\infty\!\frac{z^2}{2\tau}\phi(z)\,dz + O(z^4/\tau^2)
\end{equation}
We now take the macroscopic limit $z,\tau\rightarrow{0}$, but in such
a way that $\lim z^2/\tau=C\ne{0}$, that is, keeping $z^2$ and $\tau$
of the same order of magnitude. Then we obtain, as a Fokker-Planck
equation, a diffusion equation like (\ref{diff}):
\begin{equation}
  \label{einfock}
  \frac{\partial \rho}{\partial t} = D \frac{\partial^2 \rho}{\partial x^2}
\end{equation}
where 
\begin{equation}
  D = \frac{1}{2\tau} \int_{-\infty}^\infty\!\!\!\!\!z^2\phi(z)\,dz = \frac{\langle{z^2}\rangle}{2\tau}
\end{equation}
Note that (\ref{einfock}) depends only on the second moment of the
diffusion kernel $\phi(z)$. In no place have we made the hypothesis
that $\phi(z)$ is Gaussian so very different diffusion kernels with
the same second moment will generate the same Fokker-Planck
equation. We have lost information with respect to the distribution
$\phi(z)$: in particular, \emph{given any distribution, a Gaussian
  distribution with the same second moment will generate the same
  macroscopic distribution}.

In this sense, we also notice that the Langevin equation does indeed
make the hypothesis that $\xi(t)$ is Gaussian. It would therefore seem
that the Einstein derivation is more general than the Langevin
equation. In reality, it is not so, and the reason is the Central
Limit Theorem%
\footnote{We shall not do the derivation here, but from the Langevin
  equation one can derive the same Fokker-Planck equation as from the
  Einstein's derivation. So, despite the different hypotheses the two
  methods describe the same macroscopic phenomenon.}%
: the hypothesis that $z^2$ and $\tau$ be of the same magnitude
entails that $D$ is finite and therefore, by (\ref{einfock}), that
$\langle{Z^2}\rangle$ is finite. We are in the hypotheses of the
Central Limit Theorem, so the sum of all the jumps of the Einstein's
derivation, with distribution $\phi(z)$, will end up being Gaussian
regardless of the exact shape of $\phi(z)$.

\subsubsection{Solution of the Diffusion Equation}
One simple way to solve the diffusion equation is through the use of
the Characteristic function of the distribution $\rho$, viz.\ its
Fourier Transform. Taking the Fourier transform of (\ref{diffone}) we
get the ordinary differential equation
\begin{equation}
  \frac{d\tilde{\rho}}{dt}(\omega,t) = -D \omega^2 \tilde{\rho}(\omega,t)
\end{equation}
which has solution
\begin{equation}
  \tilde{\rho}(\omega,t) = \tilde{\rho}(\omega,0)\exp\bigl[-D\omega^2t\bigr]
\end{equation}
In the simplest case, at the beginning all the individuals are
concentrated at $x_0$, that is, $\rho(x,0)=\delta(x-x_0)$. By the
formula for the characteristic function of the Dirac distribition
(\ref{deltachar}), we have
$\tilde{\rho}(\omega,0)=\exp\bigl[-i\omega{x_0}\bigr]$. That is
\begin{equation}
  \tilde{\rho}(\omega,t) = \exp\bigl[-i\omega{x_0}-D\omega^2t\bigr]
\end{equation}
The inverse Fourier transform gives us
\begin{equation}
  \label{pussy}
  \rho(x,t) = \frac{1}{2\pi} \int_{-\infty}^\infty\!\!\!e^{i\omega{x}}\tilde{\rho}(\omega,t)\,d\omega = 
            \frac{1}{2\pi} \int_{-\infty}^\infty\!\!\!\exp\bigl[ i\omega(x-x_0)-D\omega^2{t}\bigr]\,d\omega =
            \frac{1}{\sqrt{4\pi{Dt}}} \exp\Bigl[-\frac{(x-x_0)^2}{4Dt}\Bigr]
\end{equation}
From this solution we can obtain the general solution for
$\rho(x,0)=g(x)$. Writing
\begin{equation}
  g(y) = \int_{-\infty}^\infty\!\!\!g(x)\delta(x-y)\,dy
\end{equation}
and applying superposition we have
\begin{equation}
  \rho(x,t) =  \frac{1}{\sqrt{4\pi{Dt}}} \int_{-\infty}^\infty\!\!\!\!\!g(y)\exp\Bigl[-\frac{(y-x_0)^2}{4Dt}\Bigr]\,dy
\end{equation}
With reference to the simple solution (\ref{pussy}),
Figure~\ref{difsol} shows $\rho(x,t)$ as a function of $x$ for several
values of $t$.
\begin{figure}
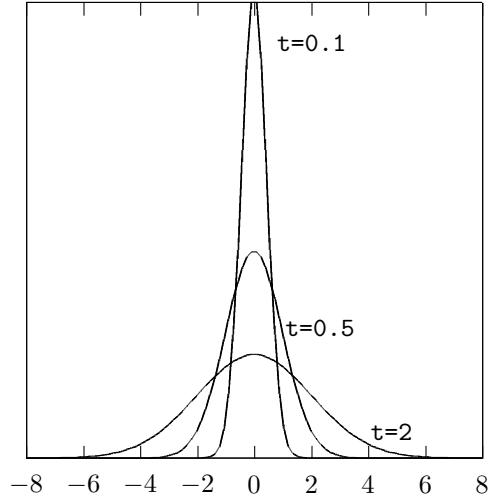

  \begin{center}
\setlength{\unitlength}{0.240900pt}
\ifx\plotpoint\undefined\newsavebox{\plotpoint}\fi

  \end{center}

  \vspace{-2em}

  \caption{\capstyle The solution of the diffusion $\rho(x,t)$ for
    $\rho(x,0)=\delta(x)$ as a function of $x$ for several values of
    $t$. The density is a Gaussian that becomes progressively more
    spread, indicating that the population covers larger and larger
    areas.}
  \label{difsol}
\end{figure}
The result is typical of diffusion processes in which the population,
initially concentrated at $x=0$ ($\rho(x,0)=\delta(x)$) spreads over
larger and larger areas. The \dqt{speed} of this diffusion is
determined by the increasing variance
$\sigma(t)=\langle{X^2}\rangle=2Dt$ which, as predicted by
(\ref{lookie}) and (\ref{einfock}), grows linearly with $t$.

\subsection{Anomalous Diffusion}
The standard diffusion process, which we have considered so far, is
characterized by the relation
\begin{equation}
  \label{difflaw}
  \langle X^2 \rangle \sim t
\end{equation}
where the notation is shorthand for
\begin{equation}
  \lim_{t\rightarrow\infty} \frac{\langle X^2 \rangle}{t} = C \ne 0
\end{equation}
The reason for this boils down to the fact that the diffusion equation
has first derivatives in time and second derivatives in space, so that
one obtain homologous quantities starting with a constant and
integrating once in time and twice in space.

To see a different kind of behavior, consider \emph{ballistic
  displacements}, that is, the motion of a particle that moves at a  constant
speed $v$ and never changes direction. For a movement along the $x$
axis, this can be modeled as a stochastic process with PDF
$P(x,t)=\delta(x-vt)$ so that
\begin{equation}
  \label{ball}
  \langle X^2 \rangle = \int_{-\infty}^\infty\!\!\!x^2\delta(x-vt)\,dx = v^2t^2\sim t^2
\end{equation}
That is, in this case
\begin{equation}
  \lim_{t\rightarrow\infty} \frac{\langle X^2 \rangle}{t^2} = C \ne 0
\end{equation}
We can see ballistic movement as a type of random walk, albeit a
not-quite-so-random one, and one that moves from the origin much
faster than the Brownian motion. Small wonder: ballistic movement
moves purposely in a fixed direction, while Brownian motion is
bounced to and fro. Note, however, that this means that ballistic
movement will explore around much less than Brownian motion: it will
stick to a trajectory and not look around at all. Just like a traveler
in a rush: you may go very far, but you miss the view.

Processes that don't follow the standard diffusion law (\ref{difflaw})
are called \emph{anomalous} \shortcite{havlin:87}. Ballistic movement
is our first example of anomalous diffusion (albeit a rather
pathological one).  The asymptotic relation between
$\langle{X^2}\rangle$ and $t$ is normally defined using the
\emph{Hurst exponent} $H$ \shortcite{hurst:65} defined by
\begin{equation}
  \label{hurst}
  \langle X^2 \rangle \sim t^{2H}
\end{equation}
Diffusion corresponds to $H=1/2$; if $H<1/2$ we have
\emph{subdiffusion}, while for $1/2<H<1$ we have
\emph{superdiffusion}. The ballistic limit (\ref{ball}) is achieved
for $H=1$ (see Figure~\ref{hurstfig}).
Note that, because of the central limit theorem, normal diffusion
($H=1/2$) is obtained under a wide family of displacements
distributions. If the displacements:
\begin{description}
\item[i)] are independent, 
\item[ii)] are identically distributed, and
\item[iii)] follow a PDF with finite mean and variance, 
\end{description}
then we can apply the CLT in its standard form, and the total distance
covered (that is, the sum of all these displacements) is a Gaussian
$\exp(-x^2/\sigma^2)$, where $\sigma^2$ is proportional to the number
of displacements, that is, $\sigma^2\sim{t}$. Then (\ref{difflaw})
follow directly from the equality $\langle{X^2}\rangle=\sigma^2$ valid
for a Gaussian. 

The hypotheses i)--iii) hint at three possible ways in which they can
be violated, resulting in three different mechanisms that can
generate anomalous diffusion.
\begin{description}
\item[i)] the displacements are not independent due to long range
  correlations: once a particle moves, it will tend to remain in
  motion (leading to superdiffusion---ballistic motion is an 
  example of this kind of process) or, contrariwise%
  \footnote{Neologism courtesy of Lewis Carroll.}%
  , once it stops it will tend to remain at rest (leading to
  subdiffusion);
\item[ii)] the distribution of the displacements is not identical,
  either because they become shorter with time (leading to
  subdiffusion) or because they become longer (leading to
  superdiffusion);
\item[iii)] the displacements are distributed according to a PDF with
  infinite variance, so that arbitrary large displacements are
  relatively likely.
\end{description}
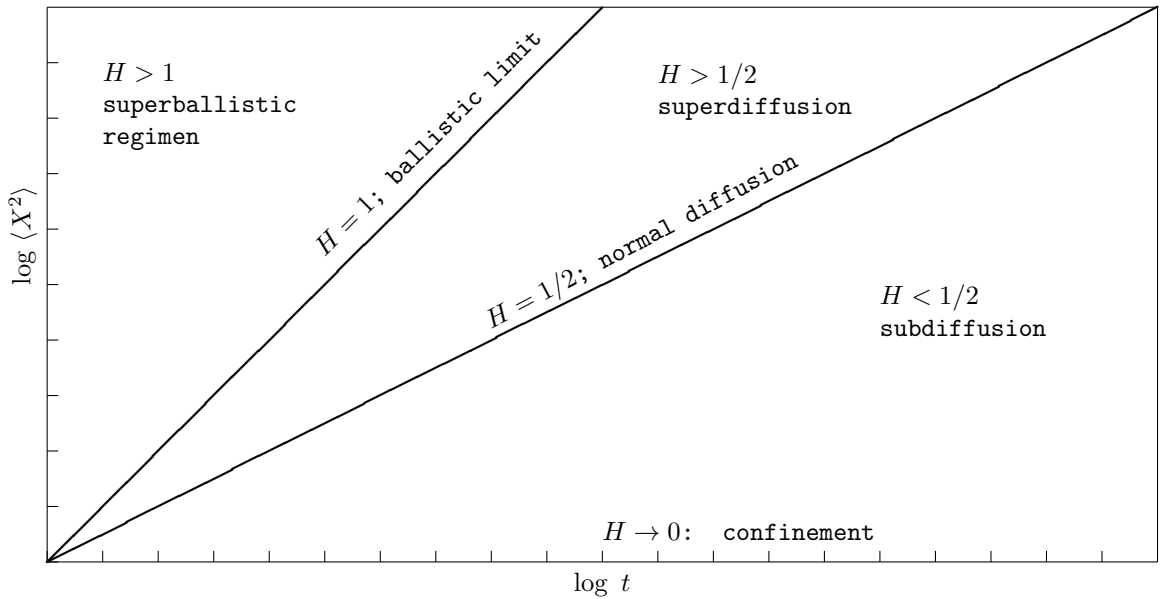
\begin{figure}[tbhp]
  \begin{center}
    \setlength{\unitlength}{2em}
    \begin{picture}(20,10.5)(0,-0.5)
      \multiput(0,0)(0,10){2}{\line(1,0){20}}
      \multiput(0,0)(20,0){2}{\line(0,1){10}}
      \multiput(0,0)(1,0){20}{\line(0,1){0.2}}
      \multiput(0,0)(0,1){10}{\line(1,0){0.2}}
      \thicklines
      \put(0,0){\line(2,1){20}}
      \put(0,0){\line(1,1){10}}
      \put(1,9){\makebox(0,0)[lt]{\parbox{8em}{$H>1$\\superballistic regimen}}}
      \put(5,5.5){\makebox(0,0){\begin{rotate}{45}$H=1$; ballistic limit\end{rotate}}}
      \put(8,4.2){\makebox(0,0){\begin{rotate}{28}$H=1/2$; normal diffusion\end{rotate}}}
      \put(11,9){\makebox(0,0)[lt]{\parbox{8em}{$H>1/2$\\superdiffusion}}}
      \put(15,5){\makebox(0,0)[lt]{\parbox{8em}{$H<1/2$\\subdiffusion}}}
      \put(10,0.4){\makebox(0,0)[lb]{\parbox{10em}{$H\rightarrow{0}$: confinement}}}
      \put(10,-0.2){\makebox(0,0)[t]{$\log\,\,{t}$}}
      \put(-0.3,5){\makebox(0,0)[t]{\begin{rotate}{90}$\log\,\,\langle{X^2}\rangle$\end{rotate}}}
    \end{picture}
  \end{center}
  \caption{\capstyle The Hurst exponent quantifies the asymptotic behavior of
    diffusive processes. The usual uncorrelated Brownian random walks
    satisfy the standard form of the central limit theorem, and have
    $H=1/2$. Subdiffusive processes have $H<1/2$. For vanishing $H$,
    the process is localized and confined: diffusion has a finite
    reach. Superdiffusion, viz.\ processes with a propagation speed
    higher than Brownian diffusion, have $H>1/2$. The ballistic limit
    is reached for $H=1$. Superballistic processes (not considered
    here) correspond to accelerated particles. L\'evy flights and
    walks, of special interest here, are superdiffusive processes.}
  \label{hurstfig}
\end{figure}

In the following, we shall consider mostly the third case but, before
digging into it, I shall give a brief example of how the first two
work. 

\example
For the case of long-term correlations, divide the trajectory into
intervals of fixed duration $\Delta{t}$. With this division, the
correlations between displacements are the same as those between
velocities. In this case, we can determine the derivative of
$\langle{X^2}\rangle$ as
\begin{equation}
  \label{kubo}
  \frac{d}{dt}\langle{X^2}\rangle = \frac{d}{dt} \int_0^t\!\!\!\int_0^t \langle{v(s)v(\tau)}\rangle\,d\tau\,ds
  = 2 \int_0^t\!\!\!\langle{v(t)v(\tau)}\rangle\,d\tau
\end{equation}
This result is known as the Taylor's formula \shortcite{taylor:21}
(also known as the Green-Kubo formula). If
$\langle{v(t)v(\tau)}\rangle$ is integrable, then the limit for
$t\rightarrow\infty$ of the integral exists, so the right-hand side of
(\ref{kubo}) is asymptotically a constant, that is, for
$t\rightarrow\infty$,
\begin{equation}
  \frac{d}{dt}\langle{X^2}\rangle \sim C \mbox{~~~or~~~}%
  \langle{X^2}\rangle\sim t
\end{equation}
and we find again a diffusive behavior. If, on the other hand, the
correlation decays slowly enough that the integral diverges, then the
CLT doesn't hold, and we observe anomalous diffusion. If, for example,
$\langle{v(t)v(\tau)}\rangle\sim(t-\tau)^{-\eta}$, with $0<\eta<1$,
then $\langle{X^2}\rangle\sim{t}^{2-\eta}$, that is, we have
superdiffusion.

~~\eoe

\example
Non-identical displacements ocurr when displacements become either
longer or shorter with time or, equivalently, as the particle gets
farther from its initial position. If we take a macroscopic point of
view---that is, if we write a diffusion-like Fokker-Planck
equation---then we can model this as a time and/or space varying
coefficient $D$. This is tantamount to saying that the obstacles to
motions become gradually larger or smaller as we get away from the
initial position. Consider a space-dependent diffusion coefficient
that varies as a power law: $D=D_0x^\theta$. This leads to a diffusion
equation:
\begin{equation}
  \frac{\partial\rho}{\partial t} = \nabla\cdot(D_0 x^\theta \nabla\rho)
\end{equation}
A rigorous derivation of the behavior of $\langle{X^2}\rangle$ under
this equation can be found in \shortcite{shaughnessy:85}; here we
shall do a simple informal derivation using dimensional analysis. The
density $\rho$ is a number of particles per unit of $x$ that is,
dimensionally, $[\rho]=[x]^{-1}$ and, consequently
\begin{equation}
  \left[ \frac{\partial \rho}{\partial t} \right] = [x]^{-1}[T]^{-1}
\end{equation}
where $T$ is the dimension of time. Similarly $[\nabla\rho]=[x]^{-2}$,
$[x^\theta\nabla\rho]=[x]^{\theta-2}$ and
$[\nabla\cdot(x^\theta\nabla\rho)]=[x]^{\theta-3}$. This equality
gives us
\begin{equation}
  [x]^{-1}[T]^{-1} = \left[ \frac{\partial \rho}{\partial t} \right] = [\nabla\cdot(x^\theta\nabla\rho)]=[x]^{\theta-3}
\end{equation}
that is, $[T]^{-1}=[x]^{\theta-2}$, $[T]=[x]^{2-\theta}$, or
$[x]=[T]^{1/(2-\theta)}$, whch leads to
\begin{equation}
  [x^2] = [x]^2 = [T]^{\frac{2}{2-\theta}} = [X^2]
\end{equation}
This dimensional equality indicates that, asymptotically,
\begin{equation}
  \langle X^2 \rangle \sim t^{\frac{2}{2-\theta}}
\end{equation}
leading, again, to anomalous diffusion. 

~~\eoe

The case of divergent moments that I shall consider closely is that
of \emph{L\'evy flights}. If the individual displacements are i.i.d.,
then we are in the conditions of the generalized Central Limit
Theorem: no matter what the individual PDF are, the sum of a large
number of them will converge to a L\'evy stable distribution. So, just
like in the finite moment case we could assume that the displacements
followed a Gaussian distribution%
\footnote{Remember that the Langevin equation, which assume Gaussian
  displacements, leads to the same macroscopic result as the Einstein
  method, which doesn't.}%
, we can now assume that they follow a L\'evy distribution which, as
seen in (\ref{levypow}), behaves like $x^{-(1+\alpha)}$ for
$t\rightarrow\infty$. For $\alpha<2$, $\langle{X^2}\rangle$ diverges
due to the \dqt{long tail} of the distribution, which makes
arbitrarily large displacements relatively frequent.

In order to frame these ideas properly, it is first necessary to study
random walks from a slightly more general point of view, that of
Continuous Time Random Walks.

\subsection{Continuous Time Random Walks}
The random walks that we have considered so far were limits of what we
can consider a discrete time scenario: we considered that jumps take
place at regular time intervals, and we take the limit of
$\Delta{t}\rightarrow{0}$, corresponding to a continuum of jumps of
length zero (this is enforced by the fact that we require
$\langle{x^2}\rangle/\tau$ to stay finite). In a \emph{Continuous Time
  Random Walk} (CTRW) we assume that the waiting time between jumps is
a random process as well, that is, that the particle will intersperse
jumps of random length with pauses of random duration. I shall
introduce the analysis of CTRW in two steps: first I shall consider
the PDF of the position of the particle after $n$ jumps, without
considering \emph{when} did these jumps occur, then the probability of
doing $n$ jumps in time $t$. This corresponds to a specific type of
CTRW, one in which the jump length is independent of the waiting time.
The result can easily be extended to the case in which waiting time
and jump length are correlated.

\bigskip

Let $Z_n$ be the length of the $n$th jump. The position of a particle
after $n$ jumps is
\begin{equation}
  X_n = \sum_{k=1}^n Z_k = X_{n-1} + Z_n
\end{equation}
This equation shows that the walk is a Markov chain. Let $Z_k$ be
i.i.d. with PDF $\phi(z)$; the function $\phi$ (the dispersal kernel)
represents the transition probability of the Markov chain. Adapting
the Chapman-Kolmogorov equation (\ref{albert}) to this discrete-time
scenario, we obtain an equation for $\rho_n(x)$, the density of
individuals after $n$ jumps:
\begin{equation}
  \rho_n(x) = \int_{-\infty}^{\infty}\!\!\!\rho_{n-1}(x-z)\phi(z)\,dz = \rho_{n-1}*\phi
\end{equation}
where $*$ denotes spatial convolution. If $\rho_0$ is the
initial density, then:
\begin{equation}
  \begin{aligned}
    \rho_1 &= \rho_0 * \phi \\
    \rho_2 &= \rho_1 * \phi = \rho_0 * \phi * \phi \\
    \vdots \\
    \rho_n &= \rho_{n-1} * \phi = \rho_0 * \overbrace{\phi * \cdots * \phi}^{n}
  \end{aligned}
\end{equation}
Considering, for the sake of simplicity, the one-dimensional case, we
can take the Fourier transform and apply (\ref{pluck}) to obtain
\begin{equation}
  \label{palooza}
  \tilde{\rho}_n(\omega) = \tilde{\rho}_0(\omega) \tilde{\phi}^n(\omega)
\end{equation}

Consider now the jump times. Let $\theta_n$ be the waiting time
between jump $n-1$ and jump $n$, and $\psi(t)$ its PDF. The time at
which the $n$th jump is taken is then
\begin{equation}
  T_n = \sum_{k=1}^n \theta_n
\end{equation}
Let $\psi^0(t)$ be the probability that no jump has ocurred by time $t$, viz.
\begin{equation}
  \psi^0(t) = \int_t^\infty\!\!\!\psi(u)\,du = 1 - \int_0^t\!\!\!\psi(u)\,du
\end{equation}
Let $P_n(t)$ be the probability of performing $n$ jumps by time
$t$. Then, clearly, $P_0(t)=\psi^0(t)$. The probability that there is a
jump at a time $u<t$ and then no further jumps until time $t$ is
$\psi(u)\psi^0(t-u)$. Integrating over all $u<t$ we have%
\footnote{The asterisk denotes here convolution in time, which has
  different integration limits than convolution in space, since $\psi$
  and $\psi^0$ can only take non-negative arguments. Strictly speaking,
  we should have used a different symbol. However, since is it usually
  clear what convolution is being used, I have preferred not to
  complicate the notation using non-standard symbols.}%
\begin{equation}
  P_1(t) = \int_0^t\!\!\!\psi(u)\psi^0(t-u)\,du = \psi*\psi
\end{equation}
Iterating this, we have
\begin{equation}
  P_n(t) = \psi^0*\overbrace{\psi* \cdots *\psi}^{n}
\end{equation}
In this case, since we have different limits and a different
convolution, one must use the Laplace transform in lieu of the
Fourier:
\begin{equation}
  \psi(s)=\int_0^\infty\!\!\!e^{-st}\psi(t)\,dt
\end{equation}
where $s\in{\mathbb{C}}$.%
\footnote{The Laplace transform has similar properties as the Fourier,
  but is more general. If $f(t)$ is a function and ${\mathcal{L}}[f]$
  is its Laplace transform (which I shall also indicate as
  $\tilde{f}$), then
  \begin{equation}
    f(t) = \frac{1}{2\pi{i}} \lim_{T\rightarrow\infty} \int_{\gamma-iT}^{\gamma+iT}\!\!\!\!{\mathcal{L}}[f](s)e^{st}\,ds
  \end{equation}
  where $\gamma$ is a real number that exceeds the real part of all
  singularities of ${\mathcal{L}}[f]$. Also:
  \begin{equation}
    \begin{aligned}
      {\mathcal{L}}\Bigl[ \int_0^t\!\!\!f(u)g(t-u)\,du\Bigr] &= {\mathcal{L}}[f]{\mathcal{L}}[g] \\
      {\mathcal{L}}\Bigl[ \int_0^t\!\!\!f(u)\,du\Bigr] &= \frac{1}{s}{\mathcal{L}}[f] \\
      {\mathcal{L}}\bigl[ e^{at}f(t)\bigr] &= {\mathcal{L}}[f](s-a) \\
      {\mathcal{L}}\bigl[1\bigr] &= \frac{1}{s}
    \end{aligned}
  \end{equation}
}%

This leads to
\begin{equation}
  \tilde{P}_n(s) = \tilde{\psi^0}(s)\tilde{\psi}^{n}(s) = \frac{1-\tilde{\psi}(s)}{s}(\tilde{\psi}(s))^{n}
\end{equation}

\separate

Consider now the combination of the two processes. The position of an
individual at time $t$ (assume $x(0)=0$) is
\begin{equation}
  x(t)=\sum_{k=0}^{N(t)} z_k
\end{equation}
where $N(t)$ is the number of jumps taken before time $t$, itself a
random variable. We are interested in finding an expression for
$\rho(t)$, the density of individuals at time $t$. If by time $t$ 
$n$ jumps have been made, then
\begin{equation}
  \rho(x,t | N(t)=n) = \rho_n(x)
\end{equation}
The value of $\rho(x,t)$ is then given by the value of $\rho_n$ for
all possible $n$, weighted by their probability:
\begin{equation}
  \rho(x,t) = \sum_{n=0}^\infty \rho_n(x)P_n(t)
\end{equation}
that is, taking the Fourier and Laplace transforms:
\begin{equation}
  \label{ooohh}
  \begin{aligned}
    \tilde{\rho}(\omega,s) &= \sum_{n=0}^\infty \tilde{\rho}_n(\omega)\tilde{P}_n(t) \\
    &= \tilde{\rho}(\omega,0) \frac{1-\tilde{\psi}(s)}{s} \sum_{n=0}^\infty \bigl[\phi(\omega)\psi(s)\bigr]^n \\
    &= \tilde{\rho}(\omega,0) \frac{1-\tilde{\psi}(s)}{s} \frac{1}{1-\phi(\omega)\psi(s)}
  \end{aligned}
\end{equation}
This is known as the \emph{Montroll-Wei$\beta$} equation. I made here
the assumption that the waiting times and the jump length are
independent, hence the product $\phi(\omega)\psi(s)$. If they are not,
then their joint probability would be expressed by a distribution
$\phi(\omega,s)$, and (\ref{ooohh}) becomes
\begin{equation}
    \tilde{\rho}(\omega,s) = \tilde{\rho}(\omega,0) \frac{1-\tilde{\psi}(s)}{s} \frac{1}{1-\phi(\omega,s)}
\end{equation}
For any distribution $\phi(\omega)$ and $\psi(s)$, (\ref{ooohh})
allows us to determine the evolution of the density of individuals by
taking the inverse Fourier/Laplace transform.

\subsubsection{Finite moments: diffusion}
The generality of the CTRW notwithstanding, if we assume that $\phi$
and $\psi$ have finite moment we still revert to the normal diffusive
behavior. In order to see this, we first rearrange (\ref{ooohh}) in a
more useful form. From (\ref{ooohh}), we write
\begin{equation}
  \label{auhh}
  s\tilde{\rho}(\omega,s) - \tilde{\rho}(\omega,0) = \tilde{\rho}(\omega,0)
    \Bigl[ \frac{1 - \psi(s)}{1 - \phi(\omega)\psi(s)} - 1 \Bigr]
\end{equation}
Express, from the same equation
\begin{equation}
  \tilde{\rho}(\omega,0) = \frac{s}{1-\psi(s)} \bigl[1 - \phi(\omega)\psi(s)\bigr]\tilde{\rho}(\omega,s)
\end{equation}
Replacing in the right-hand side of (\ref{auhh}) and simplifying we get
\begin{equation}
  \label{bebop}
  s\tilde{\rho}(\omega,s) - \tilde{\rho}(\omega,0) 
  \frac{s\psi(s)}{1-\psi(s)} \bigl[1 - \phi(\omega)\psi(s)\bigr]\tilde{\rho}(\omega,s)
\end{equation}
The quantity 
\begin{equation}
  M(s) = \frac{s\psi(s)}{1 - \psi(s)}
\end{equation}
is called the \emph{memory kernel} of the CTRW. Equation (\ref{bebop})
can in turn be rewritten in a way that separates the spatial and
temporal variables:
\begin{equation}
  \label{alula}
  \frac{1-\psi(s)}{s\psi(s)}\bigl[ s\tilde{\rho}(\omega,s) - \tilde{\rho}(\omega,o)\bigr] = 
  \bigl[1 - \phi(\omega)\psi(s)\bigr]\tilde{\rho}(\omega,s)
\end{equation} 

We now consider the macroscopic limit in space, which entails assuming
that the microscopic scale of the process is very small compared to
the scale of $x$. This means that we shall consider the limit for
$\omega\rightarrow{0}$ in the Fourier space. Similarly, the
macroscopic limit in time consists in taking the limit
$s\rightarrow{0}$ in the complex plane of the Laplace transform. If
$\phi$ is symmetric and has finite moments, then it has an expansion
$\phi(\omega)=1-\langle\phi^2\rangle\omega^2/2+o(\omega^4)$, where
$\langle\phi\rangle$ is the average displacement $\langle{X^2}\rangle$
when $X$ has PDF $\phi$. Similarly,
$\psi(s)=1-\langle\psi\rangle{s}+o(s^2)$, where $\langle\psi\rangle$
is the mean waiting time. From this we get
\begin{equation}
  \frac{1-\psi^0(s)}{s\psi(s)} \sim \frac{\langle\psi\rangle}{1-\langle\psi\rangle{s}} = \langle\psi\rangle + o(s)
\end{equation}
while
\begin{equation}
  \bigl[\phi(\omega)-1\bigr] = - \frac{\langle\phi^2\rangle}{2}\omega^2 + o(\omega^2)
\end{equation}
Putting these in (\ref{alula}) we have
\begin{equation}
  \langle\psi\rangle[s\rho(\omega,s)-\rho(\omega,0)] = \frac{\langle\phi^2\rangle}{2}\omega^2 \rho(\omega,s)
\end{equation}
that is, taking the inverse Fourier and Laplace transforms
\begin{equation}
  \frac{\partial}{\partial t}\rho(x,t) = \frac{\langle\phi^2\rangle}{2\langle\psi\rangle} \frac{\partial^2}{\partial x^2}\rho(x,t)
\end{equation}
that is, we are back to a diffusion equation that behaves like
(\ref{hurst}), with $H=1/2$.

We obtain anomalous diffusion in two ways: we can either make long pauses
(viz.\ pauses with a distribution with diverging variance) or we can make
long jumps.

\subsubsection{Long pauses}
Let us assume that $\phi(z)$ has a Gaussian distribution%
\footnote{As we have seen, any distribution, as long as it has finite
  moments, will give the same results, as we are in the hypotheses of
  the standard Central Limit Theorem.}%
, while $\psi(t)$ has a L\'evy distribution with a long tail
\begin{equation}
  \psi(t) \sim A_\alpha \left( \frac{\tau}{t} \right)^\alpha
\end{equation}
with $0<\alpha<1$. We are interested in the long term behavior of the walk, that
is, in terms of characteristic functions, in the limit
$\omega\rightarrow{0}$, $|s|\rightarrow{0}$ so we can write
\begin{equation}
  \begin{aligned}
    \phi(\omega) &= \exp\bigl(- \frac{\omega^2\sigma^2}{2} \bigr) \sim 1 - \sigma^2\omega^2 \\
    \psi(s) &= \exp\bigl( -\tau^\alpha|s|^\alpha \bigr) \sim 1 - (\tau{s})^\alpha
  \end{aligned}
\end{equation}
Introducing into (\ref{ooohh}), we get
\begin{equation}
  \tilde{\rho(\omega,s)} = \frac{1}{s} \frac{\rho(\omega,0)}{1 + K_\alpha\omega^2s^{-\alpha}}
\end{equation}
with $K_\alpha = \sigma^2/\tau^{\alpha}$.  The long term behavior of
$\langle{X^2}\rangle$ can be determined using the relation
\begin{equation}
  \langle X^2 \rangle = \lim_{\omega\rightarrow{0}} - \frac{\partial \tilde{\rho}}{\partial \omega}
\end{equation}
For $\rho(x,0)=\delta(x)$, i.e.\ $\tilde{\rho}(\omega,0)=1$, we have
\begin{equation}
  \begin{aligned}
    \langle X^2 \rangle &= \lim_{\omega\rightarrow{0}} \Bigl[
      - \frac{2}{s} K_\alpha s^{-\alpha} (1+K_\alpha\omega^2s^{-\alpha})^{-2} - \frac{8}{s}(1+K_\alpha\omega^2s^{-\alpha})^{-3}(K_\alpha s^{-\alpha}\omega^2)^2
      \\
      &= 2K_\alpha s^{-(\alpha+1)}
  \end{aligned}
\end{equation}
which, inverting the Laplace transform, gives
\begin{equation}
  \langle X^2 \rangle = \frac{2K_\alpha}{\Gamma(\alpha+1)}t^{\alpha}
\end{equation}
Since $\alpha<1$, we are in the presence of subdiffusion
($H=\alpha/2<1/2$), as could be expected given that we have
arbitrarily long pauses with relative high frequency.

\subsubsection{Long jumps}
We consider now the opposite situation: assume that $\psi(t)$ has a
distribution with finite moments (exponential, in this case, since
$t>0$) and that $\phi(z)$ has a L\'evy distribution with L\'evy
parameter $\mu$:
\begin{equation}
  \begin{aligned}
    \psi(t) &= \tau\exp\bigl(-\frac{t}{\tau}\bigr) \\
    \phi(z) &= A_\mu \left( \frac{z_0}{z} \right)^{1+\mu}
  \end{aligned}
\end{equation}
Note that in this case we consider $1<\mu<2$ for the sake of
simplicity: the results are similar for $0<\mu<1$. As before, in the
limit $\omega\rightarrow{0}$, $|s|\rightarrow{0}$, we can approximate
them as
\begin{equation}
  \begin{aligned}
    \psi(s) &\sim 1 - s\tau \\
    \phi(\omega) &\sim 1 - \sigma^\mu\omega^\mu
  \end{aligned}
\end{equation}
\begin{figure}[tbhp]
  \begin{center}
    \setlength{\unitlength}{1.5em}
    \begin{picture}(8,16)(0,0)
      \multiput(0,8)(0,8){2}{\line(1,0){8}}
      \multiput(0,8)(8,0){2}{\line(0,1){8}}
      \multiput(0,0)(0.5,0){16}{\line(1,0){0.25}}
      \multiput(0,0)(8,0){2}{
        \multiput(0,0)(0,0.5){16}{\line(0,1){0.25}}
      }
      \put(0,0){\line(1,2){8}}
      \put(9,12){\vector(-1,0){2.8}}
      \put(9.1,12){\makebox(0,0)[l]{$H=1/2$; diffusion}}
      \put(-0.1,0){\makebox(0,0)[r]{0}}
      \put(-0.1,8){\makebox(0,0)[r]{1}}
      \put(-0.1,16){\makebox(0,0)[r]{2}}
      \put(-0.3,12){\makebox(0,0)[r]{$\mu$}}
      \put(0,-0.2){\makebox(0,0)[t]{0}}
      \put(8,-0.2){\makebox(0,0)[t]{1}}
      \put(4,-0.2){\makebox(0,0)[t]{$\alpha$}}
      \put(1,15){\makebox(0,0)[l]{$H<1/2$}}
      \put(1,14){\makebox(0,0)[l]{subdiffusion}}
      \put(3,4){\makebox(0,0)[l]{$H>1/2$}}
      \put(3,3){\makebox(0,0)[l]{superdiffusion}}
    \end{picture}
  \end{center}
  \caption{\capstyle Subdiffusion and superdiffusion for long waits
    and long jumps as a function of the parameters $\alpha$ and $\mu$
    of the relation $\langle{X^2}\rangle\sim{t^{2\alpha/\mu}}$, that
    is, $H=\alpha/\mu$ (see text).}
  \label{diffme}
\end{figure}
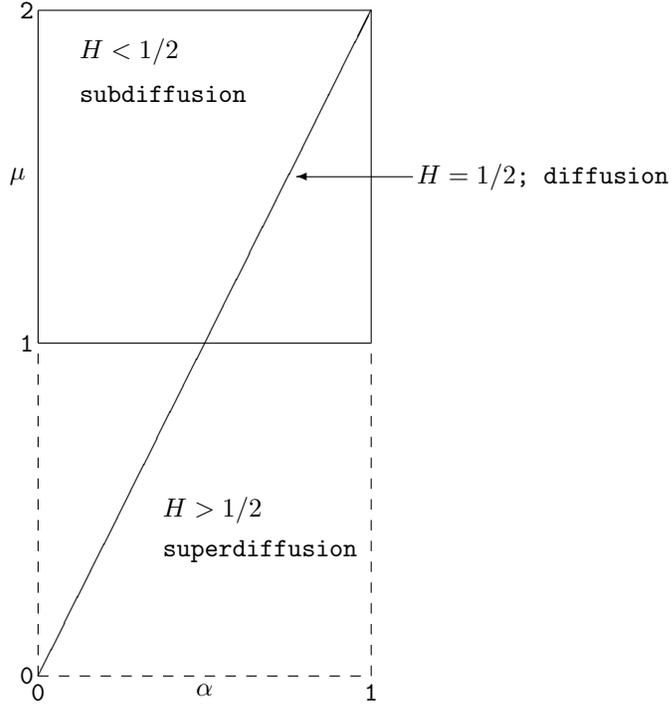

Inserting these approximations into (\ref{ooohh}) we have
\begin{equation}
  \tilde{\rho}(\omega,s) = \tau \frac{\tilde{\rho}(\omega,0)}{s + K^\mu\omega^\mu}
\end{equation}
where $K^\mu=\sigma^\mu/\tau$ or, for $\rho(x,0)=\delta(x)$, 
\begin{equation}
  \tilde{\rho}(\omega,s) = \frac{\tau}{s + K^\mu\omega^\mu}
\end{equation}
Taking the inverse Laplace transform, we have
\begin{equation}
  \tilde{\rho}(\omega,t) = \exp(-K^\mu \omega^\mu)
\end{equation}
that is, we obtain a L\'evy stable distribution, as expected from the
generalized Central Limit Theorem. Note that in this case
$\langle{X^2}\rangle\rightarrow\infty$, so we can't directly compare
this distribution with the standard diffusion. There are however several ways
to arrive at a result. The first is to use a truncated L\'evy
distribution, which is closer to real applications as in the physical
world one doesn't have arbitrarily long jumps. The second is to
extrapolate from fractional moments $\langle{X^q}\rangle$ with
$q<\mu$, which can be shown to converge and, in this case:
\begin{equation}
  \langle{X^q}\rangle \sim t^{q/\mu}
\end{equation}
which leads to a Hurst exponent $H=1/\mu>1/2$, that is, to
superdiffusion%
\footnote{An informal way to reach the same conclusion is to note that
  $K^\mu=\sigma^\mu/\tau$ so that, in order to have $K^\mu$ finite, we
  must have $\sigma^\mu\sim\tau$, that is, $\sigma^2\sim\tau^{2/\mu}$,
  leading again to $H=1/\mu$.}%
.

\subsubsection{Long waits and long jumps}
The case in which both jumps and waiting times have L\'evy distribution
can be trated similarly, leading to
\begin{equation}
  \tilde{\rho}(\omega,s) = \frac{1}{s} \frac{1}{1 + K_\alpha^\mu \omega^\mu s^{-\alpha}}
\end{equation}
By analogy with the previous cases, we see that
\begin{equation}
  \langle X^2 \rangle \sim t^{2\alpha/\mu}
\end{equation}
which entails $H=\alpha/\mu$. If $\mu>2\alpha$, then $H<1/2$, and we
have subdiffusion, if $\mu>2\alpha$ we have superdiffusion (see
figure~\ref{diffme}).

\appendix

\chapter{}
\section{\sc Random variables}
A \emph{random variable} is a mathematical object characterized by a
set $\Omega$ (the \emph{range} of the variable), which contains all
possible outcomes of the variable and a function $P_X(x)$%
\footnote{Following the standard notation, we shall use capital
  letters to indicate random variables and lowercase letters to
  indicate the values that they assume.}%
~that assigns, to each $x\in\Omega$ a value $P_X(x)\in[0,1]$ called
its \emph{probability}. The function $P_X$ is not arbitrary, but must
meet some minimal conditions. If the set $\Omega$ is finite or
countable, these conditions can be expressed simply as
\begin{description}
\item[i)] $\displaystyle \forall{x}\in\Omega.\ P_X(x)\ge0$ (positivity) 
\item[ii)] $\displaystyle \sum_{x\in\Omega} P_X(x) = 1$ (normalization)
\end{description}
If $\Omega$ is uncountable, the conditions are technically more
complex. In this case, $X$ is a continuous random variable, and $P_X$
is referred to as the \emph{probability density} function (PDF) of
$X$. The function $P_X$ in this case represents a probability only if
it is integrated over a subset of $\Omega$ of non-zero measure. In
this case, the normalization condition is
\begin{equation}
  \int_\Omega P_X(x) dx = 1
\end{equation}
If $\Omega={\mathbb{R}}$ (as we shall often assume) we have
\begin{equation}
  \int_{-\infty}^\infty P_X(x) dx = 1
\end{equation}
For continuous variables on ${\mathbb{R}}$ one can define the
\emph{cumulative probability function}, that is, the probability that
$X$ be at most $x$:
\begin{equation}
  {\mathcal{P}}(x) = {\mathbb{P}}[X\le{x}] = \int_{-\infty}^x P_X(u) du
\end{equation}
Note that
\begin{equation}
  P_X(x) = \frac{\partial}{\partial x}{\mathcal{P}}(x).
\end{equation}
From this and the positivity condition we can derive that ${\mathcal{P}}$
is monotonically non-decreasing and that
\begin{equation}
  \lim_{x\rightarrow-\infty} {\mathcal{P}}(x)=0\ \ \ \ %
  \lim_{x\rightarrow\infty} {\mathcal{P}}(x)=1
\end{equation}

In some cases, a whole function might be difficult to work with; it is
easier to work with an enumerable set of numbers that characterizes
the function completely. \emph{Statistical moments} are such
quantities. The moment of order $n$ of the variable $X$ is defined as
\begin{equation}
  \label{mom}
  \langle X^n \rangle = \int_\Omega x^n P_X(x) dx
\end{equation}
In general, given a function $f$ defined on $\Omega$, we define
\begin{equation}
  \langle f(X) \rangle = \int_\Omega f(x) P_X(x) dx
\end{equation}
The $n$th moment is obtained for $f(x)=x^n$.

The first order moment $\langle{X}\rangle$ is called the \emph{mean},
the \emph{average}, or the \emph{expected value} of $X$, while
\begin{equation}
  \sigma^2 = \langle X^2 \rangle - \langle X \rangle^2
\end{equation}
is the \emph{variance}; its square root $\sigma$ is the \emph{standard
  deviation} of $X$.

Not all distributions have finite moments, that is, the integral
(\ref{mom}) may fail to converge. If the moments are finite, then they
completely characterize the PDF. To show this, we introduce the
\emph{characteristic function} $\tilde{P}_X(\omega)$ of a PDF $P_X$:
\begin{equation}
  \label{pooh}
  \tilde{P}_X(\omega) = \langle e^{i\omega{x}}\rangle = \int_\Omega e^{i\omega{x}}P_X(x)\,dx
\end{equation}
This is simply the Fourier transform of $P_X$, so the PDF can be
recovered from its characteristic function as
\begin{equation}
  P_X(x) = \frac{1}{2\pi} \int e^{-i\omega{x}}\tilde{P}_X(\omega)\,d\omega
\end{equation}
The relation with the moments becomes evident by taking the Taylor
expansion of the exponential:
\begin{equation}
  e^{i\omega{x}} = \sum_n^\infty \frac{(i\omega{x})^n}{n!}
\end{equation}
Introducing this into (\ref{pooh}) we get
\begin{equation}
  \tilde{P}_X(\omega) = \sum_n \frac{(i\omega)^n}{n!} \int x^nP_X(x)\,dx = \sum_n \frac{(i\omega)^n}{n!} \langle X^n \rangle
\end{equation}
As a consequence, the moments of $P_X$ can be obtained by
differentiating $\tilde{P}_X$:
\begin{equation}
  \langle X^n \rangle = \lim_{\omega\rightarrow{0}} (-i)^n \frac{\partial^n}{\partial \omega^n} \tilde{P}_X(\omega)
\end{equation}

\separate

The \emph{joint probability} of two random variables $X_1$ and $X_2$,
indicated as $P_{X_1\cap{X_2}}(x_1,x_2)$ measures the simultaneous
probability that $X_1$ and $X_2$ take the values $x_1$ and $x_2$,
respectively. The \emph{conditional probability}
$P_{X_1|X_2}(x_1|x_2)$ denotes the probability that $X_1$ take value
$x_1$ conditioned to the fact that $X_2$ takes value $x_2$.  Two
variables are \emph{independent} if for all $x_1$, $x_2$
$P_{X_1|X_2}(x_1x_2)=P_{X_1}(x_1)$, that is, knowing the value of
$X_2$ does not change the distribution of $X_1$. Joint and conditional
probabilities are related through Bayes's theorem:
\begin{equation}
  P_{X_1\cap{X_2}}(x_1,x_2) = P_{X_1|X_2}(x_1|x_2)P_{X_2}(x_2) = P_{X_2|X_2}(x_2|x_1)P_{X_1}(x_1)
\end{equation}

\subsection{Useful Probability Distributions}
A variable $X$ follows a \textbf{Gaussian} (or \emph{normal})
distribution if
\begin{equation}
  P_X(x) = \frac{1}{\sigma \sqrt{2\pi}} \exp\Bigl( -\frac{(x-\mu)^2}{\sigma^2} \Bigr)
\end{equation}
(Figure~\ref{gaussian}) or, equivalently, it has characteristic function
\begin{equation}
  \label{lookie}
  \tilde{P}_X(\omega) = \int_{-\infty}^{\infty} e^{i\omega{x}} P_X(x)\,dx = \exp\Bigl( i\omega\mu - \frac{\omega^2\sigma^2}{2} \Bigr)
\end{equation}
\begin{figure}[thbp]
  \begin{center}

  \end{center}
  \caption{\capstyle The Gaussian PDF (a) and the corresponding
    cumulative distribution (b) for various values of $\sigma$; in all
    cases it is $\mu=0$.}
  \label{gaussian}
\end{figure}
The mean of the distribution is $\langle{X}\rangle=\mu$. Note that,
for $\mu=0$, the characteristic function has also the functional form
of a Gaussian, a fact that will have important consequences. In this
special case ($\mu=0$) the moments are given by
\begin{equation}
  \langle X^n \rangle = \int_{-\infty}^\infty x^n \frac{1}{\sigma \sqrt{2\pi}} \exp\Bigl( -\frac{(x-\mu)^2}{\sigma^2} \Bigr)\,dx = 
  \begin{cases}
    \frac{2^{\frac{n}{2}}\sigma^n}{\sqrt{\pi}} \Gamma\Bigl( \frac{n+1}{2} \Bigr) & \mbox{$n$ even} \\
    0 & \mbox{$n$ odd} 
  \end{cases}
\end{equation}
where $\Gamma$ is Euler's Gamma function. An important moment is 
\begin{equation}
  \langle X^2 \rangle = \sigma^2 + \langle X \rangle.
\end{equation}
One important property of the Gaussian distribution, vis-\`a-vis the
Central Limit Theorem (which we shall consider in the following), is
that it is \emph{stable}: if $X,Y$ are Gaussians, and
$a,b\in{\mathbb{R}}$, then $aX+bY$ is also Gaussian.

Let $X$ and $Y$ be two Gaussian-distributed variables with zero mean
and variance $\sigma_1^2$ and $\sigma_2^2$, respectively. Then
\begin{equation}
  \begin{aligned}
    \tilde{P}_X(\omega) &= \exp\bigl( - \frac{\omega^2\sigma_1^2}{2} \bigr) \\
    \tilde{P}_Y(\omega) &= \exp\bigl( - \frac{\omega^2\sigma_2^2}{2} \bigr) 
  \end{aligned}
\end{equation}
and consequently
\begin{equation}
  \label{ggood}
  \tilde{P}_X(\omega)\tilde{P}_Y(\omega) = \exp\bigl( - \frac{\omega^2(\sigma_1^2+\sigma_2^2)}{2} \bigr) 
\end{equation}
that is, the product of the characteristic function of two Gaussian
distributions is still the characteristic function of a Gaussian
distribution. 

\separate

The Gaussian distribution is defined for all $x\in{\mathbb{R}}$, but
many variables that one might be interested in modeling assume only
positive values in such a way that the probability that $x=0$ is $0$
and, after reaching a maximum, decreases rapidly for high values of
$x$. The most different things can be observed to have this
distribution, from the length of messages in internet fori to the
prices of hotels, or the size of particles in a collision.

All these phenomena can be modeled as following a \textbf{logonormal
  distribution}. A variable $X$ has logonormal distribution if
$\log{X}$ has normal (viz.\ Gaussian) distribution. Let $\Phi$ and
$\phi$ be the cumulative distribution and the density of a normally
distributed variable with $0$ mean and unit variance
(${\mathcal{N}}(0,1)$), and assume
$\log{X}\sim{\mathcal{N}}(\mu,\sigma)$, i.e.\ $\log{X}$ has a normal
distribution with mean $\mu$ and variance $\sigma^2$. Then
\begin{equation}
  \begin{aligned}
    P_X(x) &= \frac{d}{dx} {\mathcal{P}}_X(x) = \frac{d}{dx} {\mathbb{P}}[X\le{x}] \\
           &= \frac{d}{dx} {\mathbb{P}}[\log{X}\le\log{x}] \\
           &= \frac{d}{dx} \Phi\bigl[ \frac{\log{x}-\mu}{\sigma}\bigr] \\
           &= \phi\bigl[ \frac{\log{x}-\mu}{\sigma}\bigr] \frac{d}{dx} \bigl[ \frac{\log{x}-\mu}{\sigma}\bigr] \\
           &= \frac{1}{\sigma{x}} \phi\bigl[ \frac{\log{x}-\mu}{\sigma}\bigr] \\
           &= \frac{1}{\sqrt{2\pi}\sigma{x}} \exp\Big[- \frac{(\log{X}-\mu)^2}{2\sigma^2}\Bigr]
  \end{aligned}
\end{equation}
Figure~\ref{logogauss} shows the behavior of the logonormal PDF for various values of $\sigma$ and $\mu=0$
\begin{figure}[thp]
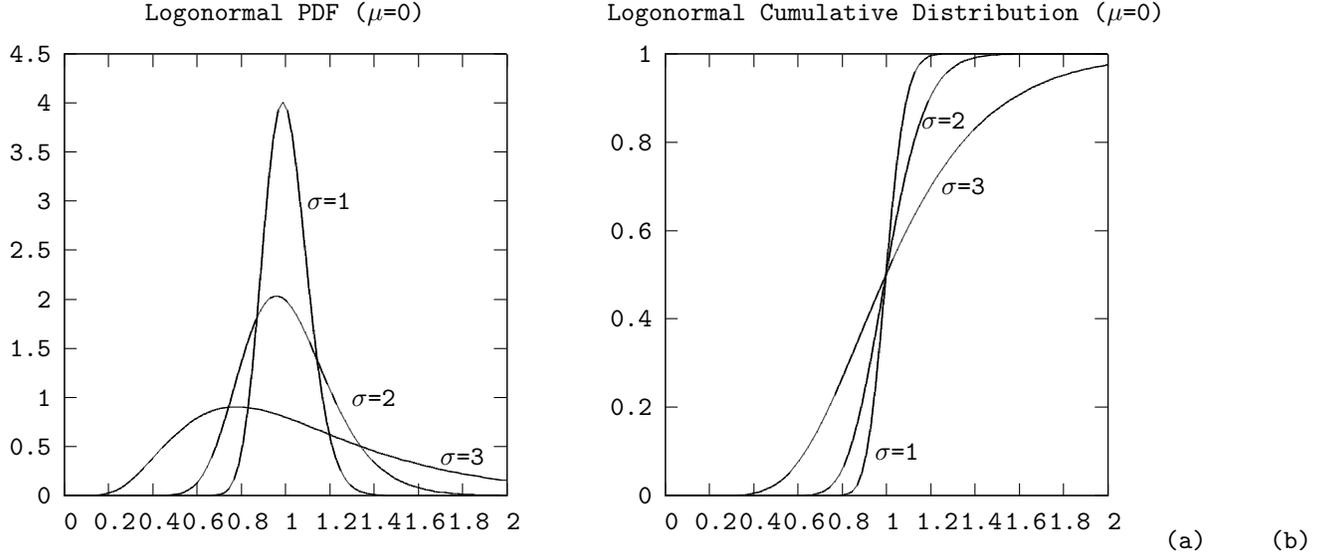

  \begin{center}

  \end{center}
  \caption{\capstyle The logonormal PDF (a) and the corresponding
    cumulative distribution (b) for various values of $\sigma$; in all
    cases it is $\mu=0$.}
  \label{logogauss}
\end{figure}
Note that $\mu$ and $\sigma$ are the mean and variance of $\log{X}$,
\emph{not} of $X$. To distinguish them, I shall indicate the mean and
the variance of $X$ as $m$ and $v$, respectively.

The moments of $X$ are given by
\begin{equation}
  \langle X^n \rangle = \int_0^\infty x^n P_X(x) dx = \exp\bigl(n\mu + \frac{n^2\sigma^2}{2}\bigr)
\end{equation}
as can be verified by replacing
$z=\frac{1}{\sigma}\bigl[\log{X}-(\mu+n\sigma^2)\bigr]$ in the
integral. From this we have
\begin{equation}
  \begin{aligned}
    m &= \langle{X}\rangle = \exp\bigl(\mu+\frac{\sigma^2}{2}\bigr) \\
    \langle{X^2}\rangle &= \exp\bigl(2\mu+2\sigma^2\bigr) \\
    v &= \langle{X^2}\rangle - \langle{X}\rangle^2 = \exp(2\mu+\sigma^2)(e^{\sigma^2}-1)
  \end{aligned}
\end{equation}
From these equality, one can derive the values of $\mu$ and $\sigma^2$ for
desired $m$ and $v$:
\begin{equation}
  \mu = \log\frac{m}{\sqrt{\displaystyle 1 + \frac{v}{m^2}}}\ \ \ %
  \sigma^2 = \log\left(1 + \frac{v}{m^2}\right)
\end{equation}
The characteristic function $\langle\exp(i\omega{x})\rangle$ is
defined, but if we try to extend it to complex variables,
$\langle\exp(s{x})\rangle$, $s\in{\mathbb{C}}$ is not defined for any
$s$ with a negative imaginary part. This entails that the
characteristic function is not analytical in the origin and,
consequently, it can't be represented as an infinite convergent
series. In particular, the formal Taylor series
\begin{equation}
  \sum_n \frac{(i\omega{x})^n}{n!} \langle{x^n}\rangle = 
  \sum_n \frac{(i\omega{x})^n}{n!} \exp\bigl(n\mu + \frac{\displaystyle n^2\sigma^2}{2}\bigr)
\end{equation}
diverges

\separate

Other positive variables follow a different distribution, one in which
the value $0$ is the most probable, and the probability decreases
sharply as $x$ increases, In these cases, the variable $x$ can be
modeled using an \textbf{exponential} distribution:
\begin{equation}
  P_X(x) = 
  \begin{cases}
    \lambda e^{-\lambda{x}} & x \ge 0 \\
    0                    & x < 0
  \end{cases}
\end{equation}
If the variable can take negative values, then
\begin{equation}
  P_X(x) = \frac{\lambda}{2} e^{-\lambda|x|}
\end{equation}
(Figure~\ref{exponential}. Its characteristic function is
\begin{figure}
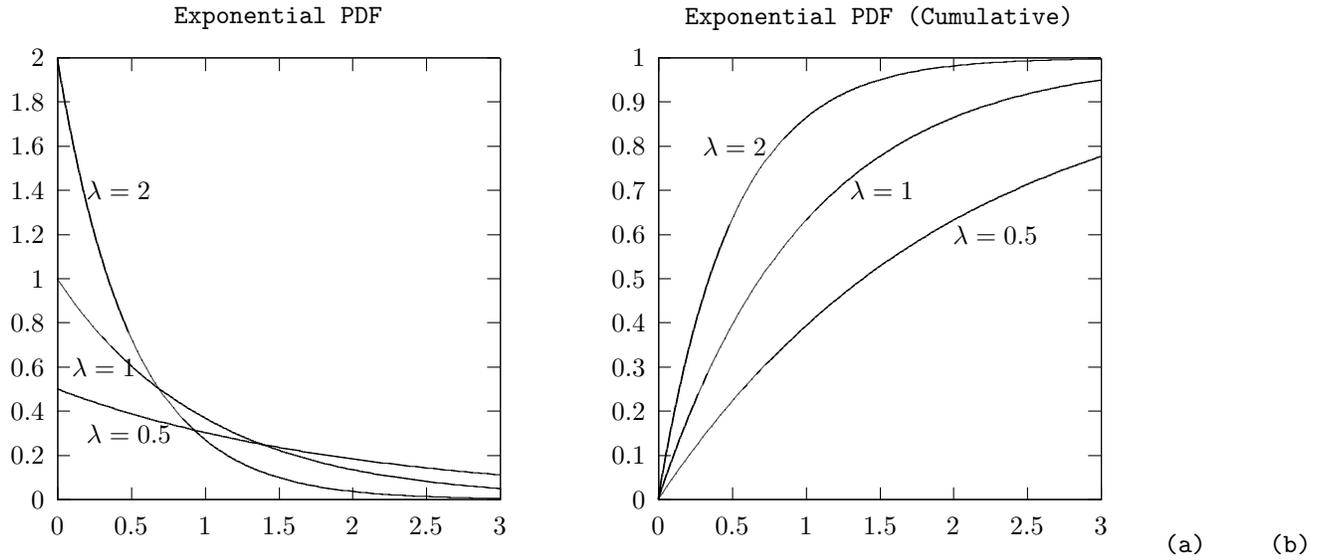

  \begin{center}

  \end{center}
  \caption{\capstyle The exponential PDF (a), an the corresponding
    cumulative distribution in (b) for various values of $\lambda$.}
  \label{exponential}
\end{figure}
\begin{equation}
  \tilde{P}_X(\omega) = \frac{\lambda^2}{\lambda^2+\omega^2}
\end{equation}
and its moments
\begin{equation}
  \langle X^n \rangle = \frac{1}{\lambda^n} \Gamma(n+1)
\end{equation}

\separate

A \textbf{uniform} or \emph{flat} distribution assigns the same
probability density to each point in $\Omega$. So, if $\Omega=[a,b]$,
\begin{equation}
  P_X(x) = 
  \begin{cases}
    \frac{1}{b-a} & a \le x \le b \\
    0             & \mbox{otherwise}
  \end{cases}
\end{equation}
The characteristic function of the uniform distribution is
\begin{equation}
  \tilde{P}_X(\omega) = \frac{e^{i\omega{b}}-e^{i\omega{a}}}{i\omega(b-a)}
\end{equation}
and its moments
\begin{equation}
  \langle X^n \rangle = \frac{1}{n+1}\frac{b^{n+1}-a^{n+1}}{b-a}
\end{equation}

\separate

A \textbf{Cauchy}, or \textbf{Lorentz} distribution has PDF
\begin{equation}
  P_X(x) = \frac{1}{\pi} \frac{\gamma}{x^2 + \gamma^2}
\end{equation}
where $\gamma$ is a positive parameter, and characteristic function
\begin{equation}
  \tilde{P}_X(\omega) = e^{-\gamma|\omega|}
\end{equation}
If one tries to compute the moments using the definition
\begin{equation}
  \langle X^n \rangle = \frac{\gamma}{\pi} \int \frac{x^n}{x^2 + \gamma^2}\,dx
\end{equation}
then, since the integrand behaves as $x^{n-2}$ for
$x\rightarrow\infty$, one observes that they diverge for
$n\ge{1}$. This limits the usefulness of this distribution as a model
of real phenomena (which typically have finite moments), and in
practice one \dqt{truncates} the distribution to a finite interval
$[a,b]$.

\separate

We have mentioned that one important property of the Gaussian
distribution is the preservation of the functional form of their
characteristic function under multiplication, as in (\ref{ggood}). The
Gaussian distribution is not the most general distribution with this
property (although it is the only one with this property \emph{and} finite
moments): it is shared by the family of \textbf{L\'evy
  distributions}. L\'evy distributions depend on four parameters:
$\alpha$ (L\'evy index), $\beta$ (skew), $\mu$ (shift), and
$\sigma$ (scale), and they are defined through their characteristic
function:
\begin{equation}
  \tilde{P}_{\alpha,\beta}(\omega;\mu,\sigma) = \int_{-\infty}^\infty e^{i\omega{x}} P_{\alpha,\beta}(x;\mu,\sigma)\,dx 
  \stackrel{\triangle}{=} \exp \left[ i\mu\omega - \sigma^\alpha |\omega|^\alpha \left( 1 - i\beta\frac{\omega}{|\omega|}\Phi \right)\right]
\end{equation}
where
\begin{equation}
  \Phi = 
  \begin{cases}
    \tan \frac{\alpha\pi}{2} & \alpha \ne 1, 0<\alpha<2 \\
    - \frac{2}{\pi} \ln |x|  & \alpha = 1
  \end{cases}
\end{equation}
the four parameters determine the shape of the distribution. Of these,
$\alpha$ and $\beta$ play a major r\^ole in this note, while
$\mu$ and $\sigma$ can be eliminated through proper scale and shift
transformations (much like mean and variance for the Gaussian
distribution):
\begin{equation}
  P_{\alpha,\beta}(x;\mu,\sigma) = \frac{1}{\sigma} P_{\alpha,\beta}(\frac{x-\mu}{\sigma}; 0, 1)
\end{equation}
From now on, I shall therefore ignore $\mu$ and $\sigma$ and refer to
the distribution as $P_{\alpha,\beta}(x)$. Note the symmetry relation
\begin{equation}
  P_{\alpha,-\beta}(x) = P_{\alpha,\beta}(x)
\end{equation}
The distributions with $\beta=0$ are symmetric, and these are the ones
that are the most relevant in this context. The closed form of
$P_{\alpha,\beta}$ is known only for a few cases. If $\alpha=2$ one
obtains the Gaussian distribution ($\beta$ is irrelevant, since
$\Phi=0$); if $\alpha=1,\beta=0$ one obtains the Cauchy distribution,
and for $\alpha=1/2,\beta=1$, the L\'evy-Smirnov distribution
\begin{equation}
  P_{1/2,1}(x) = 
  \begin{cases}
    \frac{1}{\sqrt{2\pi}} x^{-\frac{3}{2}}\exp\bigl( -\frac{1}{2x} \bigr) & x \ge 0 \\
      0 & x<0
  \end{cases}
\end{equation}
The most important property in this context is the asymptotic
behavior of $P_{\alpha,\beta}$ which is given by the power law
\begin{equation}
  \label{levypow}
  P_{\alpha,0}(x) \sim \frac{C(\alpha)}{|x|^{1+\alpha}}
\end{equation}
with
\begin{equation}
  C(\alpha) = \frac{1}{\pi} \sin \bigl( \frac{\pi\alpha}{2} \bigr) \Gamma(1+\alpha)
\end{equation}
This power law behavior entails that arbitrarily large values are
relatively probable (compared with the exponential decay of the
Gaussian). Consequently, as can be expected, $\langle{X}^2\rangle$
diverges for $\alpha<2$.

\separate

The \textbf{Dirac delta distribution} is a pathological distribution
useful in many contexts; for example, when dealing with
certainty in a probabilistic framework, or when analyzing
discrete random variables in a context created for continuous
ones. The distribution is:
\begin{equation}
  P_X(x) = \delta(x-x_0)
\end{equation}
where $\delta(\cdot)$ is the Dirac distribution. The characteristic
function of the distribution is
\begin{equation}
  \label{deltachar}
  \tilde{P}_X(\omega)=\exp(i\omega{x_0}). 
\end{equation}
The function $\delta(x)$ is zero everywhere except for $x=0$, and
\begin{equation}
  \int_{-\infty}^\infty \delta(x)\,dx = 1
\end{equation}
This property entails $\delta(ax)=\delta(x)/a$. Also
\begin{equation}
  \int_{-\infty}^\infty f(x)\delta(x-x_0)\,dx = f(x_0)
\end{equation}
from which we derive
\begin{equation}
  \langle x^n \rangle = x_0^n
\end{equation}

\separate

Unlike the previous distribution, the \textbf{binomial distribution}
is defined for discrete variables, in particular for a variable $X$
that can take two values, the first one with probability $p$, and the
second one with probability $1-p$. Suppose, for example, that we play
a game in which, at each turn, I have a probability $p$ of winning
and $1-p$ of losing (think of head-and-tails game with a tricked
coin). If we play $N$ rounds of the game, what is the probability that
I win exactly $n$ times? This turns out to be 
\begin{equation}
  P(X=n) = \left(\begin{array}{c}N\\n\end{array}\right) p^n (1-p)^{N-n}=
    \frac{N!}{n!(N-n)!}p^n (1-p)^{N-n}
\end{equation}
which is precisely the binomial distribution. Its characteristic function is
\begin{equation}
  \tilde{P}(\omega) = (1-p+pe^{i\omega})^N
\end{equation}
from which the moments can be derived. For example
\begin{equation}
  \label{binmean}
  \langle X \rangle = \lim_{\omega\rightarrow{0}} \frac{d\tilde{P}}{d\omega} = 
  \lim_{\omega\rightarrow{0}} pN e^{i\omega} (1-p+pe^{i\omega})^{N-1} = pN
\end{equation}

\separate

An important and common distribution, one that appears as a limiting
case of many finite processes, is the \textbf{Poisson
  Distribution}. Its importance will probably be more evident if we
derive it as a limiting case in some examples.

\example
Consider events that may happen at any moment in time (the events are
punctual: they have no duration). Divide the time-line in small
intervals of duration $\Delta{t}$, so short that the probability that
two or more events will take place in the same interval is
negligible. Assume that the probability that \emph{one} event take
place in $[t,t+\Delta{t})$ is constant, and proportional to the length
  of the interval:
\begin{equation}
  P(1;\Delta{t}) = \lambda\Delta{t}
\end{equation}
and, because no two events happen in the same interval,
\begin{equation}
  P(0;\Delta{t}) = 1-\lambda\Delta{t}
\end{equation}
Let $P(0;t)$ be the probability that no event has taken place up to
time $t$. Then
\begin{equation}
  P(0;t+\Delta{t}) = P(0;t)(1-\lambda\Delta{t})
\end{equation}
Rearranging the terms we get
\begin{equation}
  \frac{P(0;t+\Delta{t})-P(0;t)}{\Delta{t}}=-\lambda P(0;t)
\end{equation}
and, taking the limit for $\Delta{t}\rightarrow{0}$
\begin{equation}
  \frac{\partial}{\partial{t}} P(0;t) = -\lambda P(0;t)
\end{equation}
that is, $P(0;t)=C\exp(-\lambda{t})$ or, considering the boundary
condition $P(0,0)=1$,
\begin{equation}
  P(0;t) = e^{-\lambda{t}}
\end{equation}
This takes care of the case in which no event takes place before time $t$.
On to the general case. There were $n$ events by time $t+\Delta{t}$ if
either (1) we had $n$ events up to time $t$ and no event occurred in
$[t,t+\Delta{t}]$, or (2) there were $n-1$ events at $t$ and one event
occurred in $[t,t+\Delta{t}]$. This leads to
\begin{equation}
  P(n;t+\Delta{t}) = (1-\lambda\Delta{t})P(n;t) + \lambda\Delta{t}P(n-1;t)
\end{equation}
rearranging and taking the limit $\Delta{t}\rightarrow{0}$, we have
\begin{equation}
  \label{poissonpartial}
  \frac{\partial}{\partial t} P(n;t) + \lambda P(n;t) = \lambda P(n-1;t)
\end{equation}
In order to transform this equation into a more manageable form, we
look for a function that, multiplied by the left-hand side, transforms
it into the derivative of a product. That is, we look for a function
$\mu(t)$ such that
\begin{equation}
  \mu(t) \left[ \frac{\partial P}{\partial t} + \lambda P \right] = \frac{\partial}{\partial t} \bigl[ \mu(t) P \bigr]
\end{equation}
It is easy to verify that $\mu(t)=\exp(\lambda{t})$ fits the bill.
Equation (\ref{poissonpartial}) therefore becomes
\begin{equation}
  \frac{\partial}{\partial t} \Bigl[ e^{\lambda{t}}P(n;t) \Bigr] = e^{\lambda{t}}\lambda P(n-1;t)
\end{equation}
For $n=1$ we have
\begin{equation}
  \frac{\partial}{\partial t} \Bigl[ e^{\lambda{t}}P(1;t) \Bigr] = e^{\lambda{t}}\lambda e^{-\lambda{t}} = \lambda
\end{equation}
That is, integrating both sides and multiplying by $e^{-\lambda{t}}$
\begin{equation}
  P(1;t) = \lambda t e^{-\lambda{t}}
\end{equation}
For arbitrary $n$, I'll show by induction that
\begin{equation}
  \label{fish}
  P(n;t) = \frac{(\lambda{t})^n}{n!}e^{-\lambda{t}}
\end{equation}
We have already derived the result for $n=0$ and for $n=1$. For
arbitrary $n$, we have
\begin{equation}
  \begin{array}{lcll}
    \displaystyle \frac{\partial}{\partial t} \Bigl[ e^{\lambda{t}}P(n+1;t) \Bigl] & = & e^{\lambda{t}} \lambda P(n;t) & \\
    \displaystyle & = & e^{\lambda{t}} \lambda \frac{(\lambda{t})^n}{n!} e^{-\lambda{t}} & \mbox{(induction hypothesis)} \\
    \displaystyle & = & \lambda \frac{(\lambda{t})^n}{n!}
  \end{array}
\end{equation}
So, integrating
\begin{equation}
  e^{\lambda{t}}P(n+1;t) = \frac{\lambda}{n!} \int (\lambda{t}^n) dt = \frac{(\lambda{t})^{n+1}}{(n+1)!} + C
\end{equation}
where $C=0$ because of the initial conditions, so
\begin{equation}
  P(n+1;t) = e^{-\lambda{t}}\frac{(\lambda{t})^{n+1}}{(n+1)!}
\end{equation}

~~\eoe

The distribution that results from this example:
\begin{equation}
  \label{poisson}
  P_X(x) = e^{-x} \frac{x^n}{n!}
\end{equation}
is the Poisson distribution that, in the example, gives us the
probability that $n$ events take place in a time $x$.
Figure~\ref{poissonfig} shows the shape of this distribution as a
function of $x$ for various values of $n$.
\begin{figure}
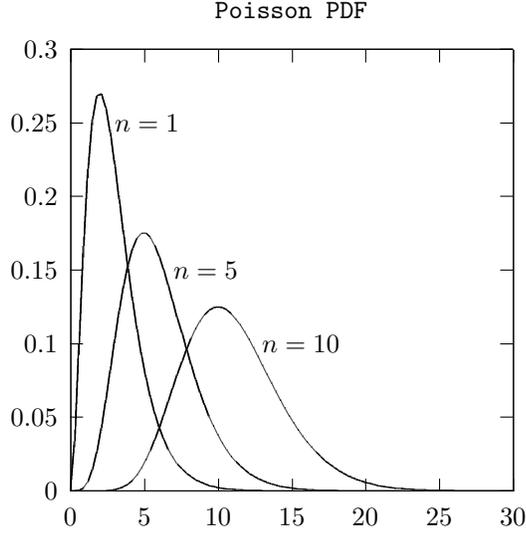

  \begin{center}
\setlength{\unitlength}{0.240900pt}
\ifx\plotpoint\undefined\newsavebox{\plotpoint}\fi

  \end{center}
  \caption{\capstyle The Poisson PDF for various values of $n$.}
  \label{poissonfig}
\end{figure}

\example
The Poisson distribution can also be seen as a limiting case of the
binomial distribution. If $p$ is the probability of success, then
$\nu=Np$ is the expected number of successful trials, as per
(\ref{binmean}). This approximation is valid for large $N$. In this
case, we have
\begin{equation}
  P(n;N) = \frac{N!}{n!(N-n)!} \left(\frac{\nu}{N}\right)^n \left(1-\frac{\nu}{N}\right)^{N-n}
\end{equation}
Taking $N\rightarrow\infty$, we have
\begin{equation}
  \begin{aligned}
    P_\nu(n) &= \lim_{N\rightarrow\infty} P(n;N) \\
            &= \lim_{N\rightarrow\infty} \frac{N\cdot(N-1)\cdots(N-n+1)}{n} \frac{\nu^n}{N^n} \left(1-\frac{\nu}{N}\right)^N \left(1-\frac{\nu}{N}\right)^{-n} \\
            &= \lim_{N\rightarrow\infty} \frac{N\cdot(N-1)\cdots(N-n+1)}{N^n} \frac{\nu^n}{n!} \left(1-\frac{\nu}{N}\right)^N \left(1-\frac{\nu}{N}\right)^{-n} \\
            &= 1 \cdot \frac{\nu^n}{n!} e^{-\nu} \cdot 1 \\
            &= \frac{\nu^n}{n!} e^{-\nu}
    \end{aligned}
\end{equation}
So, once again, we find that the number of successes has a Poisson
distribution.

~~\eoe

The characteristic function of the distribution (\ref{poisson}) is
\begin{equation}
  \tilde{P}(\omega) = e^{\lambda(e^{i\omega}-1)}
\end{equation}
from which we obtain
\begin{equation}
  \langle X \rangle = \lambda
\end{equation}

\subsection{Functions of Random Variables}
If $X$ is a random variable on $\Omega$, and
$f:\Omega\rightarrow\Omega'$, then $Y=f(X)$ is a random variable on
$\Omega'$. Here I'll consider, for the sake of simplicity, the case
$\Omega=\Omega'={\mathbb{R}}$ (all our considerations can be
generalized to arbitrary continua $\Omega$ under fairly general
conditions, essentially that $\Omega$ be a metric space). In order to
determine the distribution of $y$, I begin with a preliminary
observation. For a random variable $X$, let
${\mathbb{P}}_X[x,x+\Delta{x}]$ the probability that the value of $X$
falls in $[x,x+\Delta{x}]$. Then, for small $\Delta{x}$,
\begin{equation}
  \label{poof}
  \begin{aligned}
    {\mathbb{P}}_X[x,x+\Delta{x}] &= P(X\le x+\Delta{x}) - P(X\le x) \\
                                  &= \frac{\partial}{\partial x} P(X\le x)\Delta{x} + O(\Delta{x}^2) \\
                                  &= P_X(x)\Delta{x} + O(\Delta{x}^2)
  \end{aligned}
\end{equation}
Let now $f$ be invertible, and $g=f^{-1}$. Then
\begin{equation}
  \begin{aligned}
    P_Y\Delta{y} &= {\mathbb{P}}_Y[y,y+\Delta{y}] \\
                 &= {\mathbb{P}}_X[g(y),g(y+\Delta{y})] \\
                 &\approx {\mathbb{P}}_X\Bigl[g(y),g(y) + \left|\frac{dg}{dy}\right|\Delta{y})\Bigr] \\
                 &=  P_X(g(y))\left|\frac{dg}{dy}\right|\Delta{y}
  \end{aligned}
\end{equation}
from which we get
\begin{equation}
  P_Y(y) = P_X(g(y))\left|\frac{dg}{dy}\right|
\end{equation}
Note that equivalently one could have defined
\begin{equation}
  P_Y(y) = \int \delta(y-f(x))P_X\,dx = \langle \delta(y-f(x)) \rangle_X
\end{equation}
where the subscript on the average reminds us that we are taking the
average with respect to the distribution of $X$. From this, we can
determine the characteristic function of $Y$:
\begin{equation}
  \label{doh}
  \begin{aligned}
    \tilde{P}_Y(\omega) &= \int e^{i\omega{y}}P_Y(y)\,dy \\
                        &= \int P_X(x) \Bigl[ \int e^{i\omega{y}} \delta(y-f(x))\,dy \Bigr]\,dx \\
                        &= \int e^{i\omega{f(x)}}P_X(x)\,dx \\
                        &= \langle \exp\bigl[i\omega f(x)\bigr] \rangle_X
    \end{aligned}
\end{equation}
If $Y=aX$, then
\begin{equation}
  \label{quack}
  \tilde{P}_Y(\omega) = \langle \exp\bigl[i\omega a X\bigr] \rangle_X - \tilde{P}_X(a\omega)
\end{equation}

\separate

Consider now the sum of two random variables: $Z=X+Y$. Each value of
$Z$ can be obtained through an infinity of events: each time $X$ takes
an arbitrary value $x$, and $y$ takes a value $z-x$, $Z$ takes the
same value, namely $z$. Summing up all these possible events we obtain
\begin{equation}
  P_Z(z) = \int_{-\infty}^\infty P_X(x)P_Y(z-x)\,dx
\end{equation}
This is known as the \emph{convolution} of $P_X$ and $P_Y$, often
indicated as $P_Z=P_X*P_Y$. The properties of the Fourier transform
entail that the corresponding relation between characteristic
functions is
\begin{equation}
  \label{pluck}
  \tilde{P}_Z(\omega) = \tilde{P}_X(\omega)\tilde{P}_Y(\omega)
\end{equation}

\separate

Let $Y=\{y_1,\ldots,y_n\}$ be a set of independent and identically
distributed (i.i.d.) variables with cumulative distribution
${\mathcal{P}}_Y$ and density $P_Y$. Consider the function $\min(Y)$:
we are interested in finding its density $P_{\min}$ and cumulative
distribution ${\mathcal{P}}_{\min}$. We have:
\begin{equation}
  {\mathcal{P}}_Y(x) = {\mathbb{P}}\bigl[\min(Y)\le{x}\bigr] = 1 - {\mathbb{P}}\bigl[\min(Y)\ge{x}\bigr]
\end{equation}
We have $\min(Y)\ge{x}$ iff we have $y_i\ge{x}$ for all $i$, that is
\begin{equation}
  \begin{aligned}
    {\mathcal{P}}_{\min}(x) &= 1 - {\mathbb{P}}\Bigl[\forall y\in{Y}.y\ge{x}\Bigr] \\
                       &= 1 - {\mathbb{P}}\bigl[y\ge{x}\bigr]^n \\
                       &= 1 - \Bigl(1-{\mathbb{P}}\bigl[y\le{x}\bigr]\Bigr)^n \\
                       &= 1 - \Bigl(1-{\mathcal{P}}_Y(x)\Bigr)^n 
  \end{aligned}
\end{equation}
The density is
\begin{equation}
  \begin{aligned}
    P_{\min}(x) &= \frac{d}{dx} {\mathcal{P}}_{\min}(x) \\
               &= n\Bigl(1-{\mathcal{P}}_Y(x)\Bigr)^{n-1} \frac{d}{dx} {\mathcal{P}}_Y(x) \\
               &= n\Bigl(1-{\mathcal{P}}_Y(x)\Bigr)^{n-1} P_Y(x)
  \end{aligned}
  \label{mindense}
\end{equation}

For the function $\max(Y)$, working in a similar way, we have
\begin{equation}
  \begin{aligned}
    {\mathcal{P}}_{\max}(x) &= ({\mathcal{P}}_Y(x))^n
    P_{\max}(x) &= n({\mathcal{P}}_Y(x))^{n-1}P_Y(x)
  \end{aligned}
\end{equation}

\subsection{The Central Limit Theorem}
The Central Limit Theorem (important enough to be granted its own
acronym: CLT) is one of the fundamental results in basic probability theory
and the main reason why the Gaussian distribution is so important and
so common in modeling natural events. In a nutshell, the theorem
tells us the following: if we take a lot of random variables,
independent and identically distributed (i.i.d.), and add them up, the
result will be a random variable with Gaussian distribution. So, for
example, if we repeat an experiment many times and take the average of
the results that we obtain (the average is, normalization apart, a sum), no
matter what the characteristics of the experiment are, the resulting
average will have (more or less) a Gaussian distribution.

But, ay, there's the rub! The theorem works only in the assumption
that the moments of the distributions involved be finite. We shall see
shortly what happens if this assumption is not satisfied.

Let $X_1,\ldots,X_n$ be a set of i.i.d.\ random variables with
distribution $P_X$, zero mean, and (finite) variance $\sigma^2$. Note
that $Y=\sum_iX_i$ has zero mean and variance $n\sigma^2$, while
$Y=(\sum_iX_i)/n$ has zero mean and variance $\sigma^2/n$. It is
therefore convenient to work with the variable
\begin{equation}
  Z_n = \frac{1}{\sqrt{n}} \sum_i X_i
\end{equation}
which has zero mean and variance $\sigma^2$ independently of $n$.

\begin{theorem}
  For any distribution $P_X$ with finite mean and variance, and
  $X_1,\ldots,X_n$ i.i.d.\ with distribution $P_X$, for
  $n\rightarrow\infty$, we have $Z_n\rightarrow{Z_\infty}$, where
  $Z_\infty$ is a Gaussian random variable with zero mean and variance
  $\sigma^2$ equal to the variance of $P_X$.
\end{theorem}

\begin{proof}
  Consider the first terms of the expansion of the characteristic
  function of $P_X$:
  \begin{equation}
    \tilde{P}_X(\omega) = \int e^{i\omega{x}} P_X(x)\,dx = 1 - \frac{1}{2}\sigma^2\omega^2 + O(\omega^3)
  \end{equation}
  The characteristic function of $Y=\sum_iX_i$ is given by
  (\ref{pluck}):
  \begin{equation}
    \label{bingo}
    \tilde{P}_Y(\omega) = \prod_i \tilde{P}_{X_i}(\omega) = \bigl[ \tilde{P}_X(\omega) \bigr]^n
  \end{equation}
  (the second equality holds because the $X$s have the same
  distribution) while (\ref{quack}) with $a=1/\sqrt{n}$ gives
  \begin{equation}
      \tilde{P}_Z(\omega) = P_Y\left( \frac{\omega}{\sqrt{n}} \right) 
                          = \left[ P_Y\left( \frac{\omega}{\sqrt{n}} \right) \right]^n 
                          \approx \left(1 - \frac{\sigma^2 \omega^2}{2 n} \right)^n 
                          \stackrel{n\rightarrow\infty}{\longrightarrow} \exp(-\frac{1}{2}\sigma^2\omega^2)
  \end{equation}
  Finally, from (\ref{lookie}) we have the inverse transform
  \begin{equation}
    P_Z(z) = \frac{1}{\sigma\sqrt{2\pi}} \exp\Bigl(-\frac{z^2}{2\sigma^2}\Bigr)
  \end{equation}
\end{proof}

This theorem is true, in the form in which we have presented it, only
for distributions $X$ with finite mean and variance%
\footnote{I have assumed zero mean since, if the mean of the $X$ is
  non-zero, the mean of $Z$ goes to infinity; this doesn't represent a
  major hurdle for the theorem, which can easily be generalized by
  subtracting the mean from the variables $X$ and then adding it
  back.}%
. However, the key to the theorem is an invariance property of the
characteristic function of the Gaussian. Consider the equality
(\ref{bingo}); we can split it up as:
\begin{equation}
  \label{pringles}
  \tilde{P}_Z(\omega;n) = \bigl[ \tilde{P}_X(\omega) \bigr]^n = \bigl[ \tilde{P}_X(\omega) \bigr]^{n/2} \bigl[ \tilde{P}_X(\omega) \bigr]^{n/2} =
  \tilde{P}_Z(\omega;n/2) \tilde{P}_Z(\omega;n/2)
\end{equation}
Taking the limit $n\rightarrow\infty$, this gives us
$P_Z(\omega)=P_Z(\omega)P_Z(\omega)$. That is: the condition for a
distribution to be a central limit is that the product of two
characteristic functions have the same functional form as the original
distributions. As we have seen in (\ref{ggood}), the Gaussian
distribution does have this property. Nay: it is the \emph{only}
distribution with finite moments that has this property, hence its
appearance in the theorem in the finite moments case, and hence its
great importance in application as a model of many processes resulting
from the sum of identical sub-processes.

If we abandon the finite moment hypothesis, however, there is a more
general distribution to which (\ref{pringles}) applies: the stable
Levy distribution. So, a more general form of the CLT can be
enunciated as:

\begin{theorem}
  For any distribution $P_X$, and
  $X_1,\ldots,X_n$ i.i.d.\ with distribution $P_X$, for
  $n\rightarrow\infty$, we have 
  \begin{equation}
    \lim_{n\rightarrow\infty} \frac{1}{\sqrt{n}} \sum_{i=1}^n X_i = Z_\infty
  \end{equation}
  where $Z_\infty$ is a random variable with Levy distribution. If
  the variance of $P_X$ is finite and equal to $\sigma^2$, then
  $Z_\infty$ has a Gaussian distribution with variance $\sigma^2$.
\end{theorem}
  
\subsection{Stochastic Processes}
A \emph{stochastic process} is a set of random variables $X(t)$
indexed by a variable $t$ (commonly identified with time) that takes
value either in ${\mathbb{N}}$ or ${\mathbb{R}}^+$ (less frequently in
${\mathbb{R}}$). We indicate with $P(x,t)$ the probability that the
process take value $x$ at time $t$ (the probability density if $t$ is
continuous; I shall omit the subscript $X$ to avoid complicating the
notation), and with $P(x_2,t_2;x_1,t_1)$ the joint probability density
for the two variables $X(t_1)$ and $X(t_2)$. The multiple joint
probability density $P(x_1,t_1;\ldots,x_n,t_n)$ is defined
analogously. In the following, whenever possible, I shall use the
joint probability $P(x_2,t_2;x_1,t_1)$ to simplify the notation, but
all considerations hold for the more general multiple joint
probability.

Just as a stochastic variable is instantiated to a specific value
$x\in\Omega$ with a certain probability, so a stochastic process is
instantiated as a trajectory $X:{\mathbb{R}}\rightarrow\Omega$ (or
with a discrete series $X:{\mathbb{N}}\rightarrow\Omega$ if the
process is discrete). Each $X(t)$, for fixed $t$, is a stochastic
variable with a probability distribution that, in general, depends on
$t$. A stochastic process is \emph{stationary} if all these
distributions are the same, that is, $P(x,t)\equiv{P(x)}$,
or, equivalently, if
\begin{equation}
  P(x_1,t_1;x_2,t_2) =   P(x_1,t_1+\tau;x_2,t_2+\tau)
\end{equation}

In a stochastic process, there are two ways of computing averages: one can
compute the \emph{ensemble average} $\langle{X(t)}\rangle$, that is,
the average of the random variable $X(t)$, or the mean value along a
trajectory
\begin{equation}
  \bar{X} = \lim_{T\rightarrow\infty} \frac{1}{T} \int_0^T x(t)\,dt
\end{equation}
A process is \emph{ergodic} if the two coincide
\begin{equation}
  \langle X \rangle = \bar{X}
\end{equation}
Ergodicity is an important property for random walks: many times we are
interested in the characteristics of the motion of one individual, but
many of the equations that we shall use involve ensemble probabilities
based on a whole population. Ergodicity allows us to switch from one to
the other with impunity.

Note that in a stationary process the correlation
$\langle{X(t_1)}X(t_2)\rangle$ does not depend on $t_1$ and $t_2$
individually, but only on their difference $\tau=t_2-t_1$. Joint
probabilities are positive, symmetric
($P(x_1,t_1;x_2,t_2)=P(x_2,t_2;x_1,t_1)$) and normalized:
\begin{equation}
  \int\!\!\!\int_{\Omega^2} P(x_1,t_1;x_2,t_2)\,dx_1\,dx_2 = 1
\end{equation}
Joint probabilities can be reduced by integration
\begin{equation}
  P(x_1,t_1) = \int_\Omega P(x_1,t_1;x_2,t_2)\,dx_2
\end{equation}
and Bayes theorem can be extended to stochastic processes
\begin{equation}
  \label{extobayes}
  P(x_1,t_2 = \int_\Omega P(x_2,t_2|x_1,t_1)P(x_1,t_1)\,dx_1
\end{equation}
A process if \emph{Markov} if, for all $t_1<t_2<\cdots<t_n$,
\begin{equation}
  P(x_n,t_n|x_{n-1},t_{n-1};\ldots;x_1,t_1) = P(x_n,t_n|x_{n-1},t_{n-1})
\end{equation}
this entails that, at any time, the status of the process encodes all
the information necessary to make predictions about its future: it is
not necessary to know how the process reached that status. The Markov
property can be chained:
\begin{equation}
  \begin{aligned}
    P(x_1,t_1;x_1,t_2;x_3,t_3) &=  P(x_2,t_2;x_3,t_3|x_1,t_1) P(x_1,t_2) \\
                              &=  P(x_3,t_3|x_2,t_2)P(x_2,t_2|x_1,t_1)P(x_1,t_1)
  \end{aligned}
\end{equation}
Finally, it can be shown \cite{gardiner:85} that Markov processes must
satisfy the \emph{Chapman-Kolmogorov} equation:
\begin{equation}
  P(x_3,t_3|x_1,t_1) = \int_\Omega P(x_3,t_3|x_2,t_2)P(x_2,t_2|x_1,t_1)\,dx_2
\end{equation}
The characteristics of the Markov process is evidenced by the fact
that the probability of the transition
$(x_1,t_1)\rightarrow(x_2,t_2)\rightarrow(x_3,t_3)$ is the product of
the probabilities of the transitions $(x_1,t_1)\rightarrow(x_2,t_2)$
and $(x_2,t_2)\rightarrow(x_3,t_3)$, that is, the two transitions are
\emph{statistically independent}.

\subsection{Gaussian and Wiener processes}
\label{gausswie}
I shall provide here some details on two types of processes of
considerable importance for random walks and diffusion.

A stochastic process $X(t)$ is \emph{Gaussian} with zero mean if
$\langle{X(t)}\rangle=0$ and
\begin{equation}
  P(x_i,t_i) = \sqrt{\frac{A_{ii}}{2\pi}} \exp\Bigl( -\frac{1}{2} A_{ii} x^2 \Bigr)
\end{equation}
($A_{ii}>0$). The joint probability $P(x_1,t_1;\ldots;x_n,t_n)$ then
follows a multivariate Gaussian distribution
\begin{equation}
  P(x_1,t_1;\ldots;x_n,t_n) = \frac{\mbox{det}(\mathbf{A})^{1/2}}{(2\pi)^{n/2}} \exp\left[ -\frac{1}{2} \sum_{i,j=1}^n x_i A_{ij} x_j \right]
\end{equation}
Where $\mathbf{A}\in{\mathbb{R}}^{n\times{n}}$ is symmetric (strictly)
positive definite. The matrix $\mathbf{A}$ is a measure of the
covariance between two variables of the Gaussian process
\begin{equation}
  \langle X(t_i) X(t_j) \rangle = (\mathbf{A}^{-1})_{ij}
\end{equation}
(this is true since we assume zero mean). A process is uncorrelated if
$\langle{X(t_i)}X(t_j)\rangle=D\delta(t_i-t_j)$, in which case
$A_{ij}=D^{-1}\delta_{ij}$.

\bigskip

A \emph{Wiener process} $W$ is a process in which the variables $W(t)$
are real and with independent increments $W(t_2)-W(t_1)$ that follow a
Gaussian distribution. That is, they define a conditional probability
\begin{equation}
  P(w_2,t_2|w_1,t_1) = \frac{1}{\sigma\sqrt{2\pi(t_2-t_1)}} \exp\left[ -\frac{(w_2-w_1)^2}{2\sigma^2(t_2-t_1)} \right]
\end{equation}
from which the covariance can be computed
\begin{equation}
  \begin{aligned}
    \langle (W(t_2)-\langle{W}\rangle)(W(t_1)-\langle{W}\rangle)\rangle &= \langle (W(t_2)-W(0))(W(t_1)-W(0))\rangle \\
    &= \int_{-\infty}^\infty (w_2-w_0)\,dw_2 \int_{-\infty}^\infty\,dw_1 (w_1-w_0) P(w_2,t_2;w_1,t_1) \\
    &= \sigma^2 \min(t_1,t_2) + w_0^2
  \end{aligned}
\end{equation}
From this we get
\begin{equation}
  \label{booh}
  \langle W(t)^2 \rangle = \sigma^2 t + w_0^2
\end{equation}
Wiener processes are related to Gaussian processes, in particular to
uncorrelated (white) Gaussian processes. Let $X(t)$ be a Gaussian
process with $\langle{X(t_1)}X(t_2)\rangle=\sigma^2\delta(t_2-t_1)$,
and define a new stochastic process as the integral of $X(t)$:
\begin{equation}
  Y(t)=\int_0^t X(u)\,du
\end{equation}
then
\begin{equation}
  \begin{aligned}
    \langle Y(t_2) Y(t_1) \rangle &= \int_0^{t_2}\!\!\!\!\!du_2\int_0^{t_1}\!\!\!\!\!du_1 \langle{X(u_1)}X(u_2)\rangle \\
    &= \int_0^{t_2}\!\!\!\!\!du_2\int_0^{t_1}\!\!\!\!\!du_1 \delta(u_2-u_1)
  \end{aligned}
\end{equation}
By the properties of the Dirac function
\begin{equation}
  \int_0^{t_1}\!\!\!\!\!du_1 \delta(u_2-u_1) =
  \begin{cases}
    1 & 0<u_2<t_1 \\
    0 & \mbox{otherwise}
  \end{cases}
\end{equation}
Then
\begin{equation}
  \label{sponge}
  \langle Y(t_2) Y_(t_1) \rangle = \sigma^2 \min(t_2,t_1)
\end{equation}
which coincides with (\ref{booh}) for $w_1=0$. That is, the integral
of a Gaussian process is a Wiener process.

\end{document}